\documentclass[a4paper,12pt,floatfix]{report}
\usepackage[utf8]{inputenc}
\usepackage{amsmath, amssymb, amscd, amsthm, amsfonts,bm,subcaption,physics,mathtools}
\usepackage{graphicx}
\usepackage[labelfont=bf]{caption}
\usepackage{wrapfig} 
\usepackage{setspace} 
\usepackage{dsfont} 
\usepackage{geometry} 
\usepackage[most]{tcolorbox} 
\tcbuselibrary{breakable} 
\usepackage{xcolor}  
\usepackage{sectsty}  
\usepackage{cancel}   
\usepackage{enumitem} 
\usepackage{subfiles} 
\usepackage{placeins} 
\allsectionsfont{\sffamily}
\definecolor{linkpurple}{RGB}{152,71,155}
\definecolor{urlblue}{RGB}{20, 86, 128}
\usepackage{hyperref}
\hypersetup
{
   colorlinks=true,
    linkcolor=linkpurple,
    filecolor=magenta,      
    urlcolor=urlblue,
    citecolor=urlblue
}
\usepackage{fancyhdr}
\numberwithin{equation}{section}

\def\bec{\begin{center}}
\def\eec{\end{center}}
\def\beq{\begin{equation}}
\def\eeq{\end{equation}}
\def\bea{\begin{eqnarray}}
\def\eea{\end{eqnarray}}

\begin{document}

\begin{titlepage}

\begin{center}
    \LARGE
    \textbf{Stabilizing Quantum Simulators Of Gauge Theories Against $1/f$ Noise}

    \vspace{1cm}
    \Large
    \textbf{Bhavik Kumar}
    \vspace{1cm}
    
    \large
    \textit{A dissertation submitted for the partial fulfilment of
    BS-MS dual degree in Science}
    
    \vspace{3.5cm}

    \includegraphics[width=8cm]{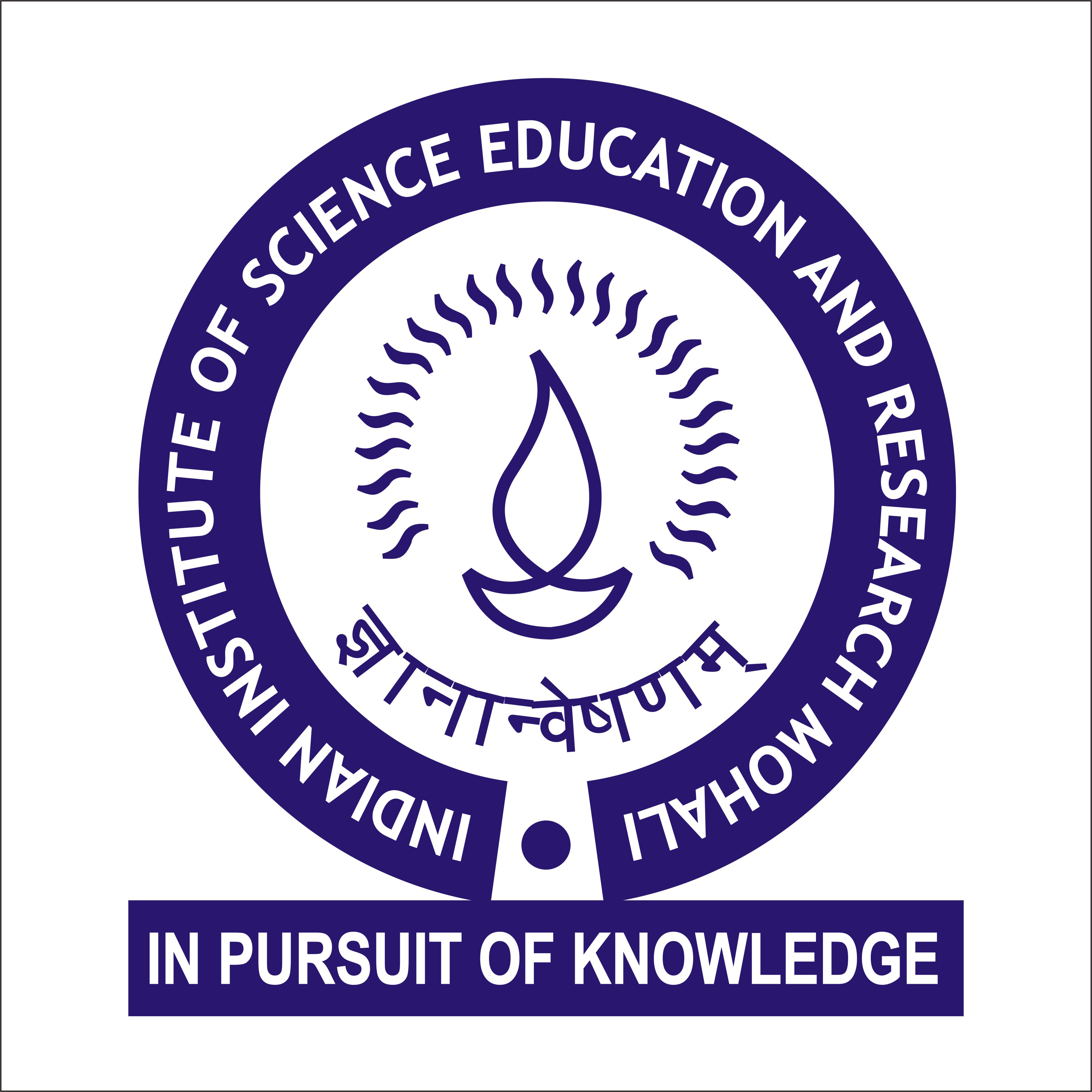}

    \large
    \textbf{Indian Institute of Science Education and Research, Mohali}\\
    \large
    \textbf{May 2, 2023}
\end{center}

\end{titlepage}
\pagenumbering{Roman}
\newpage

\begin{center}
    \textbf{\Large Certificate of Examination}
\end{center}

This is to certify that the dissertation titled \textbf{"Stabilizing Quantum Simulators Of Gauge Theories Against $1/f$ Noise"} submitted by \textbf{Bhavik Kumar} (Reg. No. MS18098) for the partial fulfillment of BS- MS Dual Degree programme of the institute, has been examined by the thesis committee duly appointed by the institute. The committee finds the work done by the candidate satisfactory and recommends that the report be accepted.

\vspace{4cm}

Dr.Manabendra Nath Bera \hspace{1.5cm} Dr.Sanjeev Kumar \hspace{1.5cm} Dr.Ambresh Shivaji

\vspace{4cm}

\begin{flushright}
    Dr. Manabendra Nath Bera
    \\
    (Local Supervisor)
    \\
    \vspace{4cm}
    Dated: May 2, 2023
\end{flushright}

\cleardoublepage
\begin{center}
    \textbf{\Large Declaration}
\end{center}
The work presented in this dissertation has been carried out by me under the guidance of Dr. Jad C. Halimeh at the Ludwig Maximilian University of Munich and Dr. Manabendra Nath Bera at the Indian Institute of Science Education and Research, Mohali.

\vspace{0.4cm}

This work has not been submitted in part or in full for a degree, a diploma, or a fellowship to any other university or institute. Whenever contributions of others are involved, every effort is made to indicate this clearly, with due acknowledgement of collaborative research and discussions. This thesis is a bonafide record of original work done by me and all sources listed within have been detailed in the bibliography.

\vspace{2cm}

\begin{flushright}
Bhavik Kumar
\\
(Candidate)
\\
Dated: May 2, 2023
\end{flushright}

In my capacity as the supervisor of the candidates project work, I certify that the above statements by the candidate are true to the best of my knowledge.

\vspace{2cm}

\begin{flushright}
Dr. Manabendra Nath Bera
\\
(Local Supervisor)
\\
Dated: May 2, 2023
\end{flushright}

\cleardoublepage
\begin{center}
\textbf{\Large Acknowledgements}
\end{center}

 I would like to express my deepest gratitude to my external supervisors, Dr. Jad C. Halimeh and Prof. Philipp Hauke for their guidance, support, and encouragement throughout the entire process of completing this thesis. As Sir Isaac Newton once famously remarked \textit{“If I have seen further, it is by standing on the shoulders of giants.”} Every result accomplished in this thesis is built on the seminal work carried out by them in the field of quantum simulation of lattice gauge theories. Their insights, feedback, and expert knowledge have been invaluable, and I am truly grateful for the time and effort they have devoted to my research. 
\\
\newline
I would also like to thank my local supervisor Dr. Manabendra Nath Bera and the members of my thesis committee, Dr. Sanjeev Kumar and Dr. Ambresh Shivaji for their valuable feedback, suggestions, and facilitation of the evaluation of this thesis. I would like to acknowledge the institute resources that have been helpful in my research and provided me with the opportunity to carry out this work.
\\
\newline
I am also grateful to my friends and family for their unwavering support and encouragement throughout my academic journey. Your love, understanding, and encouragement have been instrumental in my success, and I am honored to have you all in my life.
\\
\newline
In the end, I would like to thank me for believing in me and putting up with myself.
\newpage
\clearpage
\begin{center}
    \textbf{\Large Abstract}
\end{center}

    This work investigates the application of quantum simulation in the ongoing "second" quantum revolution, which employs various synthetic quantum matter platforms, such as ultracold atoms in optical lattices, Rydberg atoms, and superconducting qubits, to realize exotic condensed matter and particle physics phenomena with high precision and control. Gauge theories are of particular interest in modern quantum simulators as they offer a new probe of high-energy physics on low-energy tabletop devices. However, to accurately model gauge-theory phenomena on a quantum simulator, stabilizing the underlying gauge symmetry is crucial. Through this thesis we demonstrate that a recently developed experimentally feasible scheme based on linear gauge protection, initially devised to protect against coherent gauge breaking errors, can also be used to suppress incoherent errors arising from $1/f^{\beta}$ noise prominent in various quantum simulation platforms. The Bloch-Redfield formalism is introduced to model gauge violations arising due to these incoherent errors given the noise power spectrum of the environment. The efficacy of linear gauge protection in stabilizing salient features of gauge theories in quantum simulators, such as gauge invariance and exotic far from equilibrium phenomenon focusing on disorder-free localization, and quantum many-body scars against $1/f^{\beta}$ noise sources, is illustrated. These results are immediately applicable in modern analog quantum simulators and digital NISQ devices, paving the way for further development in the field of quantum simulation of lattice gauge theories.


\newpage
   

\addcontentsline{toc}{chapter}{\textbf{List of Figures}}
\listoffigures

\newpage

\addcontentsline{toc}{chapter}{\textbf{List of Tables}}
\listoftables
\newpage
\tableofcontents
\newpage
\pagenumbering{arabic}




\chapter{Introduction}\label{chap:chapter1}

\section{Quantum simulation}
\renewcommand\thefigure{\thechapter.\arabic{figure}}    
\renewcommand{\theequation}{\thechapter.\arabic{equation}}
\setcounter{equation}{0}
\setcounter{figure}{0}
\newcommand{\chapquote}[3]{\begin{quotation} \textit{#1} \end{quotation} \begin{flushright} - #2, \textit{#3}\end{flushright} }
\chapquote{``Nature isn't classical, dammit, and if you want to make a simulation of nature, you'd better make it quantum mechanical, and by golly, it's a wonderful
problem, because it doesn't look so easy."}{Feynman}{1982}
These are the inspiring words of his seminal 1982 article\cite{Feynman1982}, where he suggested that the complexities of quantum-many body physics might be addressed by \textit{"simulation"}.  It was evident in the early 1980s that simulating quantum mechanics was a very difficult challenge. The enormous amount of computer memory required to store the quantum state of a big physical system is obviously a problem. The number of parameters that describe this state increases exponentially with the system size typically referred to as the number of particles or degrees of freedom in the system. If we perform a numerical (classical) simulation of a quantum system, a linear increase in the degrees of freedom causes an exponential increase
in computational complexity. In concrete terms, if we want to store the state of a spin-1/2 chain with 40 spins, we would require about 4 terabytes of memory. Also, the number of operations needed to simulate the system's temporal evolution grows exponentially in proportion to the system's size. This exponential explosion cannot be avoided unless approximation methods (such as Monte Carlo) are applied.
\\
\newline
   Monte Carlo algorithms evaluate the system's phase space
and the integrals defined on it, such as partition functions, correlators, and expectation values
of the observables in a polynomial time with respect to the number
of components of the system. However, these methods prove to be accurate only when the functions within integrals vary slowly and do not
change significantly. In general, this does not occur in many quantum systems, especially for strongly correlated fermionic systems in condensed matter physics
and for fermionic field theories in finite-density regimes. So, in this sense,
classical simulations are severely limited by this problem, which is known in the literature as the sign problem. As a result, even for today's supercomputers, simulating quantum systems, in general, remains challenging. Richard Feynman envisioned a computer whose component elements are governed by quantum dynamics generated by a desired Hamiltonian by constructing a well-controlled system from the bottom up in order to solve this problem. Therefore, the characteristics of this quantum-engineered system can be measured, shedding light on previously unknown or difficult-to-calculate aspects of quantum many-body models. This apparatus is now known as a quantum simulator. Since classical simulations of lattice gauge theories are notoriously hard to implement, as a result, this has sparked immense theoretical and experimental efforts in the simulation of real-time dynamics of LGTs on these quantum simulators. The concept behind a QS can be described in general terms as follows:
\\
\newline
    Consider a system hamiltonian $H_{sys}$ to be simulated that can be suitably mapped to a controllable quantum system $H_{con}$
    
    \[H_{sys}\Longleftrightarrow H_{con}\]
    \begin{figure}[ht!]
    \centering
    \includegraphics[width=8cm]{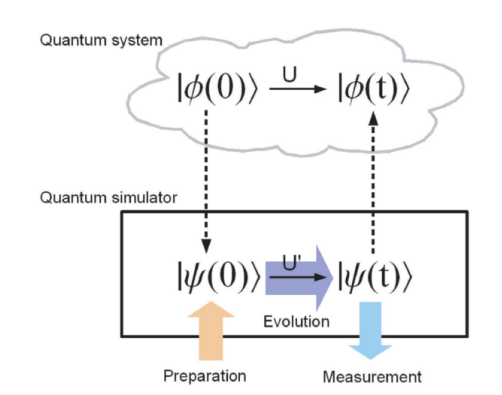}
    \caption{A schematic representation of a quantum system and a corresponding quantum simulator \cite{Georgescu_review}.}
    
\end{figure}
    Therefore, if there exists a known mapping between a system and simulator, an operator $M$ can be constructed such that we can map our initial state of the system $|\phi(0)\rangle$ to that of the simulator by taking $|\psi(0)\rangle=M|\phi(0)\rangle $ (also illustrated in fig 1). After executing the simulation procedure for time $t$, $|\psi(t)\rangle$ can be mapped back to $|\phi(t)\rangle$ via $M^{-1}$ and one can write $H_{sys}=MH_{con}M^{-1}$. The choice of mapping depends on the type of simulation and the degree of similarity between the dynamics of the system and the simulator. Quantum simulations are usually realized on \textit{"Synthetic quantum matter"}; i.e., these systems can comprise of various degrees of freedom, whose interaction between them can be microscopically controlled by external knobs up to a high degree of precision. The abovementioned approach is usually tailored for analog quantum simulators, which typically consist of ultracold atoms in optical lattices, superconducting qubits, trapped ion systems, nuclear spins, and photonic devices.\cite{Bloch2008,Georgescu_review}

    Quantum simulators can also be seen as special-purpose quantum computers implementing digital quantum evolution. The unitary evolution is implemented by a sequence of short quantum operations realized through one-qubit or two-qubit gates. This is achieved by decomposing the time evolution operator $e^{-iHt}$ using the trotter expansion.
    
    \[\mathrm{e}^{-i H t} \simeq\left(\prod_{\kappa=1}^M \mathrm{e}^{-i H_\kappa t / n}\right)^n, \quad H=\sum_{\kappa=1}^M H_\kappa,\]

    Thus the time evolution can be described as a sequence of local gates acting on a few qubits in terms of local interactions $H_{k}$.

\section{Quantum simulation of lattice gauge theories}
Recently there has been a lot of interest in the quantum simulation of lattice gauge theories\cite{Rothe_book} on experimental setups such as ultracold atoms in optical lattices and superconducting qubits due to their fundamental importance in high energy and condensed matter physics\cite{Dalmonte_review, Zohar_review,aidelsburger2021cold}. It offers a powerful approach to understanding the behavior of strongly interacting systems. The development of such simulations has the potential to shed light on the dynamics of phenomena such as quantum chromodynamics, which is the theory of strong nuclear interactions. Moreover, one of the most active fields of theoretical and experimental physics is the out-of-equilibrium dynamics of these quantum-field theories.
Gauge theories with both dynamical matter and gauge fields are particularly interesting in this endeavor. The characteristic property of gauge theories is their gauge symmetry \cite{Weinberg_book,Gattringer_book,Zee_book}, which imposes local constraints that enforce specific configurations of matter and electric fields, such as Gauss's law from quantum electrodynamics. These simulations can also offer a glimpse into how to benchmark quantum simulators since we need to have precise control of the aforementioned local constraints. Therefore, if one can control a gauge theory hamiltonian, one can simulate, with ease, other generic many-body systems.

\begin{figure}[t!]
    \centering
    \includegraphics{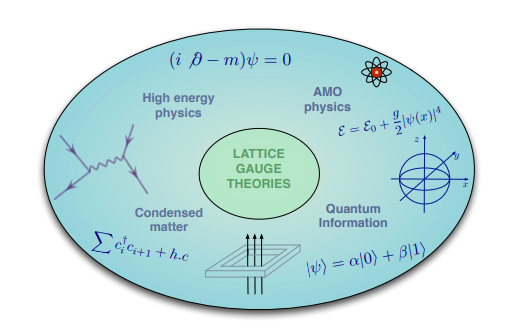}
    \caption{Quantum Lattice gauge theories lie in between different disciplines, as they describe systems in high-energy physics;
atomic, molecular and optical (AMO) physics; condensed matter physics and quantum information science. Fig adapted from \cite{Zohar_review}}
    \label{fig:my_label}
\end{figure}
\FloatBarrier

 It is also worth briefly describing a notable experiment performed by Yang et al. \cite{Yang:2020Science}, where they used a Bose-Hubbard simulator with 71 sites; apart from reaching a remarkable system size, this experiment also demonstrated the first experimental quantification of gauge invariance by simulating the real-time dynamics of a 1D LGT. In this experiment, they mapped the LGT into a system of ultracold bosons in a 1D optical lattice with 71 sites. Using high-fidelity operations, they were able to quantify Gauss law violation by extracting probabilities of
locally gauge-invariant states from correlated atom occupations.
\begin{figure}[t!]
    \centering
    \includegraphics{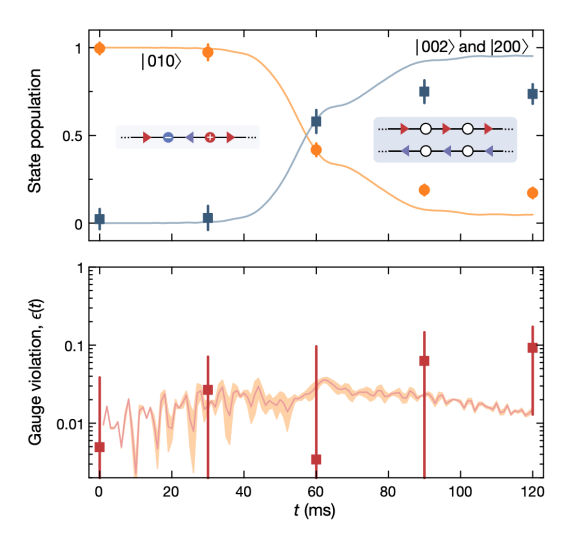}
    \caption{Plots adapted from Ref.\cite{Yang:2020Science}, illustrating the quantification of gauge invariance in an ultracold atom quantum simulator with 71 sites. The upper plot shows the population of gauge-invariant states. The probabilities are then calculated to observe gauge violation (bottom plot) as a function of time $\epsilon(t)=1-P(\text{sum of probabilities of gauge invariant states)}$}
    \label{fig:my_label}
\end{figure}

Another relevant experiment in the field of quantum simulation was performed recently
by Bernien et al.\cite{Bernien2017} In this work, the quantum many-body dynamics of an Ising-type
Hamiltonian was investigated on a 51-qubit quantum simulator based on Rydberg atoms realized by tuning 
the inter-atom distance, one can achieve the regime of the Rydberg
blockade. This prevents neighboring excitation of the atoms, making the model kinetically constrained.
This gives rise to anomalously slow dynamics in local observables, as a result, they observed persistent oscillations well beyond the relevant timescales. Although not envisioned
with this purpose, this experiment turned out to be closely related to LGTs. The authors also found that the constrained dynamics of the simulated quantum spin chains in the experiments exactly map onto
that of 1D LGTs.
\begin{figure}[t!]
    \centering
    \includegraphics{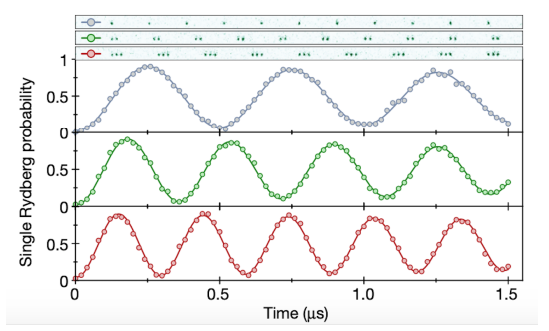}
    \caption{Observation of persistent oscillations in the many-body dynamics of a 51-qubit quantum simulation experiment realized by utilizing the properties of strongly interacting Rydberg
atoms\cite{Bernien2017}. One can also show that the
dynamics of such systems can be mapped onto the dynamics of certain 1D LGTs.}
    \label{fig:my_label}
\end{figure}
\\
\newline
In this thesis, we will mostly focus on the reliability of the quantum simulations of these gauge theories, i.e., within some prescribed error,
one should be assured that the observed physics of
the quantum simulator corresponds faithfully to that of the ideal
gauge theory Hamiltonian \cite{Hauke2012,Halimeh2020a}. Since we are in the Noisy Intermediate Quantum Era (NISQ), where fault-tolerant quantum computers are still out of reach, it becomes of central importance to design experimentally feasible error-mitigation strategies that ensure the reliability of current quantum simulators. 
\\
\newline
 The realization of LGTs still represents a big challenge for
quantum simulation, since enforcing gauge invariance is a very difficult task, gauge-breaking terms unavoidably arise due to higher orders in the perturbative mapping or experimental imperfections \cite{Halimeh2020a}. These terms allow for processes driving the system dynamics out of the \textit{physical gauge sector} of Gauss's law, in which it should stay in an ideal scenario where such terms are not present. Violations of gauge symmetries, even when they are perturbative in strength, can be quite detrimental to gauge-theory quantum simulations, leading to gauge-noninvariant dynamics that cannot be directly related to the target model \cite{Halimeh2020b,Halimeh2020c}. For example, they can generate a photon mass in QED which reduces the infinite range coulomb law to a Yukawa potential. Even if the gauge-breaking terms are small, a massless photon can emerge in a renormalized gauge theory. Also, due to the perturbative breaking of the conservation laws in the quantum-many-body system, their equilibrating dynamics are also strongly affected\cite{Zhou2021,halimeh2020staircase}.

Various methods have been proposed to suppress coherent gauge-breaking errors \cite{Hauke2013,Stannigel2014,Kuehn2014,Yang2016,Dutta2017,Halimeh2020a,Kasper2021nonabelian,Halimeh2020e,Halimeh2021stabilizing,Halimeh2021gauge,Halimeh_BriefReview}; however, going beyond these unitary gauge-breaking errors, comparatively little work has been done in the way of suppressing incoherent errors due to decoherence. Its mitigation is crucial to achieving reliable quantum simulators since decoherence poses a major roadblock to
achieving long evolution times in quantum simulations of quantum many-body models in general, whose key
properties of quantum entanglement and superposition
are particularly sensitive to interactions with the environment. Prominent examples of the detrimental effects
of decoherence on quantum many-body systems include $1/f$ -noise in superconducting quantum interference devices (SQUIDs), CMB-photon noise in superconducting
cryogenic detectors, and thermomechanical motion in microwave cavity interferometers.\cite{Bylander2011,Wang2015,Kumar2016}


\section{Outline}
This thesis is organized as follows: In the second chapter, we introduce abelian gauge theories; in particular, we will briefly cover the passage from the considered (continuum) gauge theory to a discrete version
of it: this is generally achieved based on the various works developed in
the context of lattice gauge theories (LGTs). 
\newline
\\
In Chapter 3, one of the main strategies for engineering gauge symmetries in quantum simulation is introduced, consisting of imposing energy penalty to gauge variant states. We review the recently developed scheme of \textit{linear gauge protection} that entails the addition of the linear sum of local generators or pseudo-generators (for certain gauge theories whose full generator might be too challenging to experimentally realize) of symmetry consisting of single-body or at most two body terms to controllably protect and suppress leakage out of our target gauge-invariant sector against \textbf{unitary gauge breaking errors}, up to exponentially long times and for local error terms independent of system size \cite{Halimeh2020e,Halimeh2021stabilizing,Halimeh2021gauge}. Since quantum link formulations of these gauge theories can be naturally mapped to a Bose - Hubbard Hamiltonian, which can be implemented with high precision in ultracold atoms in optical lattices, this scheme is experimentally feasible and can be implemented in analog simulators. It has also been demonstrated that this scheme yields robust results in digital devices, as the protection parameters can be implemented with single qubit gates \cite{Halimeh2021gauge}.
\newline
\\
Furthermore, we introduce the original contribution and the crux of this thesis based on the work done in \cite{kumar2022suppression}, where we finally address the open question of whether linear gauge protection, initially devised to protect against coherent errors, can be employed to protect against \textbf{incoherent errors} due to noise in an experiment. We introduce the Bloch-Redfield formalism \cite{cohen1992atom,breuer2002theory} for an open-quantum many-body system to probe the pernicious effects of $1/f^{\beta}$ noise. Furthermore, we show that this protection scheme suppresses the growth of the gauge violations due to incoherent errors with spectral form $1/f^\beta$ ($\beta>0$) as $1/V^\beta$, in $\mathrm{U}(1)$ quantum link model \cite{Chandrasekharan1997,Wiese_review,Hauke2013,Kasper2017} and $\mathbb{Z}_2$ lattice gauge theories.
\\
\newline
In chapter 4, the power of linear gauge protection is demonstrated in stabilizing and enhancing features of weak and strong ergodicity breaking, far from equilibrium phenomena occurring in gauge theories, namely \textbf{quantum many-body scars} and \textbf{disorder-free localization} \cite{Desaules2022a,Smith2017,Turner2018,Halimeh2021stabilizingDFL,Halimeh2021enhancing,Halimeh2021stabilizing} against incoherent errors arising from $1/f^\beta$ noise. This can be attributed to the fact that the dynamics of local observables are also protected within their target superselection sectors and stay close to their ideal theory dynamics.
\newline
\\
Chapter 5 provides a conclusion and an outlook highlighting the scope for future work.
\chapter{Abelian Gauge Theories}\label{chap:chapter2}
In this chapter, abelian gauge theories are discussed. We Introduce free Dirac and electromagnetic fields and promote their corresponding free field theories to a gauge theory. Then we define the covariant derivative to introduce minimal coupling in terms of the comparator or the parallel transporter. This quantity will then be useful to define U(1) gauge theory coupled to fermionic matter in (1+1) dimensions and its corresponding gauge invariance.
Then a quantization of fields is presented, paying special attention to local gauge transformations and how they act on fields. 
\section{The Dirac Field}
Dirac Fields are four component fields $\Psi(x)$ which describe the evolution of spin-1/2 particles, the four degrees of freedom being spin and helicity. The dynamics follows from the lagrangian density :
\begin{equation}
    \mathcal{L}_D=\bar{\Psi}(x)(i\not\partial-m) \Psi(x),
\end{equation}
Where $\bar{\Psi}=\Psi(x)\gamma^{0}$ with the Dirac matrices in the chiral representation. Hence we obtain the Dirac equation, which is obtained as:
\begin{equation}
    (i\not\partial-m)\Psi(x)=0
\end{equation}
We can identify the first two components of the Dirac spinor with the left-handed
components, and the two others with the right-handed ones:
\begin{equation}
    \Psi(x)=\left(\begin{array}{l}\psi_{L} \\ \psi_{R}\end{array}\right)
\end{equation}
Putting $m = 0$ we obtain two uncoupled equations whose solutions $\psi_{L}$ and
$\psi_{R}$  are massless left- and right-handed spinor fields. For $m = 0$,
plane-wave solutions with positive and negative energy are, respectively, in the
form.
\begin{equation}
  \Psi(x)=u(p)e^{-ipx}
\end{equation}
\begin{equation}
    \Psi(x)=v(p)e^{ipx}
\end{equation}
Where $p$ is the four-momentum, $v(p)$ and $u(p)$ are the 4-component spinors . A general solution is given by a superposition of plane waves. The superposition includes both positive and negative energy solutions, which are 
multiplied by coefficients $a_{s, \mathbf{p}}$ and $b_{s, \mathbf{p}}$, where the summation over $s$ runs over 
\begin{equation}
    \Psi(x)=\left.\int \frac{\mathrm{d}^3 \mathbf{p}}{(2 \pi)^3 \sqrt{2 E_{\mathbf{p}}}} \sum_{s=1,2}\left[a_{s, \mathbf{p}} u_s(p) \mathrm{e}^{-i p x}+b_{s, \mathbf{p}}^* v_s(p) \mathrm{e}^{i p x}\right]\right|_{p^0=E_{\mathbf{p}}}
\end{equation}
\section{Realizing minimal coupling over the electromagnetic field}
The four-vector potential $A_{\mu}$ whose first component is the scalar potential, and spatial components are the
vector potential determines the electromagnetic field. The electromagnetic tensor can be written as 
$$
F_{\mu \nu}=\partial_{\mu} A_{\nu}-\partial_{\nu} A_{\mu}
$$



The electromagnetic tensor is invariant under gauge transformations of the potential $A_{\mu}$ which, for a  given smooth function $\phi(x)$, can be identified as

\begin{equation}\label{eq3}
\begin{aligned}
A_{\mu} & \rightarrow A_{\mu}-\partial_{\mu} \phi, \\
F_{\mu \nu} & \rightarrow \partial_{\mu} A_{\nu}-\partial_{\mu} \partial_{\nu} \phi-\partial_{\nu} A_{\mu}+\partial_{\nu} \partial_{\mu} \phi=F_{\mu \nu} .
\end{aligned}
\end{equation}

Therefore, Lagrangian and Hamiltonian densities of the free electromagnetic field are invariant under gauge transformations.

$$
\begin{aligned}
\mathcal{L}_{E M} & =-\frac{1}{4} F_{\mu \nu} F^{\mu \nu}=\frac{1}{2}\left(\mathbf{E}^{2}-\mathbf{B}^{2}\right), \\
H_{E M} & =\frac{1}{2}\left(\mathbf{E}^{2}+\mathbf{B}^{2}\right)
\end{aligned}
$$


We now want to write a Lagrangian describing the interaction between the electromagnetic and Dirac fields. Let us start by writing the Dirac Lagrangian,

$$
\mathcal{L}_{D}=\bar{\Psi}(x)(i \not \partial-m) \Psi(x)
$$
One can see that it is symmetric under $U(1)$ global transformations of fields:  $\alpha$ is chosen to be a real number that does not depend on $x$. 

$$
\begin{aligned}
& \Psi(x) \rightarrow \Psi^{\prime}(x)=\Psi(x) \mathrm{e}^{i \alpha} \\
& \bar{\Psi}(x) \rightarrow \bar{\Psi}^{\prime}(x)=\bar{\Psi}(x) \mathrm{e}^{-i \alpha}
\end{aligned}
$$

This implies that $\mathcal{L}\left(\Psi^{\prime}, \bar{\Psi}^{\prime}\right)=\mathcal{L}(\Psi, \bar{\Psi})$, therefore  the free Dirac field is characterized by a global $U(1)$ symmetry.

The coupling of the Dirac matter field with the electromagnetic field is a consequence of local $U(1)$ transformations that  act on fields $\Psi$ as follows:

$$
\begin{aligned}
& \Psi(x) \rightarrow \Psi^{\prime}(x)=\Psi(x) \mathrm{e}^{i \alpha(x)}, \\
& \bar{\Psi}(x) \rightarrow \bar{\Psi}^{\prime}(x)=\bar{\Psi}(x) \mathrm{e}^{-i \alpha(x)} .
\end{aligned}
$$

in which $\alpha(x)$ is a real function. In the free Lagrangian, the mass term is invariant even under local transformations, while the kinetic term is not due to the presence of derivatives. In order to restore the symmetry, the derivative is replaced by the so-called covariant derivative $D_{\mu}$. Introducing a vector field, which we identify with the four-vector potential $A_{\mu}$, called the gauge field, in terms of which the covariant derivative can be defined as:

\begin{equation}\label{eq2}
D_{\mu}=\partial_{\mu}+i A_{\mu}    
\end{equation}

 We can now replace the partial derivative $\partial_{\mu}$ in the Dirac Lagrangian with the covariant derivative, implementing minimal coupling between the electromagnetic and Dirac fields with the following Local gauge transformations:

$$
\mathcal{L}_{M C D}=\bar{\Psi}(x)(i \not D-m) \Psi(x)
$$

$$
\begin{aligned}
\Psi(x) & \rightarrow \Psi^{\prime}(x)=\Psi(x) \mathrm{e}^{i \alpha(x)}, \\
D_{\mu} & \rightarrow D_{\mu}^{\prime}=\partial_{\mu}+i\left(A_{\mu}-\partial_{\mu} \alpha(x)\right) .
\end{aligned}
$$

Therefore the covariant derivative transforms as;

$$
\begin{aligned}
\left(D_{\mu} \Psi(x)\right)^{\prime}= & \left(\partial_{\mu}+i\left(A_{\mu}-\partial_{\mu} \alpha(x)\right)\left(\Psi(x) \mathrm{e}^{i \alpha(x)}\right)\right. \\
= & \mathrm{e}^{i \alpha(x)} \partial_{\mu} \Psi(x)+i \Psi(x) \mathrm{e}^{i \alpha(x)} \partial_{\mu} \alpha(x) \\
& +i A_{\mu} \Psi(x) \mathrm{e}^{i \alpha(x)}-i \Psi(x) \mathrm{e}^{i \alpha(x)} \partial_{\mu} \alpha(x) \\
= & \mathrm{e}^{i \alpha(x)}\left(\partial_{\mu}+i A_{\mu}\right) \Psi(x) \\
= & \mathrm{e}^{i \alpha(x)} D_{\mu} \Psi(x)
\end{aligned}
$$

$$
\bar{\Psi}(x) D_{\mu} \Psi(x) \stackrel{U(1)_{l o c}}{\longrightarrow} \bar{\Psi}^{\prime}(x) D_{\mu}^{\prime} \Psi^{\prime}(x)=\bar{\Psi}(x) D_{\mu} \Psi(x) .
$$

In this way, we introduce an interacting term between two fields by imposing the local symmetry of the model under $U(1)$ transformations leading to a gauge-invariant Lagrangian which reads-

$$
\mathcal{L}=\bar{\Psi}(x)(i \not D-m) \Psi(x)-\frac{1}{4} F_{\mu \nu} F^{\mu \nu}
$$

\subsection{The comparator U}
One can realize minimal coupling in an alternative manner from a geometrical perspective; we introduce this formalism because this procedure will be used to define a quantity which will be useful in the construction of gauge fields on a lattice. Consider the derivative of a Dirac field explicitly along the direction identified by the unit vector $\hat{\eta}$ :

$$
\partial_{\hat{\eta}} \Psi(x) \equiv \lim _{\epsilon \rightarrow 0} \frac{\Psi(x+\epsilon \hat{\eta})-\Psi(x)}{\epsilon}=\hat{\eta}^{\mu} \partial_{\mu} \Psi(x) .
$$

Now we know that the transformation properties of the differential operator are the same as those of the fields. However, when we transform $\Psi(x)$ with a local $U(1)$ transformation, $\partial_{\hat{\eta}} \Psi(x)$ does not transform in the same way as $\Psi(x)$ : Since we have

$$
\left(\partial_{\hat{\eta}} \Psi(x)\right)^{\prime}=\lim _{\epsilon \rightarrow 0} \frac{\mathrm{e}^{i \alpha(x+\epsilon \hat{\eta})} \Psi(x+\epsilon \hat{\eta})-\mathrm{e}^{i \alpha(x)} \Psi(x)}{\epsilon} .
$$

To circumvent this problem, we introduce a quantity indicated $U(x, y)$, called the comparator, which transforms under local $U(1)$ operations as \cite{Wiese_review} :

$$
U(x, y) \rightarrow \mathrm{e}^{i \alpha(x)} U(x, y) \mathrm{e}^{-i \alpha(y)} .
$$

One can also compute,

$$
\begin{gathered}
U(x, y) \Psi(y) \rightarrow \mathrm{e}^{i \alpha(x)} U(x, y) \mathrm{e}^{-i \alpha(y)} \mathrm{e}^{i \alpha(y)} \Psi(y) \\
=\mathrm{e}^{i \alpha(x)} U(x, y) \Psi(y) .
\end{gathered}
$$

Defining the quantity 

\begin{equation}\label{eq1}
    D_{\hat{\eta}} \Psi(x)=\lim _{\epsilon \rightarrow 0} \frac{U(x, x+\epsilon \hat{\eta}) \Psi(x+\epsilon \hat{\eta})-\Psi(x)}{\epsilon},
\end{equation}

Which we are calling the covariant derivative here; in particular, by choosing a unit vector $\hat{\mu}$ aligned with a reference axis of the Minkowski space, we obtain

$$
D_{\mu} \Psi(x)=\lim _{\epsilon \rightarrow 0} \frac{U(x, x+\epsilon \hat{\mu}) \Psi(x+\epsilon \hat{\mu})-\Psi(x)}{\epsilon} .
$$

We can immediately see that $D_{\hat{\eta}} \Psi(x)$ defined in \eqref{eq1} transforms like $\Psi(x)$, since

$$
\begin{aligned}
\left(D_{\hat{\eta}} \Psi(x)\right)^{\prime} & =\lim _{\epsilon \rightarrow 0} \frac{\mathrm{e}^{i \alpha(x)}(U(x, x+\epsilon \hat{\eta}) \Psi(x+\epsilon \hat{\eta})-\Psi(x))}{\epsilon} \\
& =\mathrm{e}^{i \alpha(x)} \lim _{\epsilon \rightarrow 0} \frac{U(x, x+\epsilon \hat{\eta}) \Psi(x+\epsilon \hat{\eta})-\Psi(x)}{\epsilon} \\
& =\mathrm{e}^{i \alpha(x)} D_{\hat{\eta}} \Psi(x),
\end{aligned}
$$

Therefore it is a gauge-covariant quantity. Now we show that the covariant directional derivative $D_{\mu}$ coincides with the covariant derivative defined in \eqref{eq2} with the gauge field $A_{\mu}$. Under the assumption that $U(x, y)$ is unitary, there exists a function, $\phi(x, y)$, such that $U(x, y)=\mathrm{e}^{i \phi(x, y)}$, and let us impose that $U(x, x)=1$, so $U^{-1}(x, y)=U(y, x)$. We consider its derivatives with respect to $y$ and call them $\partial \phi(x, y) / \partial y^{\mu}=\tilde{A}_{\mu}$. Therefore a first order approximation in $\epsilon$ of $U(x, x+\hat{\eta} \epsilon)$ can be written down as:

$$
U(x, x+\epsilon \hat{\eta}) \simeq 1+i \epsilon \hat{\eta}^{\mu} \tilde{A}_{\mu}
$$

The covariant derivative can now be written as follows:

$$
\begin{aligned}
D_{\hat{\eta}} \Psi(x) & =\lim _{\epsilon \rightarrow 0} \frac{\left(1+i \epsilon \hat{\eta}^{\mu} \tilde{A}_{\mu}\right) \Psi(x+\epsilon \hat{\eta})-\Psi(x)}{\epsilon} \\
& =\lim _{\epsilon \rightarrow 0} \frac{\Psi(x+\epsilon \hat{\eta})-\Psi(x)+i \epsilon \hat{\eta}^{\mu} \tilde{A}_{\mu} \Psi(x+\epsilon \hat{\eta})}{\epsilon} \\
& =\hat{\eta}^{\mu}\left(\partial_{\mu}+i \tilde{A}_{\mu}\right) \Psi(x),
\end{aligned}
$$

Therefore the two definitions for the covariant derivative are equivalent. With the introduction of the field $\tilde{A}_{\mu}$ the form of $U(x, y)$ is 

$$
U(x, y)=\exp \left\{i \int_{x}^{y} \mathrm{~d} x^{\mu} \tilde{A}_{\mu}\right\} .
$$

 Looking at the transformation properties of $U(x, x+\epsilon \hat{\eta})$, and also neglecting second-order terms, one can obtain.

$$
\begin{aligned}
U(x, x+\epsilon \hat{\eta}) & \rightarrow \mathrm{e}^{i \alpha(x)} U(x, x+\epsilon \hat{\eta}) \mathrm{e}^{-i \alpha(x+\epsilon \hat{\eta})} \\
& \simeq \mathrm{e}^{i \alpha(x)}\left(1+i \epsilon \hat{\eta}^{\mu} \tilde{A}_{\mu}\right)\left(1-i \epsilon \hat{\eta}^{\mu} \partial_{\mu} \alpha(x)\right) \mathrm{e}^{-i \alpha(x)} \\
& =1+i \epsilon \hat{\eta}^{\mu}\left(\tilde{A}_{\mu}-\partial_{\mu} \alpha(x)\right)+O\left(\epsilon^{2}\right)
\end{aligned}
$$

Therefore we can infer that transforming the comparator with is equivalent to transforming the field $A_{\mu}$ with \eqref{eq3}. We can, therefore, identify the field that appears above with the four-vector potential $A_{\mu}$.
\section{Quantization of the Dirac and electromagnetic field}
If we are given the Dirac field $\Psi(x)$ with the Lagrangian density $\mathcal{L}=\bar{\Psi}\left(i \gamma^{\mu} \partial_{\mu}-m\right) \Psi$, its canonically conjugate momentum can be defined as follows:

$$
\Pi_{\Psi}(x)=\frac{\partial \mathcal{L}}{\partial\left(\partial_{0} \Psi(x)\right)}=i \Psi^{\dagger}(x)
$$

The canonical quantization of fields consists of promoting the classical fields $\Psi(x)$ and $\Pi_{\Psi}(x)$ to field operators $\hat{\Psi}(x)$ and $\hat{\Pi}_{\Psi}(x)$ whose components satisfy the equal-time anticommutation relations,

$$
\begin{aligned}
& -i\left\{\hat{\Psi}_{a}\left(x^{0}, \mathbf{x}\right), \hat{\Pi}_{\Psi, b}\left(x^{0}, \mathbf{x}^{\prime}\right)\right\} \\
& =\left\{\hat{\Psi}_{a}\left(x^{0}, \mathbf{x}\right), \hat{\Psi}_{b}^{\dagger}\left(x^{0}, \mathbf{x}^{\prime}\right)\right\}=\delta^{(3)}\left(\mathbf{x}-\mathbf{x}^{\prime}\right) \delta_{a, b}
\end{aligned}
$$

By recalling the wave expansion of the field $\Psi(x)$,

$$
\Psi(x)=\left.\int \frac{\mathrm{d}^{3} \mathbf{p}}{(2 \pi)^{3} \sqrt{E_{\mathbf{p}}}} \sum_{s=1,2}\left[a_{s, \mathbf{p}} u_{s}(p) \mathrm{e}^{-i p x}+b_{s, \mathbf{p}}^{*} v_{s}(p) \mathrm{e}^{i p x}\right]\right|_{p^{0}=E_{\mathbf{p}}},
$$

the quantization procedure is equivalent to replace the coefficients $a_{\mathbf{p}, s}$ and $b_{\mathbf{p}, s}$ with the operators $\hat{a}_{\mathbf{p}, s}$ and $\hat{b}_{\mathbf{p}, s}$ which obey the following anticommutation relations

$$
\begin{array}{r}
\left\{\hat{a}_{\mathbf{p}, s}, \hat{a}_{\mathbf{q}, r}^{\dagger}\right\}=(2 \pi)^{3} \delta^{(3)}(\mathbf{p}-\mathbf{q}) \delta_{s, r}, \\
\left\{\hat{b}_{\mathbf{p}, s}, \hat{b}_{\mathbf{q}, r}^{\dagger}\right\}=(2 \pi)^{3} \delta^{(3)}(\mathbf{p}-\mathbf{q}) \delta_{s, r},
\end{array}
$$

All other anticommutators vanish.

 The states of these Hilbert spaces on which these operators act have Fock space's structure. Namely, they can be characterized by the number of particles with a certain momentum and spin. We define a vacuum vector in $\mathcal{H} ,|0\rangle$, such that-

$$
\begin{aligned}
& \hat{a}_{\mathbf{p}, s}|0\rangle=0, \\
& \hat{b}_{\mathbf{p}, s}|0\rangle=0, \quad \forall \mathbf{p}, s .
\end{aligned}
$$

Operators $\hat{a}_{\mathbf{p}, s}$ refers to particles, $\hat{b}_{\mathbf{p}, s}$ to antiparticles. By considering operators $\hat{a}_{\mathbf{p}, s}^{\dagger}$ and $\hat{b}_{\mathbf{p}, s}^{\dagger}$ we have that

$$
\begin{aligned}
\sqrt{2 E_{\mathbf{p}}} \hat{a}_{\mathbf{p}, s}^{\dagger}|0\rangle & =|\mathbf{p}, s\rangle, \\
\sqrt{2 E_{\mathbf{p}}} \hat{b}_{\mathbf{p}, s}^{\dagger}|0\rangle & =\left|\mathbf{p}^{*}, s\right\rangle, \quad \forall \mathbf{p}, s
\end{aligned}
$$

in which the * indicates that we are referring to antiparticles. Therefore $\hat{a}_{\mathbf{p}, s}^{\dagger}$ and $\hat{b}_{\mathbf{p}, s}^{\dagger}$ act as creation operators of a particle and an antiparticle with momentum $\mathbf{p}$ and spin $s$. A state with $n$ particles, constructed by repeated applications of $\hat{a}^{\dagger}$ creation operators, reads

$$
\begin{aligned}
|\Omega\rangle & =\left|\mathbf{p}_{1}, s_{1} ; \ldots ; \mathbf{p}_{n}, s_{n}\right\rangle \\
& =\left(2 E_{\mathbf{p}_{1}} \ldots 2 E_{\mathbf{p}_{n}}\right)^{1 / 2} \hat{a}_{\mathbf{p}_{1}, s_{1}}^{\dagger} \ldots \hat{a}_{\mathbf{p}_{n}, s_{n}}^{\dagger}|0\rangle
\end{aligned}
$$

in this state one particle has momentum $\mathbf{p}_{1}$ and spin $s_{1}$, and so on. In the same way, a state including antiparticle could be obtained by applying $\hat{b}^{\dagger}$ operators. Operators $\hat{a}_{\mathbf{p}, s}$ and $\hat{b}_{\mathbf{p}, s}$ destroy a particle and an antiparticle, respectively, with momentum $\mathbf{p}$ and spin $s$, if the state contains such a particle, otherwise, they annihilate it.

In the formalism of canonical quantization, the Hamiltonian density operator can be written as.

$$
\hat{H}=\hat{\Pi}_{\Psi} \hat{\Psi}-\mathcal{L}=\hat{\bar{\Psi}}\left(-i \gamma^{i} \partial_{i}+m\right) \hat{\Psi}
$$

 When we quantize fields, one can show that the operator $\hat{\Psi}^{\dagger}(x) \hat{\Psi}(x)$ is the generator for global and local $U(1)$ transformations of fields. Given a real function $\alpha(x)$, we can define the following operator

$$
T=\exp \left\{i \int \mathrm{d}^{3} \mathbf{x} \alpha(x) \hat{\Psi}^{\dagger}(t, \mathbf{x}) \hat{\Psi}(t, \mathbf{x})\right\}
$$

The field transformation reads as:

$$
\hat{\Psi}(y) \rightarrow T^{\dagger} \hat{\Psi}(y) T=\hat{\Psi}(y) \mathrm{e}^{i \alpha(y)}
$$

This result can be demonstrated by computing the commutator,

$$
\begin{aligned}
& {\left[\int \mathrm{d}^{3} \mathbf{x} \alpha(x) \hat{\Psi}^{\dagger}(t, \mathbf{x}) \hat{\Psi}(t, \mathbf{x}), \hat{\Psi}(t, \mathbf{y})\right] } \\
= & \int \mathrm{d}^{3} \mathbf{x} \alpha(x)\left[\hat{\Psi}^{\dagger}(t, \mathbf{x}) \hat{\Psi}(t, \mathbf{x}), \hat{\Psi}(t, \mathbf{y})\right] \\
= & \int \mathrm{d}^{3} \mathbf{x} \alpha(x)\left(\hat{\Psi}^{\dagger}(t, \mathbf{x}) \hat{\Psi}(t, \mathbf{x}) \hat{\Psi}(t, \mathbf{y})-\hat{\Psi}(t, \mathbf{y}) \hat{\Psi}^{\dagger}(t, \mathbf{x}) \hat{\Psi}(t, \mathbf{x})\right) \\
= & \int \mathrm{d}^{3} \mathbf{x} \alpha(x)\left(\hat{\Psi}^{\dagger}(t, \mathbf{x}) \hat{\Psi}(t, \mathbf{x}) \hat{\Psi}(t, \mathbf{y})+\hat{\Psi}^{\dagger}(t, \mathbf{x}) \hat{\Psi}(t, \mathbf{y}) \hat{\Psi}(t, \mathbf{x})\right. \\
& \left.-\delta^{(3)}(\mathbf{x}-\mathbf{y}) \hat{\Psi}(t, \mathbf{x})\right) \\
= & -\alpha(y) \hat{\Psi}(t, \mathbf{y})
\end{aligned}
$$

$$
\begin{aligned}
& T^{\dagger} \hat{\Psi}(y) T= \\
& \exp \left\{-i \int \mathrm{d}^{3} \mathbf{x} \alpha(x) \hat{\Psi}^{\dagger}(x) \hat{\Psi}(x)\right\} \hat{\Psi}(y) \exp \left\{i \int \mathrm{d}^{3} \mathbf{x} \alpha(x) \hat{\Psi}^{\dagger}(x) \hat{\Psi}(x)\right\} \\
& =\mathrm{e}^{i \alpha(y)} \hat{\Psi}(y),
\end{aligned}
$$
In writing the above expression, we have used an additional result $\mathrm{e}^{X} Y \mathrm{e}^{-X}=\mathrm{e}^{c} Y$, given two operators $X$ and $Y$, and a real number $c$, satisfying $[X, Y]=c Y$.

 For the case of the EM field, the dynamical variables of the system described by $\mathcal{L}$ are the components of the four-vector potential $A_{\mu}$, and their conjugate momenta are defined by the relation.

$$
\Pi_{\mu}=\frac{\partial \mathcal{L}}{\partial\left(\partial_{0} A_{\mu}\right)}
$$

One also observes  that the momentum $\Pi_{0}$ is zero since $\partial_{0} A_{0}$ does not appear in $\mathcal{L}$. Therefore no commutation relation involving the temporal component of the four-vector potential can be imposed. Hence we fix a particular gauge and take a function $\phi(x)$ such that the condition  $\partial_{0} \phi=A_{0}$ holds, and we perform a gauge transformation with this function $\phi(x)$, which becomes,

$$
A_{\mu}^{\prime}=A_{\mu}-\partial_{\mu} \phi(x), \quad A_{0}^{\prime}=0
$$

Promoting $A_{i}$ and $\Pi_{i}$ to field operators $\hat{A}_{i}$ and $\hat{\Pi}_{i}$ which obey the following equal time commutation relation to implement quantization:

$$
\left[\hat{A}_{i}(t, \mathbf{x}), \hat{\Pi}_{j}\left(t, \mathbf{x}^{\prime}\right)\right]=i \delta\left(\mathbf{x}-\mathbf{x}^{\prime}\right) \delta_{i j}
$$

which becomes,

$$
\left[\hat{A}_{i}(t, \mathbf{x}), \hat{E}_{j}\left(t, \mathbf{x}^{\prime}\right)\right]=-i \delta\left(\mathbf{x}-\mathbf{x}^{\prime}\right) \delta_{i j}
$$

Now we define the following operator, which performs the gauge transformation of $\hat{A}_{i}(t, \mathbf{x})$. 

$$
\begin{aligned}
\hat{R}[\phi] & \equiv \exp \left\{-i \int \mathrm{d} \mathbf{z} \phi(\mathbf{z}) \nabla \cdot \hat{\mathbf{E}}(t, \mathbf{z})\right\} \\
& =\exp \left\{i \int \mathrm{d} \mathbf{z} \nabla \phi(\mathbf{z}) \cdot \hat{\mathbf{E}}(t, \mathbf{z})\right\} \\
& =\exp \left\{-i \int \mathrm{d} \mathbf{z} \nabla^{j} \phi(\mathbf{z}) \hat{E}_{j}(t, \mathbf{z})\right\} .
\end{aligned}
$$
The transformation is:

$$
\begin{aligned}
& \hat{R}^{\dagger}[\phi(\mathbf{z})] \hat{A}_{i}(t, \mathbf{x}) \hat{R}[\phi(\mathbf{z})] \\
= & \exp \left\{i \int \mathrm{d} \mathbf{z} \nabla^{j} \phi(\mathbf{z}) \hat{E}_{j}(t, \mathbf{z})\right\} \hat{A}_{i}(t, \mathbf{x}) \exp \left\{-i \int \mathrm{d} \mathbf{z} \nabla^{j} \phi(\mathbf{z}) \hat{E}_{j}(t, \mathbf{z})\right\} \\
= & -\int \mathrm{d} \mathbf{z} \nabla^{j} \phi(\mathbf{z}) \frac{\delta}{\delta \hat{E}_{i}(t, \mathbf{x})} \hat{E}_{j}(t, \mathbf{z})+\hat{A}_{i}(t, \mathbf{x}) \\
= & -\int \mathrm{d} \mathbf{z} \nabla^{j} \phi(\mathbf{z}) \delta(\mathbf{x}-\mathbf{z}) \delta_{i j}+\hat{A}_{i}(t, \mathbf{x}) \\
= & -\nabla^{i} \phi(\mathbf{x})+\hat{A}_{i}(t, \mathbf{x})=\hat{A}_{i}(t, \mathbf{x})-\partial_{i} \phi(\mathbf{x})
\end{aligned}
$$

This result suggests we implement gauge transformations.

$$
\hat{A}_{i}(t, \mathbf{x}) \rightarrow \hat{A}_{i}(t, \mathbf{x})-\partial_{i} \phi(\mathbf{x}) \equiv \hat{A}_{i}(t, \mathbf{x})-\nabla^{i} \phi(\mathbf{x})
$$

\subsection{Gauss' law and gauge transformations}
Since the physics of the system must be gauge-invariant, physical states must also be gauge-invariant. Therefore, for a free electromagnetic field, given a state $|\Phi\rangle$ the following local condition must therefore hold:

$$
\hat{R}[\phi(\mathbf{z})]|\Phi\rangle=|\Phi\rangle
$$

it is equivalent to,

$$
\nabla \cdot \hat{\mathbf{E}}(t, \mathbf{z})|\Phi\rangle=0 \quad \forall t, \mathbf{z}
$$

This is the second quantization counterpart of the classical Maxwell equation and is now a constraint that selects physical states by requiring their gauge invariance.

If we now consider a system in which the Dirac and the electromagnetic fields interact with each other, the Hilbert space will be the tensor product of the Hilbert spaces on which the two fields act: physical states like $|\Omega\rangle_{\text {Dirac }}|\Phi\rangle_{\text {EM }}$ must be invariant for transformations in the form-

$$
|\Omega\rangle_{\text {Dirac }}|\Phi\rangle_{\text {EM }} \rightarrow(T \otimes \hat{R}[\phi(\mathbf{z})])|\Omega\rangle_{\text {Dirac }}|\Phi\rangle_{\text {EM }}
$$

and this condition equivalently reads,

$$
\left(\nabla \cdot \hat{\mathbf{E}}(x)-\hat{\Psi}^{\dagger} \hat{\Psi}(x)\right)|\Omega\rangle_{\text {Dirac }}|\Phi\rangle_{\text {EM}}=0 .
$$

This equation is equivalent to the classical Gauss' law in presence of free charges,

$$
\nabla \cdot \mathbf{E}(\mathbf{x})=\rho(\mathbf{x})
$$

Now, let us define the comparator in the quantized theory:

$$
\begin{aligned}
\hat{U}(t ; \mathbf{x}, \mathbf{y}) & \equiv \exp \left\{-i \int_{\mathbf{x}}^{\mathbf{y}} \mathrm{d} \mathbf{z} \cdot \hat{\mathbf{A}}(t, \mathbf{z})\right\} \\
& =\exp \left\{i \int_{\mathbf{x}}^{\mathbf{y}} \mathrm{d} z^{i} \hat{A}_{i}(t, \mathbf{z})\right\} .
\end{aligned}
$$

The comparator transforms as follows:

$$
\begin{aligned}
& \hat{U}(t ; \mathbf{x}, \mathbf{y}) \rightarrow \hat{W}^{\dagger}[\phi(\mathbf{z})] \hat{U}(t ; \mathbf{x}, \mathbf{y}) \hat{W}[\phi(\mathbf{z})] \\
= & \exp \left\{i \int_{\mathbf{x}}^{\mathbf{y}} \mathrm{d} z^{i}\left(\hat{A}_{i}(t, \mathbf{z})-\partial_{i} \phi(\mathbf{z})\right)\right\} \\
= & \mathrm{e}^{i \phi(\mathbf{x})} \hat{U}(t ; \mathbf{x}, \mathbf{y}) \mathrm{e}^{-i \phi(\mathbf{y})} .
\end{aligned}
$$

Recovering the transformation rule similar to that for the classical comparator. Let us now write the covariant derivative using the comparator for the quantized fields: we define;

$$
\hat{D}_{\hat{\eta}} \hat{\Psi}(x)=\lim _{\epsilon \rightarrow 0} \frac{\hat{U}(x, x+\epsilon \hat{\eta}) \hat{\Psi}(x+\epsilon \hat{\eta})-\hat{\Psi}(x)}{\epsilon}
$$

Therefore we obtain the derivative $\hat{D}_{\mu}$, choosing the unit vector aligned along the $\mu$ axis in the Minkowski space, whose properties of gauge transformations are similar to that as of quantized fields. The Lagrangian for the interaction theory is,

$$
\hat{\mathcal{L}}=\hat{\bar{\Psi}}(i \hat{\not D}-m) \hat{\Psi}-\frac{1}{4} \hat{F}_{\mu \nu} \hat{F}^{\mu \nu}
$$

with $\hat{F}_{\mu \nu}=\partial_{\mu} \hat{A}_{\nu}-\partial_{\nu} \hat{A}_{\mu}$ and $\hat{\not D}=\gamma^{\mu} \hat{D}_{\mu}$, while the Hamiltonian is-

$$
\hat{H}=\hat{\bar{\Psi}}\left(-i \gamma^{i} \hat{D}_{i}+m\right) \hat{\Psi}+\frac{1}{2}\left(\hat{\mathbf{E}}^{2}+\hat{\mathbf{B}}^{2}\right)
$$
This operator is invariant under the gauge transformation $(T \otimes \hat{R}[\phi(\mathbf{z})])$ 
and represents the quantum counterpart of the classical local transformations defined in the previous sections.

\section{Gauge Theory On A lattice}
The inclusion of fermion fields on a lattice is highly non-trivial. If we "naively" introduce lattice fermions in which, as with scalar fields, we allow the field to exist on each site of the lattice, one encounters the well-known fermion doubling problem\cite{Rothe_book}. In the continuum limit, a single fermion field corresponds to multiple particles, indicating the appearance of unphysical fermions known as "doublers."
The Nielsen-Ninomiya theorem asserts that any lattice action with fermions satisfying the usual requirements of locality, translational invariance, hermiticity of the corresponding hamiltonian, and chirality (required to reproduce the Standard Model) will inevitably encounter doublers. Any method that makes sense for introducing fermions on a lattice must violate one of these assumptions. Utilizing staggered fermions \cite{Kogut1975} is a prevalent method among the various available options, which has recently  gained a lot of interest in quantum simulations.
\subsection{Staggered fermions}
 We now define a field that describes spinless fermions in a $(1+1)$-dimensional space: we represent it with a two-component Dirac spinor,

$$
\chi(x)=\left(\begin{array}{l}
\chi^{1}(x) \\
\chi^{2}(x)
\end{array}\right)
$$

And the Dirac matrices for this model are,

$$
\gamma^{\prime 0} \equiv \beta=\left(\begin{array}{cc}
1 & 0 \\
0 & -1
\end{array}\right), \quad \gamma^{\prime 1}=\left(\begin{array}{cc}
0 & 1 \\
-1 & 0
\end{array}\right)
$$
The Dirac Hamiltonian operator is now-

$$
\mathcal{H}=-i \beta \gamma^{\prime 1} \frac{\partial}{\partial x^{1}}+m \beta \equiv-i \alpha^{1} \frac{\partial}{\partial x^{1}}+m \beta
$$

with,

$$
\alpha^{1}=\left(\begin{array}{ll}
0 & 1 \\
1 & 0
\end{array}\right)
$$

It follows that the equations of motion for the two spinor components are

$$
\begin{aligned}
i \partial_{t} \chi^{1}\left(t, x^{1}\right) & =-i \partial_{x^{1}} \chi^{2}\left(t, x^{1}\right)+m \chi^{1}\left(t, x^{1}\right) \\
i \partial_{t} \chi^{2}\left(t, x^{1}\right) & =-i \partial_{x^{1}} \chi^{1}\left(t, x^{1}\right)-m \chi^{2}\left(t, x^{1}\right)
\end{aligned}
$$

Let us now work in a discrete one-dimensional space, with spacing $a$ between points; henceforth, $x$ will be the integer number labeling the site of the lattice, so the relation with the continuous coordinate is $x^{1}=a x$. The equations of motion in the discrete space are,

$$
\begin{aligned}
i \partial_{t} \chi_{x}^{1} & =-\frac{i}{2 a}\left(\chi_{x+1}^{2}-\chi_{x-1}^{2}\right)+m \chi_{x}^{1} \\
i \partial_{t} \chi_{x}^{2} & =-\frac{i}{2 a}\left(\chi_{x+1}^{1}-\chi_{x-1}^{1}\right)-m \chi_{x}^{2} .
\end{aligned}
$$

To avoid the fermion doubling problem, we introduce the solution performed by Kogut and Susskind. Let us start by giving up the locality in the definition of the fermion field, and by defining a new one-component field ${\psi}_{x}$ by taking the first component of $\chi$ on even sites and the second component on odd sites, such that:

$$
{\hat{\psi}}_{x}= \begin{cases}\chi_{x}^{1} & \text { if }(-1)^{x}=1 \\ \chi_{x}^{2} & \text { if }(-1)^{x}=-1\end{cases}
$$

This means that on even sites there are positive energy solutions, while negative energy ones correspond to odd sites. Hence, the definition of ${\psi}_{x}$ that the equation of motion is

$$
i \partial_{t} {\psi}_{x}=-\frac{i}{2 a}\left({\psi}_{x+1}-{\psi}_{x-1}\right)+m(-1)^{x} {\psi}_{x}
$$

 and the new Hamiltonian is, 

\begin{equation}\label{eq7}
  \begin{aligned}
H_{\text {stagg }} & =\sum_{x} {\psi}_{x}^{\dagger}\left[-\frac{i}{2 a}\left({\psi}_{x+1}-{\psi}_{x-1}\right)+m(-1)^{x} {\psi}_{x}\right] \\
& =-\frac{i}{2 a} \sum_{x} {\psi}_{x}^{\dagger} {\psi}_{x+1}+\frac{i}{2 a} \sum_{x} {\psi}_{x}^{\dagger} {\psi}_{x-1}+m \sum_{x}(-1)^{x} {\psi}_{x}^{\dagger} {\psi}_{x} \\
& =-\frac{i}{2 a} \sum_{x} {\psi}_{x}^{\dagger} {\psi}_{x+1}+\frac{i}{2 a} \sum_{x} {\psi}_{x+1}^{\dagger} {\psi}_{x}+m \sum_{x}(-1)^{x} {\psi}_{x}^{\dagger} {\psi}_{x} \\
& =-\frac{i}{2 a} \sum_{x} {\psi}_{x}^{\dagger} {\psi}_{x+1}+\text { H.c. }+m \sum_{x}(-1)^{x} {\psi}_{x}^{\dagger} {\psi}_{x} .
\end{aligned}  
\end{equation}

 As in the continuous case, $H_{\text {stagg }}$ is symmetric under global $U(1)$ transformations of fields in the form

$$
{\psi}_{x} \rightarrow \mathrm{e}^{i \alpha} {\psi}_{x}
$$

with $\alpha$ a real constant. When we quantize the theory, field functions $\hat{\psi}_{x}$ become field operators which satisfy anticommutation relations:

$$
\left\{\hat{\psi}_{x}, \hat{\psi}_{x^{\prime}}\right\}=\left\{\hat{\psi}_{x}^{\dagger}, \hat{\psi}_{x^{\prime}}^{\dagger}\right\}=0, \quad\left\{\hat{\psi}_{x}, \hat{\psi}_{x^{\prime}}^{\dagger}\right\}=\delta_{x x^{\prime}}
$$

The above transformations are implemented on lattice field operators by converting the continium transformations in the discrete, one-dimensional space case, and the result is,

$$
\hat{\psi}_{x} \rightarrow \prod_{y} \mathrm{e}^{-i \alpha \hat{\psi}_{y}^{\dagger} \hat{\psi}_{y}}\hat{\psi}_{x} \prod_{z} \mathrm{e}^{i \alpha \hat{\psi}_{z}^{\dagger} \hat{\psi}_{z}}=\mathrm{e}^{i \alpha} \hat{\psi}_{x}
$$


\subsection{Minimal coupling and gauge transformations on the lattice}

As in the continuum field model, the coupling of fermions field with a gauge field is introduced in order to promote the $U(1)$ symmetry of the Hamiltonian \eqref{eq7} from global to local. We will use the comparator transformation rule defined in the previous section,

$$
U(x, y) \rightarrow \mathrm{e}^{i \alpha(x)} U(x, y) \mathrm{e}^{-i \alpha(y)} .
$$

Now, let us recall the kinetic term in $H_{\text {stagg }}$,

$$
H_{k i n}=-\frac{i}{2 a} \sum_{x} \hat{\psi}_{x}^{\dagger} \hat{\psi}_{x+1}+\text { H.c., }
$$

and observe that it transforms under a local $U(1)$ transformation as follows

$$
H_{k i n} \rightarrow H_{k i n}^{\prime}=-\frac{i}{2 a} \sum_{x} \mathrm{e}^{-i \alpha_{x}} \mathrm{e}^{i \alpha_{x+1}} \hat{\psi}_{x}^{\dagger} \hat{\psi}_{x+1}+\text { H.c., }
$$

with $\alpha_{x}$ a real function. To implement the local $U(1)$ symmetry on $H_{\text {stagg }}$, we have to define the comparator on the lattice links between the lattice sites. We consider the vector potential to be defined in the continuum space, in the gauge $A_{0}=0$; the only non-vanishing component is $A^{1}\left(x^{1}\right) \equiv A\left(x^{1}\right)$. A general definition for the comparator on the lattice is

$$
U(x, x+1)=\mathrm{e}^{-i a A\left(x^{1 }\right)}
$$

Now, let us define a link variable for the vector potential which replaces the function $A\left(x^{1}\right)$ : henceforth, we set the lattice spacing $a=1$ and define the variable $A_{x, x+1}$ by imposing \cite{Wiese_review}

$$
U(x, x+1)=\mathrm{e}^{-i A_{x, x+1}},
$$

and, given a real function $\alpha_{x}$, it transforms according to the rule

$$
U(x, x+1) \rightarrow \mathrm{e}^{i \alpha_{x}} U(x, x+1) \mathrm{e}^{-i \alpha_{x+1}} .
$$

The gauge-invariant expression for the kinetic term $H_{k i n}$ is, therefore

$$
H_{k i n G}=-\frac{i}{2 a} \sum_{x} \hat{\psi}_{x}^{\dagger} U_{x, x+1} \hat{\psi}_{x+1}+\text { H.c.. }
$$

We now quantize the vector potential by promoting $A_{x, x+1}$ to a field operator and define the electric field $\hat{E}_{x, x+1}$ as its conjugate variable: the algebra commutation rule they satisfy is, namely
\begin{equation}\label{eq4}
    \left[\hat{A}_{x, x+1}, \hat{E}_{x^{\prime}, x^{\prime}+1}\right]=-i \delta_{x x^{\prime}}
\end{equation}

Once we define the comparator in the quantized theory as $\hat{U}_{x, x+1}=\mathrm{e}^{-i \hat{A}_{x, x+1}}$, the commutation relation between the electric field and the comparator is
\begin{equation}\label{eq5}
    \left[\hat{E}_{x^{\prime}, x^{\prime}+1}, \hat{U}_{x, x+1}\right]=\delta_{x x^{\prime}} \hat{U}_{x, x+1}
\end{equation}

Therefore we can write the gauge invariant Hamiltonian density for the quantized theory as,
\begin{figure}[t!]
    \centering
    \includegraphics[scale=0.7]{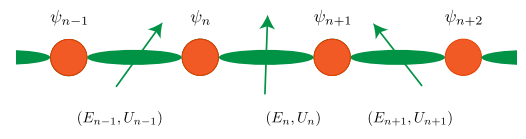}
    \caption{Schematic of an abelian gauge theory in the Kogut-Suskind formulation on a 1D lattice describing staggered fermions.}
    \label{fig:my_label}
\end{figure}
\begin{equation}\label{eq6}
  \hat{H}=-\frac{J}{2} (\sum_{x} \hat{\psi}_{x}^{\dagger} \hat{U}_{x, x+1} \hat{\psi}_{x+1}+\text { H.c. })+m \sum_{x}(-1)^{x} \hat{\psi}_{x}^{\dagger} \hat{\psi}_{x}+\frac{g^{2}}{2} \sum_{x} \hat{E}_{x, x+1}^{2} .  
\end{equation}
 Where $m$ is the fermion mass and $g$ is the coupling constant for the electric field. Since we are studying a one-dimensional system, we do not consider a magnetic field energy term. The phase factor $(-1)^{x}$ in the mass term is due to the use of staggered fermions. This model is the lattice formulation
of Quantum Electrodynamics, also known as the Schwinger model. Despite its
simplicity, it shares a lot of interesting physics with (3 + 1)D $SU(3)$
quantum chromodynamics, describing the strong
interactions of the Standard Model, such as confinement,
nontrivial $\theta$ vacuum, or chiral symmetry breaking and
anomaly.
\begin{figure}[t!]
    \centering
    \includegraphics[scale=0.6]{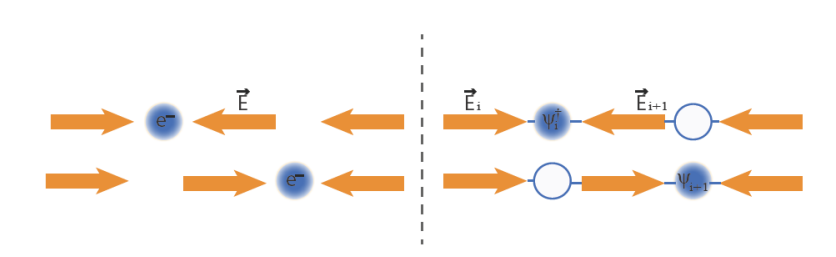}
    \caption{ A hopping of a fermion in the Schwinger model causes the field to reciprocate this
change and flip its direction encapsulating the constraint imposed by Gauss's law.}
    \label{fig:my_label}
\end{figure}
The Hilbert space on which the Hamiltonian acts is the tensor product of the Hilbert spaces relative to each site and each link. In particular, we observe that the spaces relative to links are infinite-dimensional since they contain the eigenstates of the electric field, which forms an infinite orthonormal basis.

The generators of gauge transformations are the discrete version of the operators.

$$
\hat{\Psi}^{\dagger} \hat{\Psi}(x)-\nabla \cdot \hat{\mathbf{E}}(x)
$$

 With an additional term due to the use of staggered fermions. We define the operators on the lattice sites 

$$
\hat{G}_{x}=\hat{\psi}_{x}^{\dagger} \hat{\psi}_{x}-\left(\hat{E}_{x, x+1}-\hat{E}_{x-1, x}\right)+\frac{1}{2}\left[(-1)^{x}-1\right]
$$

in which $\left(\hat{E}_{x, x+1}-\hat{E}_{x-1, x}\right)$ is the discrete version of the electric field divergence. The local gauge transformations for the Hamiltonian are

$$
\hat{H} \rightarrow \prod_{x} \mathrm{e}^{-i \alpha_{x} G_{x}} \hat{H} \prod_{y} \mathrm{e}^{i \alpha_{y} G_{y}}
$$

with $\alpha_{x}$ a real-valued function defined on lattice sites. The gauge invariance of the Hamiltonian is encoded in the following commutation relation:

$$
\left[\hat{H}, \hat{G}_{x}\right]=0 \quad \forall x
$$

In conclusion, we have defined the Hamiltonian for a quantized Abelian gauge theory on a one-dimensional lattice, in which the gauge field is coupled with a two-component Dirac field. 
\newline
\\
At this juncture we also emphasize the necessity of a 
suitable formalism to establish the connection between the lattice gauge
theory and the system that potentially will be able to be used as a quantum
simulator: One of the ways to achieve this is by using the quantum link model formalism \cite{Chandrasekharan1997,Wiese_review}, which can be used to encode our dynamical gauge fields coupled to matter using an equivalent spin-based model, since our simulator has access to only a finite-dimensional Hilbert space. We introduce this in the next chapter.


\chapter{Stabilizing lattice gauge theories in quantum simulators}
\textit{Parts of this chapter including the main results and discussion appear in ref.\cite{kumar2022suppression}}
\section{How to engineer gauge symmetries?}

Currently, there are two primary approaches to achieve quantum simulations of gauge theories. The first approach employs Gauss's law to eliminate redundant degrees of freedom, meaning it expresses the state of electric fields in terms of charge configurations or vice versa . This approach offers advantages in terms of efficient resource utilization and ensures the conservation of Gauss's law by encoding it directly into the simulator. However, a drawback is that the resulting effective Hamiltonian may involve complex long-range interactions, necessitating careful parameter tuning in the Hamiltonian describing the simulator.

The second approach retains both matter and gauge fields as dynamic degrees of freedom in the quantum simulator, allowing for more flexibility in realization, such as through degenerate perturbation theory. This approach has been successfully implemented in various experimental platforms. In this chapter and throughout this work, our focus will be on stabilization schemes based on the second approach\cite{Halimeh_BriefReview,Hauke2013}, as Gauss's law is not imposed \emph{a priori}. This allows for testing of fundamental questions related to the emergence of gauge invariance in nature.

\subsection{The Quantum Link Model}

One of the ways to implement local gauge invariance on the lattice using a finite-dimensional local Hilbert space is by constructing a  quantum link model. In this model, one replaces the comparator and the electric field operators with two finite and discrete operators which satisfy the same commutation relation \eqref{eq5}.

We fix the dimension $n$ of the link Hilbert spaces and consider $n$-dimensional spin operators $\hat{S}^{j}$, with $j=1,2,3$ and $n=2 S+1$. The raising operator for the eigenvectors of $\hat{S}^{3}$ is $\hat{S}^{+}=\hat{S}^{1}+i \hat{S}^{2}$, and obeys the commutation relation $\left[\hat{S}^{3}, \hat{S}^{+}\right]=\hat{S}^{+}$.

Therefore, one can define the following operators on each link,

$$
\begin{aligned}
& \hat{U}_{i, i+1}\rightarrow \frac{\hat{S}_{i, i+1}^{+}}{\sqrt{S(S+1)}}, \\
& \hat{E}_{i, i+1}\rightarrow \hat{S}_{i, i+1}^{3}
\end{aligned}
$$

The electric field is represented by a discrete operator whose spectrum is $\{-S, \ldots,+S\}$, Henceforth the QLM formulation of  QED Hamiltonian in (1+1)D can be written as,

\begin{equation} \label{eq9}
    \begin{aligned}
    & H_{Q L M}= 
    -\frac{J}{2\sqrt{S(S+1)}}\sum_{i} (\hat{\psi}_{i}^{\dagger} \hat{S}_{i, i+1}^{+} \hat{\psi}_{i+1}+\text { H.c. })+m \sum_{i}(-1)^{i} \hat{\psi}_{i}^{\dagger} \hat{\psi}_{i}+\frac{g^{2}}{2} \sum_{i} (\hat{S}_{i, i+1}^{3})^{2} .
    \end{aligned}
\end{equation}

Let us define the generators of gauge transformations in the Quantum Link Model, which are

$$
G_{i}=\hat{\psi}_{i}^{\dagger} \hat{\psi}_{i}-\left(\hat{S}_{i, i+1}^{3}-\hat{S}_{i-1, i}^{3}\right)+\frac{1}{2}\left[(-1)^{i}-1\right]
$$

The system is $U(1)$ gauge invariant, since $\left[G_{i}, H_{Q L M}\right]=0$ . The validity of the commutation rule \eqref{eq5} in the passage from the continuous field model to the QLM is important since it guarantees the theory's $U(1)$ gauge invariance.

Also, in the QLM, we define the physical states as those states $|\Psi\rangle$ satisfying Gauss's Law:

$$
G_{i}|\Psi\rangle=0 \quad \forall i
$$
Equation \eqref{eq9} converges to
the lattice Schwinger model in the Kogut–Susskind limit $S\rightarrow \infty$. It has been shown that even for small spin lengths, one can make reliable extrapolations to the Schwinger model \cite{Zache2021achieving}. 
This feature
has also been observed in other truncation schemes
of the gauge field \cite{MariCarmen2019}. The QLM formulation has the advantage that it preserves the canonical commutation relations between the electric field and the comparator.
\subsection{Jordan-Wigner and the particle-hole transformation}
To make the QLM of Eq.\eqref{eq9} accessible to a quantum simulator setup, we first map the fermionic fields to spin-1/2 degrees
of freedom, using a Jordan-Wigner transformation,
\begin{equation}
\begin{aligned}
\psi_i^{\dagger} & =\mathrm{e}^{i \pi \sum_{k<i}\left(\sigma_k^z+1\right) / 2} \sigma_i^{+}, \\
\psi_i & =\mathrm{e}^{-i \pi \sum_{k<i}\left(\sigma_k^z+1\right) / 2} \sigma_i^{-}, \\
\psi_i^{\dagger} \psi_i & =\frac{\sigma_i^z+1}{2} .
\end{aligned}
\end{equation}

Then, the Hamiltonian that we aim to simulate reads,
  \begin{equation} \label{spin s}
    \begin{aligned}
    & H_{Q L M}= 
    -\frac{J}{2\sqrt{S(S+1)}}\sum_{i} (\hat{\sigma_i^{+}}\hat{S}_{i, i+1}^{+} \hat{\sigma}_{i+1}^{-}+\text { H.c. })+\frac{m}{2} \sum_{i}(-1)^{i}\hat{\sigma}_i^z+\frac{g^{2}}{2} \sum_{i} (\hat{S}_{i, i+1}^{z})^{2} .
    \end{aligned}
\end{equation}
with pauli $\hat{\sigma}_i^z$ associated with the matter fields. The tunneling process $J$, when expressed in spin language, transforms into an assisted flip-flop process that connects the matter and gauge fields. This process describes the creation and annihilation of an 'electron-positron' pair, accompanied by the flipping of the electric field to adhere to Gauss's law. This is similar to a ferromagnetic $XY$ interaction but also involves a flip of an additional spin degree of freedom. The second part in $H_{QLM}$ is simply a staggered,
transverse magnetic field. In one dimension, we can remove the various alternating
signs by the basis transformation, also known as a particle-hole transformation (corresponding to a staggered rotation about the $x$ axis).
\begin{equation}
\begin{aligned}
{\sigma}_i^z \rightarrow(-1)^i \sigma_i^z & , & {\sigma}_i^y \rightarrow(-1)^i \sigma_i^y\\
{S}_{i-1, i}^z \rightarrow(-1)^i S_{i-1, i}^z & , & {S}_{i-1, i}^y \rightarrow(-1)^i S_{i-1, i}^y
\end{aligned}
\end{equation}
Hence $H_{QLM}$ and the corresponding symmetry generators can now be written as follows,
\begin{figure}[t!]
    \centering
    \includegraphics[scale=0.7]{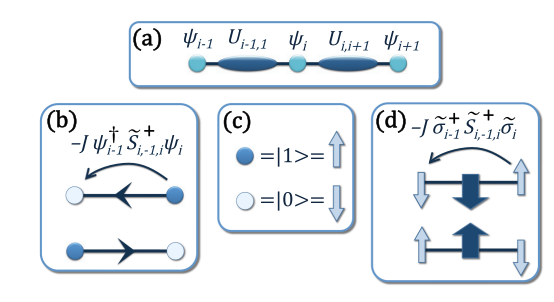}
    \caption{Schematic of the lattice Schwinger model of QED in the Kogut-Susskind formulation. This model can be mapped to an equivalent quantum link formulation via spin operators, to make the model accessible to a quantum simulator.}
    \label{fig:my_label}
\end{figure}
\begin{equation} \label{spin s}
    \begin{aligned}
    & H_{Q L M}= 
    -\frac{J}{2\sqrt{S(S+1)}}\sum_{i} (\hat{\sigma_i^{-}}\hat{S}_{i, i+1}^{+} \hat{\sigma}_{i+1}^{-}+\text { H.c. })+\frac{m}{2} \sum_{i}\hat{\sigma}_i^z+\frac{g^{2}}{2} \sum_{i} (\hat{S}_{i, i+1}^{z})^{2} .
    \end{aligned}
\end{equation}

\begin{align}\label{eq:GjU1}
\hat{G}_j=(-1)^i\bigg(\hat{S}^z_{i-1,i}+\hat{S}^z_{i,i+1}+\frac{\hat{\sigma}^z_i+1}{2}\bigg).
\end{align}
\subsection{Energy penalty scheme}

The realization of LGTs on \textit{synthetic quantum matter} describing Hamiltonian dynamics on a Hilbert space represents specific challenges, primarily because of the constraints imposed due to Gauss's Law described in the previous chapter.

So let us consider an experimental implementation of our Abelian gauge theory described by the Hamiltonian $\hat{H}_0$, which is mapped into our quantum simulator's microscopic degrees of freedom. The corresponding gauge symmetry is generated by the operator $\hat{G}_j$, where $j$ denotes a site on a lattice of length $L$ and the gauge invariance of $\hat{H}_0$ is encoded in the commutation relations $\big[\hat{H}_0,\hat{G}_j\big]=0,\,\forall j$. Therefore we can block-diagonalize our Hamiltonian in a common eigenbasis defined by the set of gauge invariant states, $\{\ket{\psi}\}$ satisfying: $\hat{G}_j\ket{\psi}=g_j\ket{\psi},\,\forall j$. A set of these eigenvalues $\mathbf{g}=(g_1,g_2,\ldots,g_L)$ over the volume of the system defines a unique gauge superselection sector, the projector onto which is $\hat{\mathcal{P}}_\mathbf{g}$. We then select a  \textit{target} or \textit{physical} gauge superselection sector $\mathbf{g}_\text{tar}=(g_1^\text{tar},g_2^\text{tar},\ldots,g_L^\text{tar})$ in which we wish to restrict the dynamics in an experiment. Gauge
invariance restricts dynamics within a gauge sector. So in case our initial state lies in a gauge invariant sector, then the system will remain in the given sector if the dynamics are only generated due to $\hat{H}_0$.
\newline
\\
However, due to unitary or incoherent errors arising out of experimental imperfections in equipment or higher orders of perturbative mapping, the dynamics generate gauge violations that will spread across various gauge sectors, leading to the complete departure from faithful gauge-theory dynamics beyond certain timescales, for the case of coherent errors arising in the implementation of the U(1) QLM given by Eq. \ref{spin s} , these can take the following form,
\begin{align}\label{eq:H1}
\lambda\hat{H}_1=\lambda\sum_{j=1}^L\bigg[\hat{\sigma}^-_j\hat{\sigma}^-_{j+1}+\hat{\sigma}^+_j\hat{\sigma}^+_{j+1}+\frac{\hat{s}^x_{j,j+1}+\hat{s}^z_{j,j+1}}{\sqrt{S(S+1)}}\bigg],
\end{align}
The measure of gauge violation that we will use throughout this chapter is quantified as follows-

\begin{equation}\label{gvt}
\begin{aligned}
\epsilon(t) &=\operatorname{Tr}\left\{\rho(t) \sum_{\mathbf{g}} P_{\mathbf{g}}\left[\mathbf{g}-\mathbf{g}_{\mathrm{tar}}\right]^2\right\} \\
&=\sum_j \operatorname{Tr}\left\{\rho(t)\left[G_j-g_j^{\mathrm{tar}}\right]^2\right\} \\
&=\sum_j\left\langle\left[G_j-g_j^{\mathrm{tar}}\right]^2\right\rangle .
\end{aligned}
\end{equation}
where $\rho(t)$ is the density matrix of the time-evolved system at time $t$.

\begin{figure}[t!]
    \centering
    \includegraphics{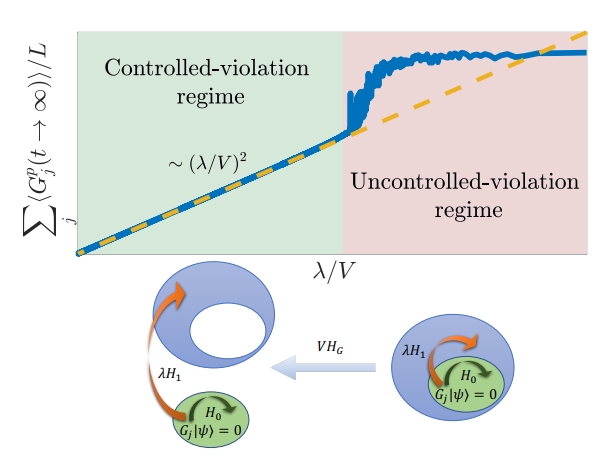}
    \caption{ Fig. adapted from \cite{Halimeh2020a}. Schematic describing the addition of energy penalties $\propto{VH_{G}}$ to suppress the undesired processes $\propto{\lambda H_{1}}$ that takes the dynamics out of the gauge invariant subspace(green bubble). For a large enough value of $V$, the gauge invariant sector is energetically isolated, where one achieves a controlled violation which scales as $(\lambda / V)^{2}$ persisting for infinite times.}
    \label{fig:my_label}
\end{figure}
One of the main strategies for engineering gauge symmetries in quantum simulation consists of imposing an energy penalty to gauge variant states, such that the low-energy physics takes place only on the gauge invariant Hilbert space \cite{halimeh2020g}. This is achieved as follows. One considers a Hamiltonian with the following form:
$$
H=V \hat{H}_{G}+\lambda \hat{H}_{1}, \quad H_{G}:=\sum_{j} G_{j}^{2}
$$
\begin{figure}[!htb]
    \centering
    \begin{subfigure}[b]{0.49\textwidth}
        \centering
        \includegraphics[width=\textwidth]    {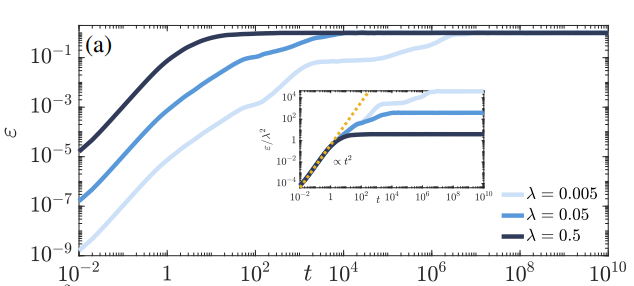}
         
    \end{subfigure}
    \begin{subfigure}[b]{0.49\textwidth}
        \centering
        \includegraphics[width=\textwidth]  {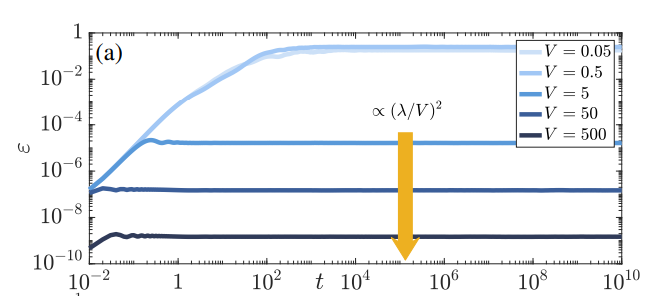}

    \end{subfigure} 
    \caption{Figs adapted from ref.\cite{Halimeh2020a} showing the perturbative growth of gauge violations arising from coherent gauge breaking errors according to Eq.\ref{gvt}(left), and the corresponding addition of energy penalties to suppress them for different values of protection strength $V$ (right). the gauge violations are indefinetly suppressed by $(\lambda / V)^{2}$.  }
    \label{gv1}
\end{figure}

with $V>0$. Thus, in the limit $V \gg \lambda$ all states that violate Gauss law have an energy $E \geq V$ and consequently, low-energy physics is restricted to the gauge-invariant sector of the Hilbert space. The dynamics in this regime is driven by $H_{1}$, which can be gauge variant. For Abelian LGTs, such as the U(1) and $\mathbb{Z}_{2}$ theories, when gauge-variant terms are present, it was found that gauge violation accumulates perturbatively at short times which grow as $\lambda^2t^2$ before proliferating at very long times. As shown in Fig.\ref{gv1}, by imposing an energy penalty on processes that drive the dynamics away from the initial gauge-invariant sector, such proliferation can be suppressed indefinitely.

The numerical results obtained through exact diagonalization in Fig 3.2 and 3.3 were corroborated by analytical arguments in time-dependent perturbation theory in Ref.\cite{Halimeh2020a}. It was rigorously shown that for any unitary symmetry
that is broken on a scale $\propto{\lambda}$, the opening of a gap due to the protection term $\hat{H}_G$ generates a deformed gauge symmetry that is perturbatively close in $(\lambda / V)^{2}$  to the original one.  Furthermore, the numerics indicated that the results were largely independent of system size\cite{Halimeh2020a}.
\subsection{Linear gauge protection}

\begin{figure}[!htb]
    \centering
    \begin{subfigure}[b]{0.49\textwidth}
        \centering
        \includegraphics[width=\textwidth]{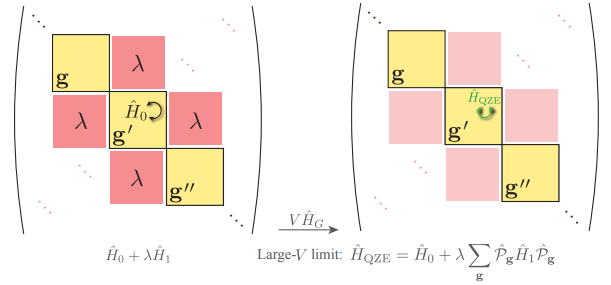}

    \end{subfigure}
    \begin{subfigure}[b]{0.49\textwidth}
        \centering
        \includegraphics[width=\textwidth]{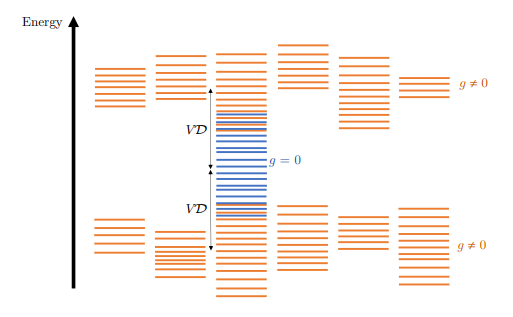}

    \end{subfigure} 
    \caption{Schematic showing linear gauge protection in the case of a faulty gauge-theory implementation $\hat{H}_0+\lambda\hat{H}_1$ . Yellow blocks indicate gauge superselection sectors $\mathbf{g}=(g_1,g_2,\ldots,g_L)$ of the gauge symmetry, with projectors $\hat{\mathcal{P}}_\mathbf{g}$. Upon adding the linear gauge protection $V\hat{H}_G$~\eqref{eq:HG} based on the local generator $\hat{G}_j$, gauge violations are suppressed and a renormalized gauge theory emerges that hosts the same gauge symmetry as $\hat{H}_0$, and which faithfully reproduces the dynamics up to a timescale linear in $V$ at the least, as predicted from the concept of the quantum Zeno dynamics(left), the sequence $c_j$ is chosen in such a way as to spectrally push all the gauge violating states up/down in energy by $V$ (right). Figure is adapted from Ref.~\cite{Halimeh2021stabilizing,Halimeh2021gauge}. }
\end{figure}

In the previous section, we introduced the energy penalty scheme, where quadratic protection terms were added to obtain a renormalized gauge theory. However, this scheme is experimentally very difficult to realize since $G_{j}^{2}$
contains \textit{2 body terms}, which requires extensive parameter tuning to realize such multispecies interactions (in case $G_{j}$ is composed of
single-body terms). Moreover, quadratic protection necessarily energetically isolates a given target superselection sector from all its counterparts. This is inapplicable in applications where the dynamics span multiple sectors simultaneously, such as Disorder-free localization, which will be discussed in the following chapter. To circumvent this issue, the concept of linear gauge protection consisting of only single-body terms was introduced in Ref.~\cite{Halimeh2021gauge}. Now, our protection terms become-
\begin{align}\label{eq:HG}
    V\hat{H}_G=V\sum_jc_j\hat{G}_j,
\end{align}
where $V$ is the protection strength. The sequence $c_j$ can be chosen to be rational and satisfying the condition 
\begin{equation}
    \sum_jc_j\big(g-g_j^\text{tar}\big)=0\iff g_j=g_j^\text{tar},\,\forall j.
\end{equation}
In other words, the sequence $c_j$ has to be chosen in such a way that the target sector is spectrally/energetically isolated from other gauge superselection sectors
In this case, the sequence is said to be \textit{compliant}. For a volume-independent and sufficiently large $V$, the gauge violation is controlled up to times exponential in $V$, proven by the Gauge-Protection theorem in Ref.\cite{Halimeh2021gauge}.

Although $V$ is volume-independent, the sequence $c_j$ would have to grow (not faster than) exponentially with system size in order to satisfy the compliance condition. This renders the compliant sequence somewhat inconvenient for large-scale gauge-theory quantum simulators such as those realized in recent cold-atom setups \cite{Yang:2020Science,Zhou2021}.

However, reality turns out to be more forgiving, and even simple noncompliant sequences such as $c_j=(-1)^j$ can give excellent protection in the target sector against gauge errors up to all accessible evolution times in both finite systems \cite{Halimeh2020e} and the thermodynamic limit \cite{vandamme2021reliability}. 

This can be explained through the coherent quantum Zeno effect \cite{facchi2002quantum,facchi2004unification,facchi2009quantum,burgarth2019generalized}, which guarantees that upon adding the protection term~\eqref{eq:HG} an effective Zeno Hamiltonian $\hat{H}_Z=\hat{H}_0+\lambda\hat{\mathcal{P}}_{\mathbf{g}_\text{tar}}\hat{H}_1\hat{\mathcal{P}}_{\mathbf{g}_\text{tar}}$ emerges that faithfully reproduces the dynamics of the faulty gauge theory $\hat{H}_0+\lambda\hat{H}_1+V\hat{H}_G$ up to timescales linear in $V$ in a worst-case scenario \cite{Halimeh2020e}.


\subsection{$\mathbb{Z}_{2}$ LGT and local-pseudo generators}
Some gauge theories do not have a local generator for their gauge symmetry that is as simple as others. The $\mathbb{Z}_{2}$ lattice gauge theory, which has been the focus of numerous recent experiments, is a prime example \cite{Schweizer2019,aidelsburger2021cold}. As described by the Hamiltonian, 
\begin{align}\label{eq:Z2LGT}
\hat{H}_0=J \sum_{j=1}^{L}\big(\hat{a}_j^{\dagger} \hat{\tau}_{j, j+1}^z \hat{a}_{j+1}+\text{H.c.}\big)-h \sum_{j=1}^L \hat{\tau}_{j, j+1}^x,
\end{align}
where the hardcore bosonic ladder operators $\hat{a}_j,\hat{a}^\dagger_j$ on site $j$ represent the annihilation and creation of matter obeying the mixed commutation relations, which can be described as follows-
\begin{equation}
\begin{gathered}
{\left[a_i, a_j\right]=\left[a_i^{\dagger}, a_j^{\dagger}\right]=\left[a_i, a_j^{\dagger}\right]=0 \quad \text { for } i \neq j,} \\
\left\{a_i, a_i\right\}=\left\{a_i^{\dagger}, a_i^{\dagger}\right\}=0, \\
\left\{a_i, a_i^{\dagger}\right\}=I .
\end{gathered}
\end{equation}
These commutation relations represent a system of bosons with strong but very short-range repulsion, making any state with multiple occupancy on a single site energetically unfavourable.

Because they have a simple tensor product structure similar to ordinary bosons, and the Hilbert space for a finite system is finite-dimensional, these hardcore bosonic operators are easier to treat numerically. 
As a result, they can be precisely mapped onto a system of spin-1/2 on a lattice without utilising the non-local Jordan-Wigner transformation as in the case of fermions, by identifying $\hat{a}^\dagger_j=\hat{S}^{+}_j$.  

The electric (gauge) field on the link between sites $j$ and $j+1$ is represented by the Pauli operator $\hat{\tau}^x_{j,j+1}$ ($\hat{\tau}^z_{j,j+1}$), where the electric field strength is given by $h$. The generator of the $\mathbb{Z}_2$ gauge symmetry of Hamiltonian~\eqref{eq:Z2LGT} is given by
\begin{align}\label{eq:GjZ2}
    \hat{G}_j=(-1)^{\hat{a}_j^\dagger\hat{a}_j}\hat{\tau}^x_{j-1,j}\hat{\tau}^x_{j,j+1},
\end{align}
and its eigenvalues are $\pm1$, where, due to the $\mathbb{Z}_2$ gauge symmetry, $\hat{G}_j^2=\hat{\mathds{1}}_j$. Unlike the generator~\eqref{eq:GjU1} of the $\mathrm{U}(1)$ quantum link model~\eqref{eq:U1QLM}, which is composed of one-body terms, the generator~\eqref{eq:GjZ2} of the $\mathbb{Z}_2$ lattice gauge theory~\eqref{eq:Z2LGT} is a three-body term that mixes matter and gauge degrees of freedom. This renders it significantly impractical in experimental implementations. One can then utilize the concept of the local pseudogenerator \cite{Halimeh2021stabilizing}, where in this case it takes the form
\begin{align}\label{eq:Lpg}
    \hat{W}_j=\hat{\tau}^x_{j-1,j}\hat{\tau}^x_{j,j+1}+2g_j^\text{tar}\hat{a}_j^\dagger\hat{a}_j.
\end{align}

Note that even though $\big[\hat{H}_0,\hat{G}_j\big]=0,\,\forall j$, on account of the $\mathbb{Z}_2$ gauge symmetry of Hamiltonian~\eqref{eq:Z2LGT}, $\big[\hat{H}_0,\hat{W}_j\big]\neq0$. However, when working in the target sector $\mathbf{g}_\text{tar}$, then $\hat{W}_j$ and $\hat{G}_j$ are indistinguishable. It is interesting to note that the local symmetry associated with $\hat{W}_j$ contains the $\mathbb{Z}_2$ gauge symmetry generated by $\hat{G}_j$. In fact, one can prove for a given Hamiltonian $\hat{H}'$ that $\big[\hat{H}',\hat{W}_j\big]=0\Rightarrow\big[\hat{H}',\hat{G}_j\big]=0$.
\begin{figure}[t!]
    \centering
    \includegraphics[scale=0.8]{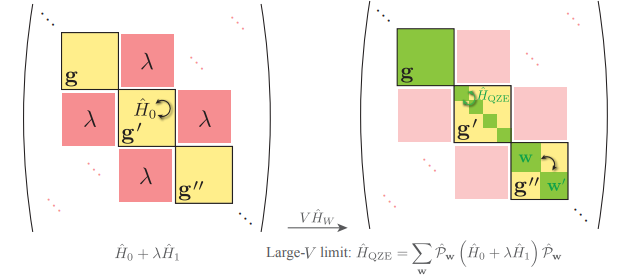}
    \caption{The use of the local pseudogenerator $\hat{W}_j$ (described in Equation~\eqref{eq:Lpg}) in linear gauge protection leads to an enhanced symmetry in the resulting gauge theory. This enhanced symmetry has superselection sectors labeled by $\mathbf{w}$, which are represented by green blocks, and each sector is associated with a projector $\hat{\mathcal{P}}\mathbf{w}$. Figure is adapted from Ref.~\cite{Halimeh2021enhancing}}
    \label{fig gZ2}
\end{figure}
 In particular, $\hat{W}_j$ is identical to the full local generators $\hat{G}_j$ in the target sector, but not necessarily outside of it \cite{Halimeh2021stabilizing}. Formally, they satisfy the relation
\begin{align}
    \hat{W}_j\ket{\phi}=g_j^\text{tar}\ket{\phi}\iff\hat{G}_j\ket{\phi}=g_j^\text{tar}\ket{\phi}.
\end{align}
One can then extend the principle of linear gauge protection to one in terms of the local pseudogenerator, with protection term 
\begin{align}\label{eq:HW}
    V\hat{H}_W=V\sum_jc_j\hat{W}_j,
\end{align}
where the same rules apply for the sequence $c_j$ as in the case of Eq.~\eqref{eq:HG}. Note that even though $\hat{H}_0$ commutes with $\hat{G}_j$, it generally does not commute with $\hat{W}_j$, with the latter associated with a local symmetry richer than that generated by $\hat{G}_j$ \cite{Halimeh2021enhancing}.
The coherent gauge breaking terms for the case of $\mathbb{Z}_2$ LGT given by Eq. \eqref{eq:Z2LGT}, inspired by the recent experiment of Ref.\cite{Schweizer2019}, takes the following form-
\begin{align}\nonumber
	\lambda\hat{H}_1=\,\lambda\sum_{j=1}^{L}\Big\{&\Big[\hat{a}_j^\dagger\hat{a}_{j+1}\big(\eta_1\hat{\tau}^+_{j,j+1} +\eta_2\hat{\tau}^-_{j,j+1}+1\big)+\mathrm{H.c.}\Big]\\\label{eq:H1}
	&+\big(\eta_3\hat{n}_j-\eta_4\hat{n}_{j+1}+1\big)\hat{\tau}^z_{j,j+1}\Big\},
\end{align}
where the coefficients $\eta_{1\ldots4}$ are real numbers that depend on a driving parameter employed in the Floquet setup of the experiment.

The resulting Zeno Hamiltonian when protecting with Eq.~\eqref{eq:HW} is $\hat{H}_Z=\hat{\mathcal{P}}_{\mathbf{g}_\text{tar}}\big(\hat{H}_0+\lambda\hat{H}_1\big)\hat{\mathcal{P}}_{\mathbf{g}_\text{tar}}$, under which the dynamics of the faulty gauge theory $\hat{H}_0+\lambda\hat{H}_1+V\hat{H}_W$ can be faithfully reproduced up to times at least linear in $V$ \cite{Halimeh2021stabilizing}.

\begin{table}[t!]
    \centering
   \begin{tabular}{||c|c|c||c|c|c||}
    \hline$\hat{n}_j$ & $\hat{\tau}_{j-1, j}^x$ & $\hat{\tau}_{j, j+1}^x$ & $\hat{G}_j$ & $\hat{W}_j\left(g_j^{\operatorname{tar}}=-1\right)$ & $\hat{W}_j\left(g_j^{\operatorname{tar}}=+1\right)$ \\
    \hline \hline 0 & -1 & -1 & +1 & +1 & +1 \\
    \hline 0 & -1 & +1 & -1 & -1 & -1 \\
    \hline 0 & +1 & -1 & -1 & -1 & -1 \\
    \hline 0 & +1 & +1 & +1 & +1 & +1 \\
    \hline 1 & -1 & -1 & -1 & -1 & +3 \\
    \hline 1 & -1 & +1 & +1 & -3 & +1 \\
    \hline 1 & +1 & -1 & +1 & -3 & +1 \\
    \hline 1 & +1 & +1 & -1 & -1 & +3 \\
    \hline
\end{tabular}
    \caption{ Table showing eigenvalues $g_j$ and $w_j$ of the full generator $G_j$
and the local pseudogenerator $W_j$ for the different possible local configurations of the fields and matter sites $j$.}
    \label{}
\end{table}

In terms of purely unitary errors, extensive numerical simulations in exact diagonalization (ED) and infinite matrix product states (iMPS) based on the time-dependent variational principle have shown that for a compliant or properly chosen noncompliant sequence, linear gauge protection in the full local generator or the local pseudogenerator leads to stabilized gauge-theory dynamics up to all accessible evolution times with the gauge violation settling at a timescale $\propto1/V$ into a plateau of value $\propto\lambda^2/V^2$ \cite{Halimeh2020e,vandamme2021reliability,Halimeh2021stabilizing,vandamme2021suppressing}. Importantly, the linear gauge protection terms~\eqref{eq:HG} and~\eqref{eq:HW} are composed of single and two-body terms at most, and they are local, which renders them experimentally highly feasible.
\section{Suppressing incoherent errors due to $1/f$ noise in quantum simulators of gauge theories }\label{sec:BRE}
Up until this point, the above-mentioned energy penalty schemes have been only discussed in the context of closed systems since the perturbative errors obey unitary dynamics. However, when one is also concerned with our target quantum-many body system of the quantum simulator interacting with the environment/bath, it can be useful to reinterpret the linear gauge protection scheme in terms of \textit{energy gap protection}. EGP has been shown to mitigate errors for, e.g., by encoding logical qubits
into stabilizer codes in the context of adiabatic quantum computing. The composite Hilbert space corresponding to the system and the environment is now given by $\mathcal{H}_{sys}\otimes\mathcal{H}_{env}$. Hence the full dynamics is now governed by-
\begin{equation}
    H=H_{sys}\otimes\mathds{I}_{env}+\mathds{I}_{sys}\otimes H_{env}+\sqrt{\gamma}\sum_{\alpha}\hat{A}_{\alpha}\otimes \hat{B}_{\alpha}
\end{equation}
where $\hat{A}_{\alpha}$ and $\hat{B}_{\alpha}$ are system and bath operators, respectively, with system-environment coupling strength $\gamma$. Our objective in using EGP is to suppress this coupling. Particularly we focus here on $1/f$ noise, a decohering process with a noise power spectrum 
\begin{align}\label{eq:spectral}
S(\omega)=\frac{\gamma}{\lvert\omega\rvert^\beta},
\end{align}
where $\omega$ is the frequency, and $0<\beta<2$. This type of noise is ubiquitous in nature, especially in condensed matter systems in quasi-equilibrium (for $\beta\approx1$) and electronic equipment, but this signal can also be found in biological systems, music, and even in economics. In particular, as mentioned above, it is present in SQUIDs, which can lead to adverse effects on quantum simulation platforms based on superconducting qubits \cite{Yoshihara2006,Kakuyanagi2007,Bialczak2007,Bylander2011,Wang2015,Kumar2016}. 
\\
\newline
The underlying bath or microscopic model generating this noise can be approximated as a set of classically modeled random telegraph noise sources , which is used to classically model an environment for solid-state devices
where the system is considered to be interacting with a bistable fluctuator. Meaning, one treats the bath operator as a classical time-dependent parameter 
 that randomly flips between two
values $B(t)=\pm 1$ with a switching rate $r$ describing a system-bath coupling for a single qubit system. The Hamiltonian that incorporates the aforementioned noise source exhibits an exponential decay in the correlation function of the fluctuating quantity $B(t)$, described by $e^{-r |t|}$. Therefore the corresponding noise spectrum is a Lorentzian function that reads as
\begin{eqnarray}
\nonumber
S(\omega,r) &=& \int_{-\infty}^{+\infty} dt e^{i\omega t}~ \overline{B(t) B(0)}  \\
&=& \int_{-\infty}^{+\infty} dt e^{i\omega t} e^{-r |t|} = \frac{1}{\pi} \frac{r}{\omega^2 + r^2} .
\label{rtn1}
\end{eqnarray}
where $\overline{B(t) B(0)}$ denotes the classical statistical average.
In order to approximate the $1/f^{\alpha}$ spectrum, the power spectrum of a single Random-Telegraph Noise (RTN) is integrated over  switching rates $r$ using a probability distribution which serves as the density of transition rates.
\begin{eqnarray}
S_{1/f^{\alpha}}(\omega) = \int_{r_1}^{r_2} S(\omega,r) p_{\alpha}(r) dr,
\label{rtn2}
\end{eqnarray}

To emulate a $1/f^{\alpha}$ noise spectrum, the distribution of switching rates, $p_{\alpha}(r)$, is assumed to be proportional to $1/r^{\alpha}$. Upon performing the integration as shown in Equation (\ref{rtn2}), the resulting spectrum exhibits $1/\omega^{\alpha}$ behavior within a specific frequency range, satisfying the condition 
$r_1 \le{\omega} \le {r_2}$.
\\
\newline
Since the spectral density of $1/f$, noise is mostly concentrated in
the low-frequency range; therefore, the energy gap needed to suppress this noise is feasible and within reach of the current state-of-the-art quantum simulators. Hence one can anticipate that EGP can be used to suppress $1/f$ noise too.
\section{Computing dissipative dynamics}
To theoretically model the time evolution of the state of an open quantum system. We must treat such a state within the density matrix formalism since decoherence makes any time-evolved state more and more mixed. A common approach is to write a Markovian quantum master equation (QME) in Lindblad form which generalizes Schrodinger equation to systems which is in contact with the environment\cite{10.1063/1.522979} ,
\begin{equation}\label{lind}
    \frac{d}{d t} \hat{\rho}_{\mathrm{S}}=-\frac{i}{\hbar}\left[\hat{H}_{\mathrm{S}}, \hat{\rho}_{\mathrm{S}}\right]+\sum_{i} \gamma_{i}\left(\hat{L}_{i} \hat{\rho}_{\mathrm{S}} \hat{L}_{i}^{\dagger}-\frac{1}{2}\left\{\hat{L}_{i}^{\dagger} \hat{L}_{i}, \hat{\rho}_{\mathrm{S}}\right\}\right)
\end{equation}

The Lindblad operators $\hat{L}_{i}$ and their corresponding damping rates $\gamma_i$ describe the dissipation channels in open systems. A common approach is to use the Lindblad equation (Equation \eqref{lind}) and select suitable Lindblad operators that represent the desired dissipation channels. This method has the advantage of maintaining positivity. However, it may result in a loss of physical intuition about the sources of dissipation.

Another approach - the one we will adopt throughout this thesis, takes into account systems with varying energy biases and eigenstates that couple to an environment in an established manner (through a physically motivated system-environment interaction operator). In such systems, it is often advantageous to derive a master equation from more fundamental physical principles and relate it to, for example, the noise-power spectrum of the environment. Since we \textit{a priori} know the noise power spectrum of the environment, we employ the Bloch--Redfield formalism \cite{cohen1992atom,breuer2002theory} to derive a master equation from a microscopic perspective.

The Bloch-Redfield formalism is a classic instance of an approach for deriving a master equation from a microscopic system. Under the assumption of weak system-environment coupling, a perturbative master equation for the system alone can be derived from a combined system-environment perspective. A benefit of this method is that the dissipation processes and rates are derived directly from environmental properties. In the following section, we provide a concise derivation of our model in the framework of Bloch-Redfield formalism which forms the basis of all our results throughout this thesis and discuss its implementation in QuTiP. \cite{Johansson2012,Johansson2013}.

\subsection{Bloch-Redfield master equation}

We consider a system $\hat{H}_S$ coupled to a bath (the environment) $\hat{H}_{B}$ with the interaction Hamiltonian 
\begin{equation}
    \hat{H}_{SB}=\sqrt{\gamma}\sum_{\alpha}\hat{A}_{\alpha}\otimes \hat{B}_{\alpha}, 
\end{equation}

In general, the system operators $\hat{A}_{\alpha}$ do not preserve Gauss's law and is responsible for the gauge violating transitions in our system characterized by the transition frequencies $\omega$ , hence $S(\omega)$ acts like an effective transition rate out of our gauge invariant subspace. 
\newline
\\
Under the assumption of weak system-environment coupling, we obtain a master equation in terms of system operators and correlation functions that characterize the statistical properties of the bath. We denote tilde on quantities written in the interaction picture
Going into the interaction picture with respect to $\hat{H}_S+\hat{H}_B$ via the operators $\hat{U}_S=e^{-it\hat{H}_S}$ and $\hat{U}_B=e^{-it\hat{H}_B}$, we start by writing the von-Neumann equation
\begin{align}
d_t\hat{\tilde{\rho}}_{SB}(t)=-i\left[\hat{\tilde{H}}_{SB}(t), \hat{\tilde{\rho}}_{S B}(t)\right].    
\end{align}
After substituting the integrated solution into the equation of motion for the combined system, we can obtain the evolution of the reduced density matrix of the system in the interaction picture as
\begin{align}
d_t \hat{\tilde{\rho}}(t)=-\operatorname{Tr}_{B}\left\{\left[\hat{\tilde{H}}(t), \int_{0}^{t} d s\left[\hat{\tilde{H}}(s), \hat{\tilde{\rho}}_{S B}(s)\right]\right]\right\},
\end{align}
where $\hat{\tilde{H}}(t)=\sqrt{\gamma}\sum_{\alpha}\hat{\tilde{A}}_{\alpha}(t)\otimes {\hat{\tilde{B}}_{\alpha}(t)}$. After the change of variables $\tau=t-s$, we get
\begin{align}\label{eq:vn}
d_t \hat{\tilde{\rho}}(t)=-\operatorname{Tr}_{B}\left\{\left[\hat{\tilde{H}}(t), \int_{0}^{t} d s\left[\hat{\tilde{H}}(t-\tau), \hat{\tilde{\rho}}_{S B}(t-\tau)\right]\right]\right\}.
\end{align}

We further proceed to use a Born approximation where we assume the state of the composite system is always uncorrelated and hence can be factorized as $\hat{\tilde{\rho}}_{SB}=\hat{\tilde{\rho}}(t)\otimes{\hat{\rho}_{B}}$, also assuming the bath is much larger than the system in question.
Further, we introduce the Markov approximation, where we assume that the bath has a very short correlation time $\tau_{B}$, i.e., that the correlation function,
\begin{equation}
    C_{\alpha \nu}(\tau)=\gamma \operatorname{Tr}_{B}\left[\hat{\tilde{B}}_{\alpha}(t) \hat{\tilde{B}}_{\nu}(t-\tau) \hat{\rho}_{\tilde{B}}\right]=\gamma\left\langle \hat{\tilde{B}}_{\alpha}(\tau) \hat{\tilde{B}}_{\nu}(0)\right\rangle
\end{equation}
 decays rapidly with some characteristic timescale $\left|C_{\alpha \nu}(\tau)\right| \sim e^{-\tau / \tau_{B}}$. In the limit of $\tau_B\to0$ and replacing $\hat{\tilde{\rho}}(t-\tau)$ with $\hat{\tilde{\rho}}(t)$, which is possible due to the fact that correlation function is negligible for $\tau\gg\tau_B$, and under the assumption that $t\gg\tau_B$, one obtains a memory-less evolution of the density matrix. 
It then becomes also a good approximation to extend the integration to infinity as the integrand vanishes sufficiently fast for $\tau\gg\tau_B$, making it a fully Markovian equation. 
These approximations ensure the trace-preserving nature of the density matrix throughout the time evolution.
\newline
\\
However, the master equation that is obtained is still often times known to give rise to evolution which is not completely positive. Therefore, a secular approximation which is also known as the rotating wave approximation is then used to make the evolution of the resulting dynamical map completely positive (CPTP) \cite{Albash_2012,Davies1976,1974CMaPh..39...91D}. Writing Eq.~\eqref{eq:vn} in terms of system operators and bath correlation functions, one obtains after evaluating the partial trace
\begin{align}\nonumber
d_t\hat{\tilde{\rho}}(t)&=- \sum_{\alpha \nu} \int_{0}^{\infty} d \tau\bigg\{C_{\alpha \nu}(\tau)\Big[\hat{\tilde{A}}_{\alpha}(t) \hat{\tilde{A}}_{\nu}(t-\tau)  \hat{\tilde{\rho}}(t)-\\\nonumber
&\hat{\tilde{A}}_{\alpha}(t-\tau)  \hat{\tilde{\rho}}(t) \hat{\tilde{A}}_{\nu}(t)\Big]
+C_{\alpha \nu}(-\tau)\Big[ \hat{\tilde{\rho}}(t) \hat{\tilde{A}}_{\alpha}(t-\tau) \hat{\tilde{A}}_{\nu}(t)\\\label{eq:bre4}
& -\hat{\tilde{A}}_{\alpha}(t)  \hat{\tilde{\rho}}(t) \hat{\tilde{A}}_{\nu}(t-\tau)\Big]\bigg\}.
\end{align}
Going into the frequency domain and expanding in the eigenbasis of the system Hamiltonian $\hat{H}_S$, we can write the operators acting on the system as
\begin{align}\nonumber
\hat{\tilde{A}}_{\alpha}(t)&=\sum_{m, n} e^{-i\left(\epsilon_{m}-\epsilon_{n}\right) t}\left|\epsilon_{n}\right\rangle\left\langle\epsilon_{n}\left|\hat{A}_{\alpha}\right| \epsilon_{m}\right\rangle\left\langle\epsilon_{m}\right|\\\label{eq:Bre5}
&=\sum_{m,n} A_{mn}(\omega) e^{-i \omega_{mn} t},
\end{align}
where we have defined the transition frequencies $\omega_{mn}=\epsilon_{m}-\epsilon_{n}$. 
In the Schr\"odinger picture, we obtain the master equation in matrix form after substituting Eq.~\eqref{eq:Bre5} into Eq.~\eqref{eq:bre4} as
\begin{align}\nonumber\label{eqn shr}
d_t\rho_{a b}(t) & =-i \omega_{a b} \rho_{a b}(t)- \sum_{\alpha, \nu} \sum_{c, d} \int_{0}^{\infty} d \tau  \bigg\{C_{\alpha \nu}(\tau)\Big[\delta_{b d}\\ \nonumber
&\times\sum_{n} A_{a n}^{\alpha} A_{n c}^{\nu} e^{i \omega_{c n} \tau}-A_{a c}^{\alpha} A_{d b}^{\nu} e^{i \omega_{c a} \tau}\Big]\\\nonumber
&+C_{\alpha \nu}(-\tau)\Big[\delta_{a c} \sum_{n} A_{d n}^{\alpha} A_{n b}^{\nu} e^{i \omega_{n d} \tau}\\
&-A_{a c}^{\alpha} A_{d b}^{\nu} e^{i \omega_{b d} \tau}\Big]\bigg\} \rho_{c d}(t).
\end{align}

 The spectral function $S_{\alpha \nu}(\omega)$, after neglecting a small energy shift arising due to the imaginary part in the Fourier transform of $C_{\alpha \nu}(\tau)$ can be written as:
\begin{align}\label{eq:sfn}
S_{\alpha \nu}(\omega)=2\int_{0}^{\infty}d \tau e^{i\omega\tau}C_{\alpha \nu}(\tau).
\end{align}
Further substituting the expression for spectral function of Eq.~\eqref{eq:sfn} in the above Eqn. under the assumptions of vanishing cross correlations between different environment operators acting at different particle sites, i.e, ${C}_{\nu \alpha}(\tau)={C}_{\alpha \nu}(\tau)=\delta_{\alpha \nu}{C}_{\nu}(\tau)$, one can show that the final form of the Bloch--Redfield master equation, describing the evolution of the reduced density matrix for the system, after employing the Born, Markov, and the secular approximation  can be written explicitly as,
\begin{align}\label{eq:redfield}
d_t\rho_{a b}(t)=-i \omega_{a b} \rho_{a b}(t)+\sum_{c, d} R_{a b c d} \rho_{c d}(t),
\end{align}
where $R_{a b c d}$ is the Bloch--Redfield relaxation tensor, which can be written in matrix form with $\hat{A}_{\alpha}$ assumed to be Hermitian for ease of numerical implementation,
\begin{align}\nonumber
R_{a b c d}=&-\frac{1}{2} \sum_{\alpha}\bigg[\delta_{b d} \sum_{n} A_{a n}^{\alpha} A_{n c}^{\alpha} S_{\alpha}(\omega_{c n})\\\nonumber
&-A_{a c}^{\alpha} A_{d b}^{\alpha} S_{\alpha}(\omega_{c a})+\delta_{a c} \sum_{n} A_{d n}^{\alpha} A_{n b}^{\alpha} S_{\alpha}(\omega_{d n})\\
&-A_{a c}^{\alpha} A_{d b}^{\alpha} S_{\alpha}(\omega_{d b})\bigg].
\end{align}
The Redfield tensor contains all the information about the dissipative processes that arise due to the coupling of the system with the bath degrees of freedom.
\\
\newline
One requirement for the validity of the Bloch--Redfield approach is the smallness of the Bloch--Redfield decay rates that describe the effective incoherent coupling between two eigenlevels $i$ and $f$  against the relevant transition frequencies $\omega_{if}$ \cite{PhysRevA.80.022303}. 
The Bloch--Redfield decay rates, also known as the golden rule rates, are defined as 
$\Gamma_{if}\propto{\sum_{\mathbf{\alpha}}\left|\left\langle i\left|\hat{A}_{\alpha}\right| f\right\rangle\right|^2 S_\alpha\left(\omega_{if}\right)}$. 
We checked for the numerical models that the condition $\Gamma_{if}\ll\omega_{if}$ was always satisfied. 
In particular, 
as the system operators $\hat{A}_{\alpha}$ violate Gauss's law, the relevant incoherent transitions happen on large energy scales of order $V$, where the noise spectrum becomes weak, thus further solidifying our approach for employing this formalism.

As $1/f$ noise and other types of decoherence can drastically undermine performance in an experimental setup, it becomes important to find ways that may ameliorate its effect. Left unchecked, decoherence can lead to a fast buildup in the gauge violation, which renders the quantum simulation of true gauge-theory dynamics unfaithful \cite{halimeh2020f,halimeh2020g}.

\subsection{Numerical implementation in QuTiP}
The Bloch-Redfield tensor can be computed in QuTiP using 
the function qutip.bloch\_redfield tensor.
There are three required arguments: The system Hamiltonian $H$, a list of operators through which to the bath $A_{\alpha}(\omega)$, and a list of spectral density functions $S_{\alpha}(\omega)$ corresponding to those operators. The spectral density functions are callback functions that accept a single (angular) frequency argument. This function also returns eigen kets and a basis, which comes in handy while calculating the time evolution in the instantaneous eigenbasis of the hamiltonian.

The evolution of a wavefunction or density matrix, according to the Bloch-Redfield master equation, can be calculated using the 
ode solver available in QuTiP by taking in arguments the Redfield tensor, the list of eigenkets , the initial state (as a ket or density matrix), and a list of times for which to evaluate the time-evolved density matrices.

When implementing BR theory numerically, it is important to keep track of the required computational resources since, as the size of the
system increases, solving density operator master equations can become computationally demanding. Calculating the BlochRedfield tensor $R_{abcd}$ becomes substantially more demanding. This is
because each system coupling operator $A_{\alpha}$ may contribute to any of the $d^{2}$ 
transitions in the eigenbasis of $H_S$, where $d$ is the dimension of the Hilbert space of the system.
Therefore, when constructing a Redfield tensor from $N$ coupling operators, we may need to perform a number of
operations that scales with $N^{2}\cross d^{2}$.

\section{Results and discussion}\label{sec:results}

We now present our numerical results on the quench dynamics of gauge theories subjected to $1/f$ noise, which we have computed using the exact diagonalization toolkit QuTiP \cite{Johansson2012,Johansson2013}. In all cases, we prepare our system in an initial state $\hat{\rho}_0$ in the target gauge sector $\mathbf{g}_\text{tar}$, and monitor its quench dynamics in the presence of $1/f$ noise with and without linear gauge protection. In particular, we will focus on the dynamics of the gauge violation,
\begin{align}\label{eq:viol}
\varepsilon(t)&=\frac{1}{L}\sum_{j=1}^{L}\Tr\Big\{\hat{\rho}(t)\big(\hat{G}_j-g_j^\text{tar}\big)^2\Big\},
\end{align}
where $\hat{\rho}(t)$ is the time-evolved density operator of the system at time $t$, in addition to calculating the dynamics of relevant local observables. Due to the large evolution times we investigate, we restrict our system size to $L=4$ sites due to computational overhead, and we employ periodic boundary conditions.

\begin{figure}[t!]
	\centering
	\includegraphics[scale=0.45]{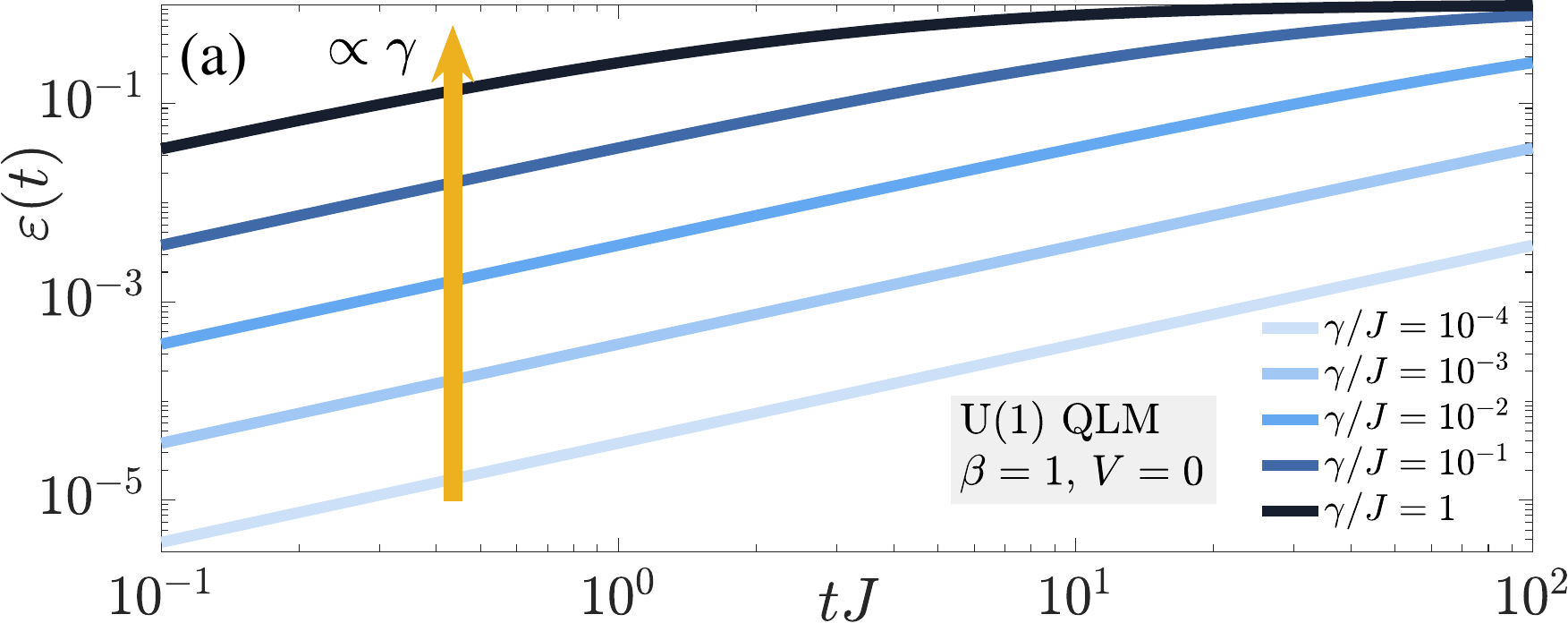}\\
	\vspace{1.1mm}
	\includegraphics[scale=0.45]{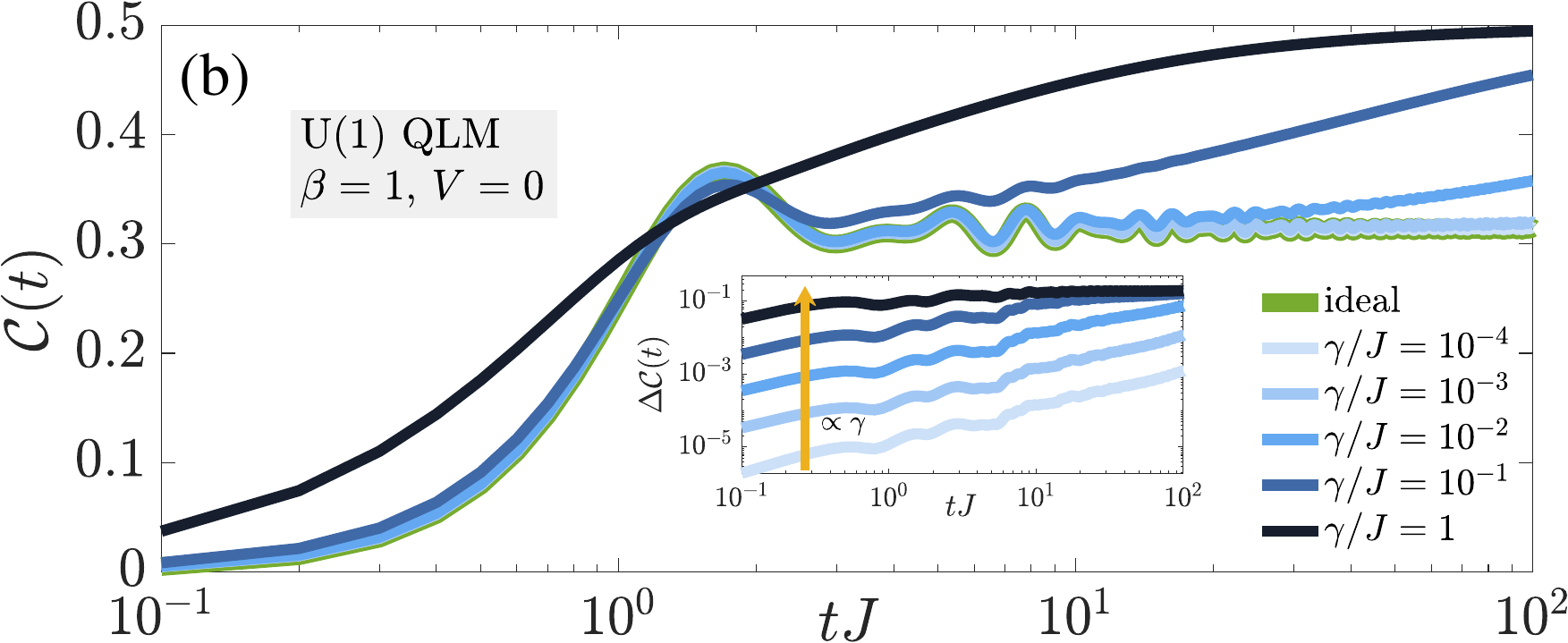}
	\caption{(Color online). (a) Quench dynamics of the gauge violation~\eqref{eq:viol} and (b) the chiral condensate~\eqref{eq:CC} in the presence of incoherent errors generated by the noise spectral function $S(\omega)=\gamma/|\omega|$ for various values of system-environment coupling strength $\gamma$ with quantum jump operators $\hat{A}_j^{m}=\hat{\sigma}^{x}_{j}$ and $\hat{A}_{j,j+1}^{g}=\hat{s}^{x}_{j,j+1}$ for matter and gauge fields, respectively, and without adding any protection terms, i.e., $V=0$. Here, the quench Hamiltonian is the $\mathrm{U}(1)$ quantum link model~\eqref{eq:U1QLM}, and the initial state is the gauge-invariant vacuum, with all matter sites empty while the local electric fields on odd (even) links point down (up). For both quantities, we see that errors evolve $\propto\gamma t$, and already small values of $\gamma$ significantly undermine gauge-theory dynamics.}
	\label{fig:U1_beta1_V0} 
\end{figure}

\subsection{$\mathrm{U}(1)$ quantum link model}
We first consider the spin-$1/2$ formulation of the $\mathrm{U}(1)$ quantum link model, 
\begin{align}\label{eq:U1QLM}
\hat{H}_0=\sum_{j=1}^{L}\left[J\left(\hat{\sigma}_j^- \hat{s}^+_{j,j+1}\hat{\sigma}_{j+1}^{-}+\text {H.c.}\right)+\frac{\mu}{2} \hat{\sigma}_j^z\right],
\end{align}
where on site $j$ the matter field is represented by the Pauli operator $\hat{\sigma}^z_j$, with $\mu$ denoting the fermionic mass, the gauge (electric) field on the link between sites $j$ and $j+1$ is denoted by the spin-$1/2$ operator $\hat{s}^+_{j,j+1}$ ($\hat{s}^z_{j,j+1}$), $L$ is the total number of sites with periodic boundary conditions enforced, and the overall energy scale is set by the coupling strength $J=1$. note that we have dropped the electric field term, since for the spin-$1/2$ case this only accounts for a constant energy shift.The generator of the $\mathrm{U}(1)$ gauge symmetry of Hamiltonian~\eqref{eq:U1QLM} is given by
\begin{align}\label{eq:GjU1}
\hat{G}_j=(-1)^j\bigg(\hat{s}^z_{j-1,j}+\hat{s}^z_{j,j+1}+\frac{\hat{\sigma}^z_j+1}{2}\bigg).
\end{align}
The model~\eqref{eq:U1QLM} is a quantum link formulation \cite{Chandrasekharan1997} of lattice quantum electrodynamics in $1+1$D, and is experimentally very relevant as it has been the subject of recent large-scale cold-atom quantum simulations \cite{Yang2016,Zhou2021}.

We now prepare the system in a vacuum state, which is one of two doubly degenerate eigenstates of Hamiltonian~\eqref{eq:U1QLM} at $\mu/J\to\infty$. This initial state is in the target sector $g_j^\text{tar}=0,\,\forall j$, i.e., $\Tr\{\hat{\rho}_0\hat{G}_j\}=0,\,\forall j$, where its sites host no matter and the local electric fields are in a staggered formation. We then quench this vacuum state with $\hat{H}_0+V\hat{H}_G$ at $\mu/J=0.5$ in the presence of $1/f$ noise with power spectrum~\eqref{eq:spectral} and jump operators $\hat{A}_j^{m}=\hat{\sigma}^{x}_{j}$ and $\hat{A}_{j,j+1}^{g}=\hat{s}^{x}_{j,j+1}$, which couple the matter and gauge fields to the environment, respectively. Let us first consider the case without protection, i.e., $V=0$, shown in Fig.~\ref{fig:U1_beta1_V0} setting $\beta=1$. The dynamics of the gauge violation~\eqref{eq:viol} is shown for various values of the system-environment coupling strength $\gamma$ in Fig.~\ref{fig:U1_beta1_V0}(a). At early times, the violation grows $\propto\gamma t$, as can be shown in time-dependent perturbation theory, until it begins to settle into a maximal violation plateau at a timescale $\propto1/\gamma$. We observe similar behavior in the chiral condensate, a measure of how strongly the dynamics spontaneously breaks the chiral symmetry associated with fermions in the theory, 
\begin{align}\label{eq:CC}
\mathcal{C}(t)&=\frac{1}{L} \sum_{j=1}^{L}\Tr\big\{\hat{\rho}(t)\hat{\psi}_j^{\dagger}\hat{\psi}_{j}\big\} =\frac{1}{2}+\frac{1}{2L} \sum_{j=1}^{L}\Tr\big\{\hat{\rho}(t)\hat{\sigma}^z_j\big\},
\end{align}
shown in Fig.~\ref{fig:U1_beta1_V0}(b). The error with respect to the ideal case, shown in the inset, grows $\propto\gamma t$ before settling into a maximal value at late times for sufficiently large $\gamma$. These results demonstrate the pernicious effect of $1/f$ noise on quantum simulations of gauge theories when left unprotected.

\begin{figure}[t!]
	\centering
	\includegraphics[scale=0.45]{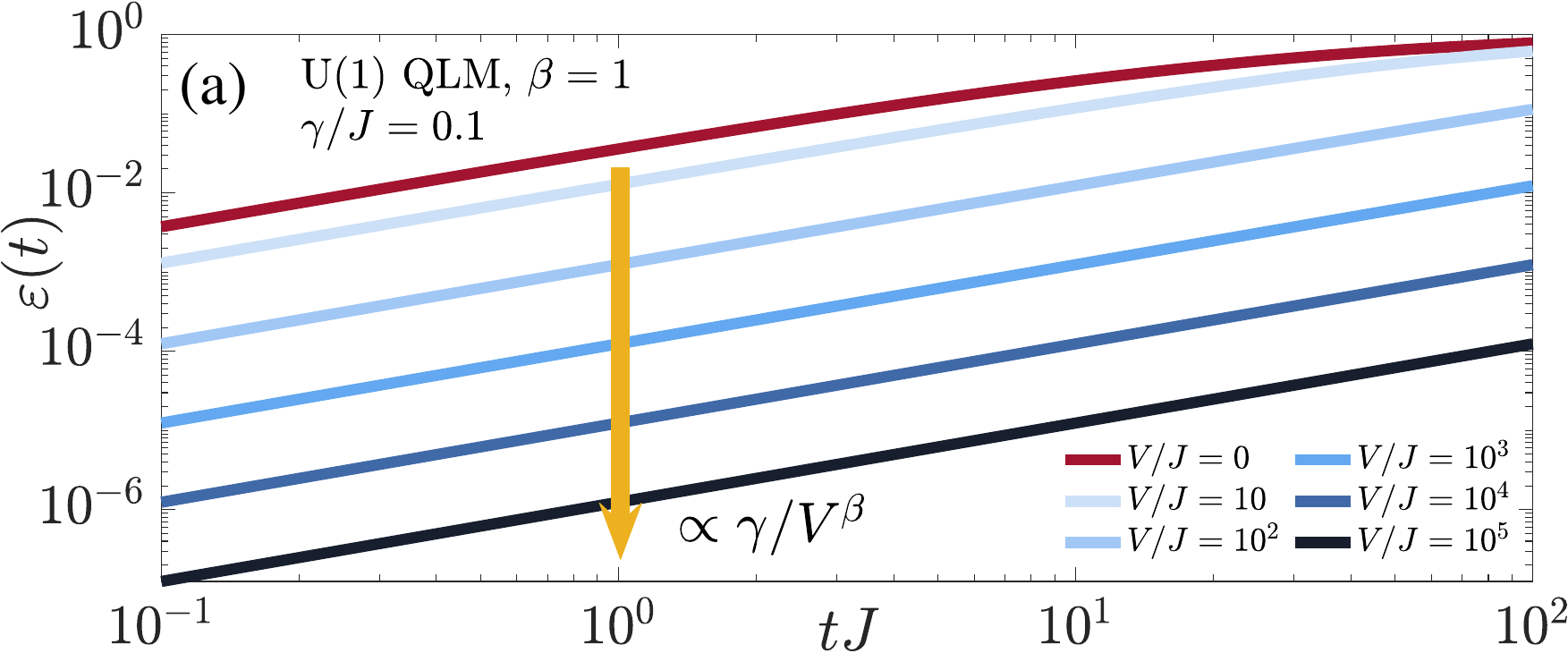}\\
	\vspace{1.1mm}
	\includegraphics[scale=0.45]{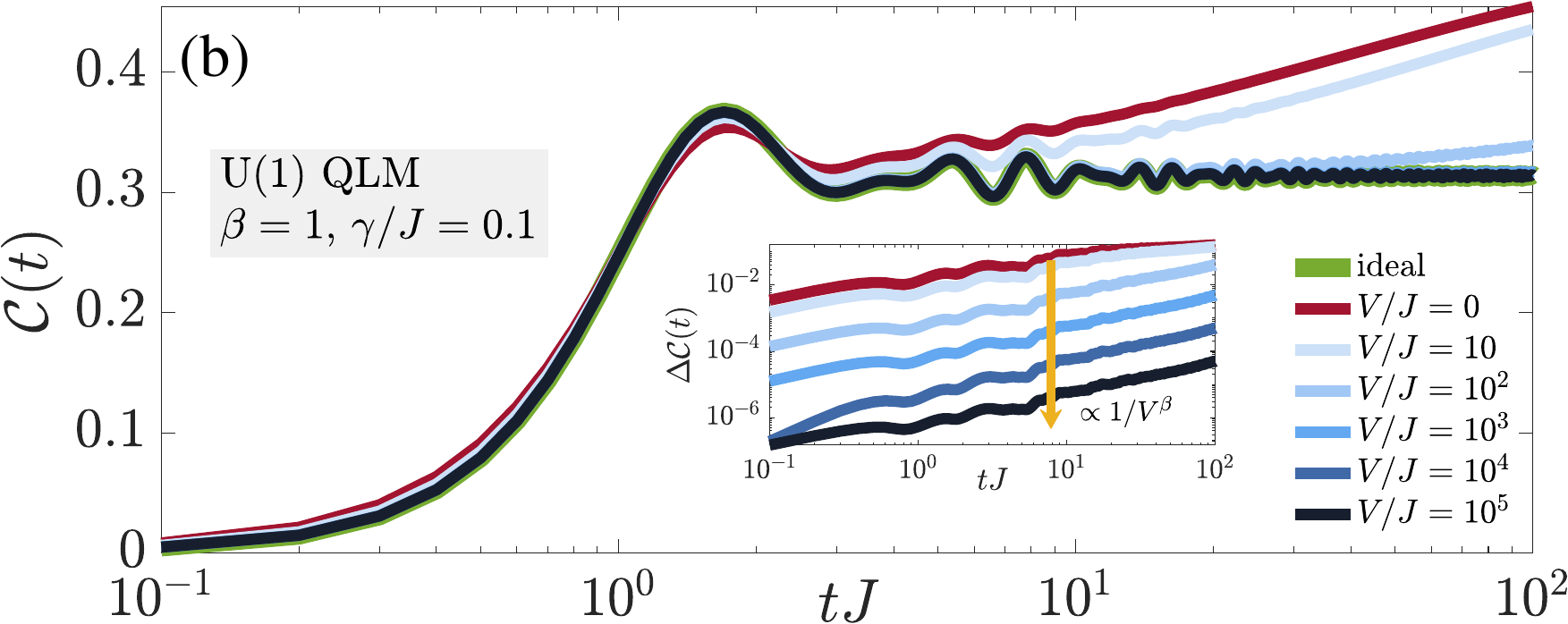}
	\caption{(Color online)(a) Quench dynamics of the gauge violation~\eqref{eq:viol} and (b) the chiral condensate~\eqref{eq:CC} in the presence of incoherent gauge-breaking errors generated by the noise spectral function $S(\omega)=\gamma/|\omega|$ at fixed system-environment coupling strength $\gamma=0.1J$ and with the linear gauge protection term~\eqref{eq:HG} turned on at various values of the protection strength $V$. We employ the compliant sequence $c_j\in\{-115,116,-118,122\}/122$. As we switch on the gauge protection, the growth of the gauge violation is suppressed as $\epsilon(t)\propto{\gamma t/V}$ until it starts to plateau at a timescale $\propto{V/(J\gamma)}$, extending the coherent lifetime of a potential experiment linearly in $V$. Similar conclusions can be drawn for the chiral condensate where in the presence of linear gauge protection, the ideal-theory dynamics is reproduced up to a timescale $\propto{V/(J\gamma)}$ with a deviation $\propto{\gamma/V}$ as shown in the inset.}
	\label{fig:U1_beta1_gamma0.1} 
\end{figure}

We now repeat the same quench protocol as in Fig.~\ref{fig:U1_beta1_V0}, but with fixed $\gamma{=}0.1J$ and the addition of the gauge protection~\eqref{eq:HG} at strength $V$, with the compliant sequence chosen to be $c_j=\{-115,116,-118,122\}/122$. The corresponding dynamics of the gauge violation is shown in Fig.~\ref{fig:U1_beta1_gamma0.1}(a), where we see a robust suppression in the growth of the gauge violation such that $\varepsilon(t)\propto\gamma t/V$ at short times, in agreement with time-dependent perturbation theory; see section~\ref{app:TDPT}. This suppression is also seen in the dynamics of the chiral condensate, shown in Fig.~\ref{fig:U1_beta1_gamma0.1}(b). Indeed, whereas the unprotected case (red curve) quickly and significantly diverges from the ideal case (green curve), at sufficiently large $V$ the agreement with the ideal case is excellent. The inset shows the deviation from the ideal case for the various considered values of $V$, where we find that the error grows roughly $\propto\gamma t/V$. These results show, therefore, that linear gauge protection extends the timescale of the dynamics during which one can perturbatively connect to a gauge theory from $\propto1/\gamma$ to $\propto V/(J\gamma)$. Even though linear gauge protection does not suppress the gauge violation into a long-lived plateau of constant value as it does in the case of purely coherent errors \cite{Halimeh2020e}, this is nevertheless a positive result that can allow one to enhance the achievable coherent evolution times significantly, and which can thus be of significant benefit to current and near-term gauge-theory quantum simulators.

\begin{figure}[t!]
	\centering
	\includegraphics[scale=0.45]{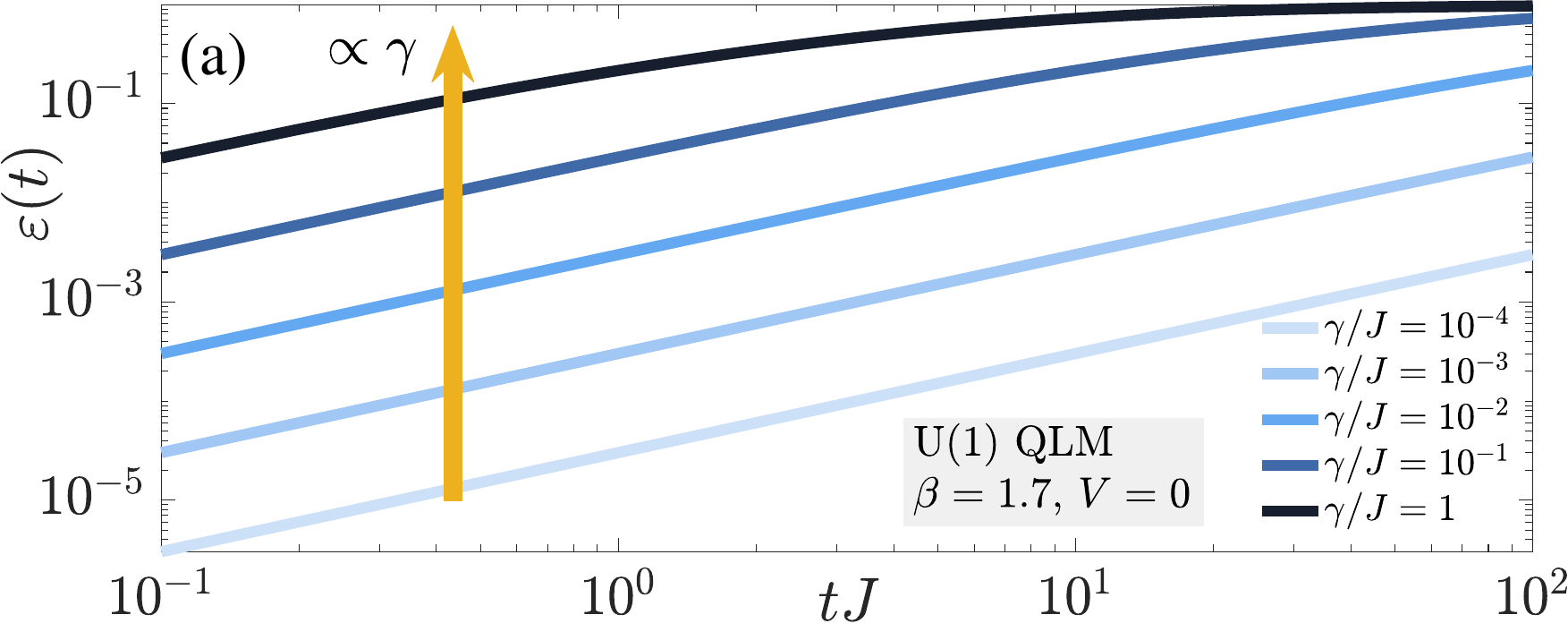}\\
	\vspace{1.1mm}
	\includegraphics[scale=0.45]{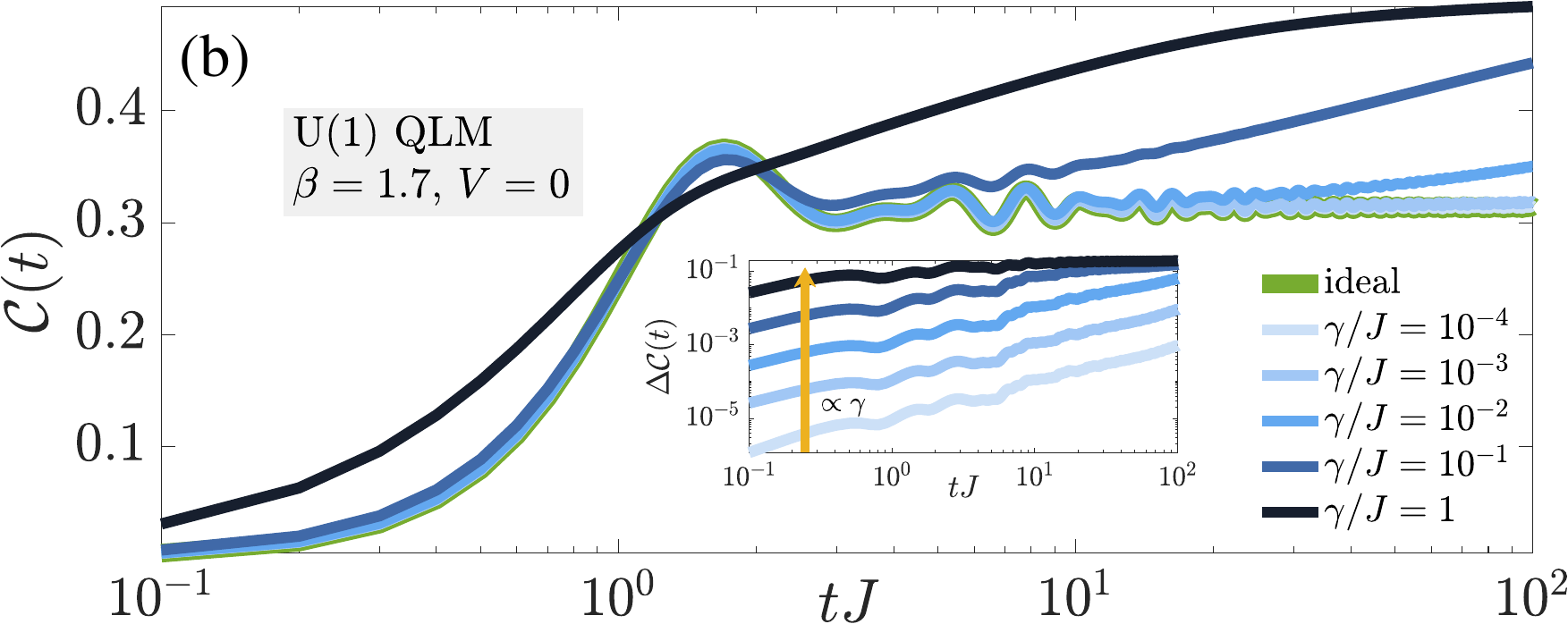}
	\caption{(Color online). Same as Fig.~\ref{fig:U1_beta1_V0} but for the noise spectral function  $S(\omega)=\gamma/|\omega|^{\beta}$ where $\beta=1.7$. The qualitative picture is identical to that of $\beta=1$ in Fig.~\ref{fig:U1_beta1_V0} for both the gauge violation and the chiral condensate, with only insignificant quantitative differences in these quantities between different values of $\beta$.}
	\label{fig:U1_beta1.7_V0} 
\end{figure}

\begin{figure}[t!]
	\centering
	\includegraphics[scale=0.45]{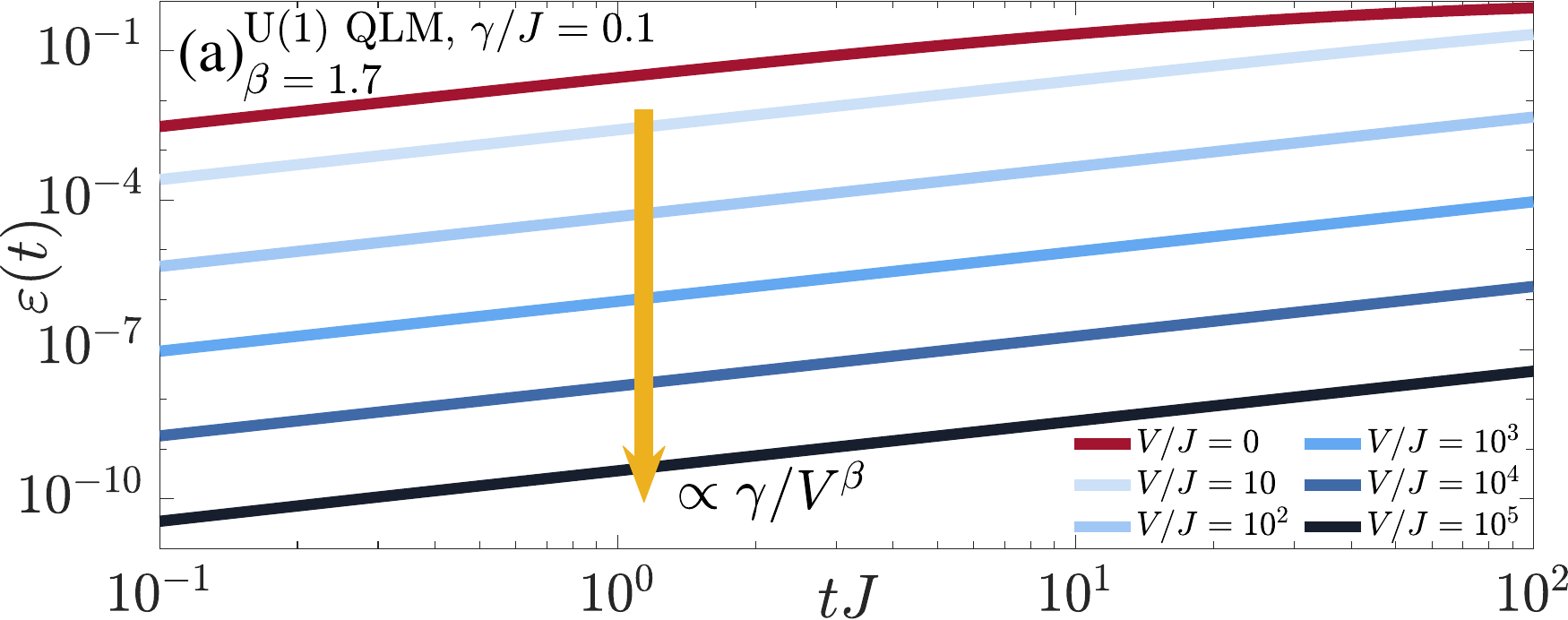}\\
	\vspace{1.1mm}
	\includegraphics[scale=0.45]{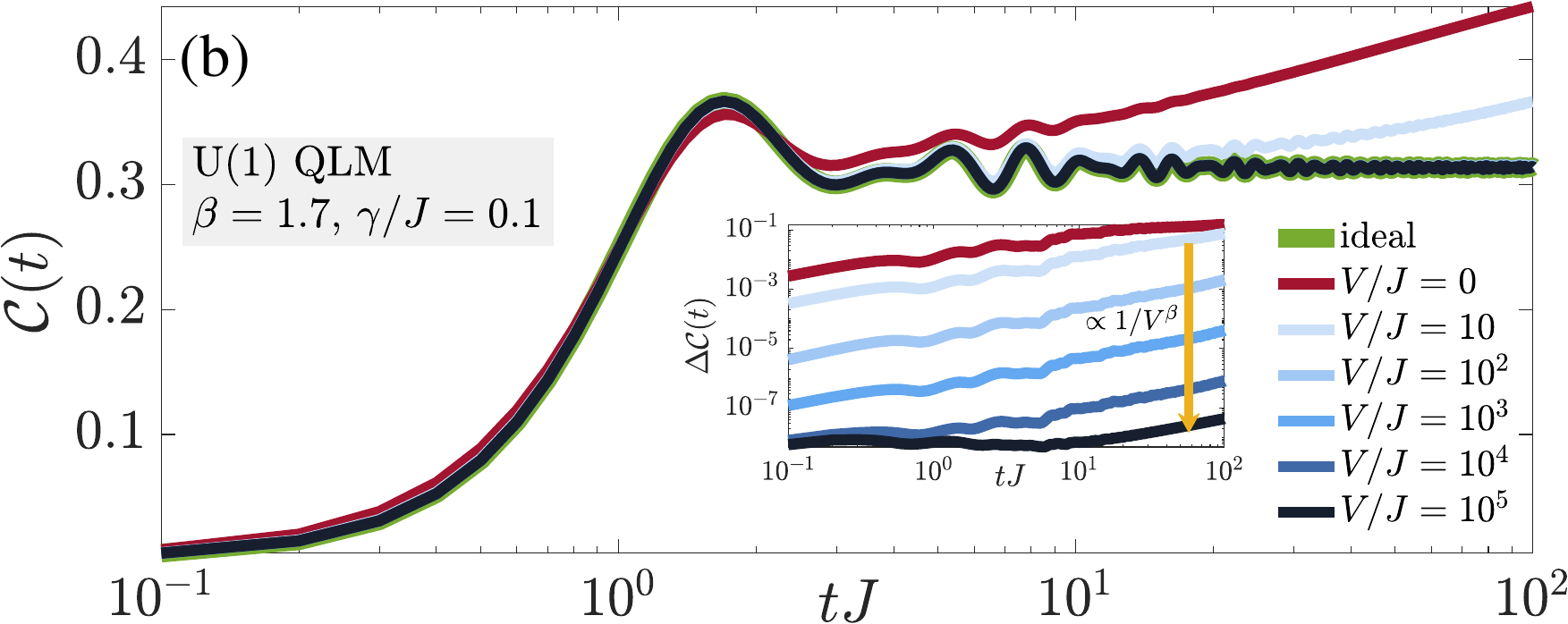}
	\caption{(Color online). Same as Fig.~\ref{fig:U1_beta1_gamma0.1}, but where $\beta=1.7$ in the noise spectral function  $S(\omega)=\gamma/|\omega|^{\beta}$. A qualitative difference arises whereby the gauge violation and the deviation of the chiral condensate from the ideal case both grow $\propto\gamma t/V^{1.7}$ instead of $\propto\gamma t/V$, showing that linear gauge protection suppresses errors more for a larger value of $\beta$.}
	\label{fig:U1_beta1.7_gamma0.1} 
\end{figure}
Let us now investigate the case of a fractional coefficient $\beta$ in the spectrum $S(\omega)=\gamma/\lvert\omega\rvert^\beta$. For this purpose, we repeat the above quench protocols for $\beta=1.7$. The protection-free case is shown in Fig.~\ref{fig:U1_beta1.7_V0}. The result is qualitatively similar to that of $\beta=1$ in Fig.~\ref{fig:U1_beta1_V0}. Indeed, the gauge violation grows $\propto\gamma t$ until a timescale $\propto1/\gamma$, where it begins to settle into a maximal-violation plateau, as can be seen for large enough values of $\gamma$; see Fig.~\ref{fig:U1_beta1.7_V0}(a). This type of behavior is replicated in the chiral condensate, as depicted in Fig.~\ref{fig:U1_beta1.7_V0}(b), where the deviation from the ideal case grows $\propto\gamma t$ at short times before beginning to plateau at $t\propto1/\gamma$. We can thus conclude that the effect of $\beta$ is merely quantitative in the case of no protection.

Upon employing gauge protection, the qualitative picture changes significantly. The gauge violation grows $\propto\gamma t/V^{1.7}$, as shown in Fig.~\ref{fig:U1_beta1.7_gamma0.1}(a) at fixed $\gamma=0.1J$. In other words, the suppression in the growth of the gauge violation directly depends on $\beta$, with greater suppression at larger $\beta$. This also happens in the case of the chiral condensate, shown in Fig.~\ref{fig:U1_beta1.7_gamma0.1}(b). We find that even though the unprotected case vastly deviates from the ideal one ($\gamma=V=0$), upon adding linear gauge protection, the chiral condensate faithfully reproduces the ideal case up to all accessible evolution times at sufficiently large $V$, with the deviation from the ideal case $\propto\gamma t/V^{1.7}$ (see inset).

This behavior can be explained in the following way. The spectral function of the considered decoherence process is $S(\omega)=\gamma/\lvert\omega\rvert^\beta$, where the relevant frequencies $\omega$ governing the system dynamics are those that create transitions between the target gauge sector and the other gauge sectors. Upon switching on the linear gauge protection, the undesired sectors are energetically separated from the target gauge sector proportionally to $V$. Hence, the relevant transition frequencies are on the order $\omega\sim V$. The strength of the spectral function thus scales as ${S}(\omega)\sim\gamma/ V^\beta$ and becomes weaker as $V$ increases.

It is worth noting that we have also checked that our conclusions hold for different jump operators, quench parameters (different values of $\mu/J$), and initial states, as well as for noncompliant sequences. See ~\ref{app:supp} for supplemental numerical results.

\subsection{$\mathbb{Z}_2$ lattice gauge theory}

\begin{figure}[t!]
	\centering
	\includegraphics[scale=0.45]{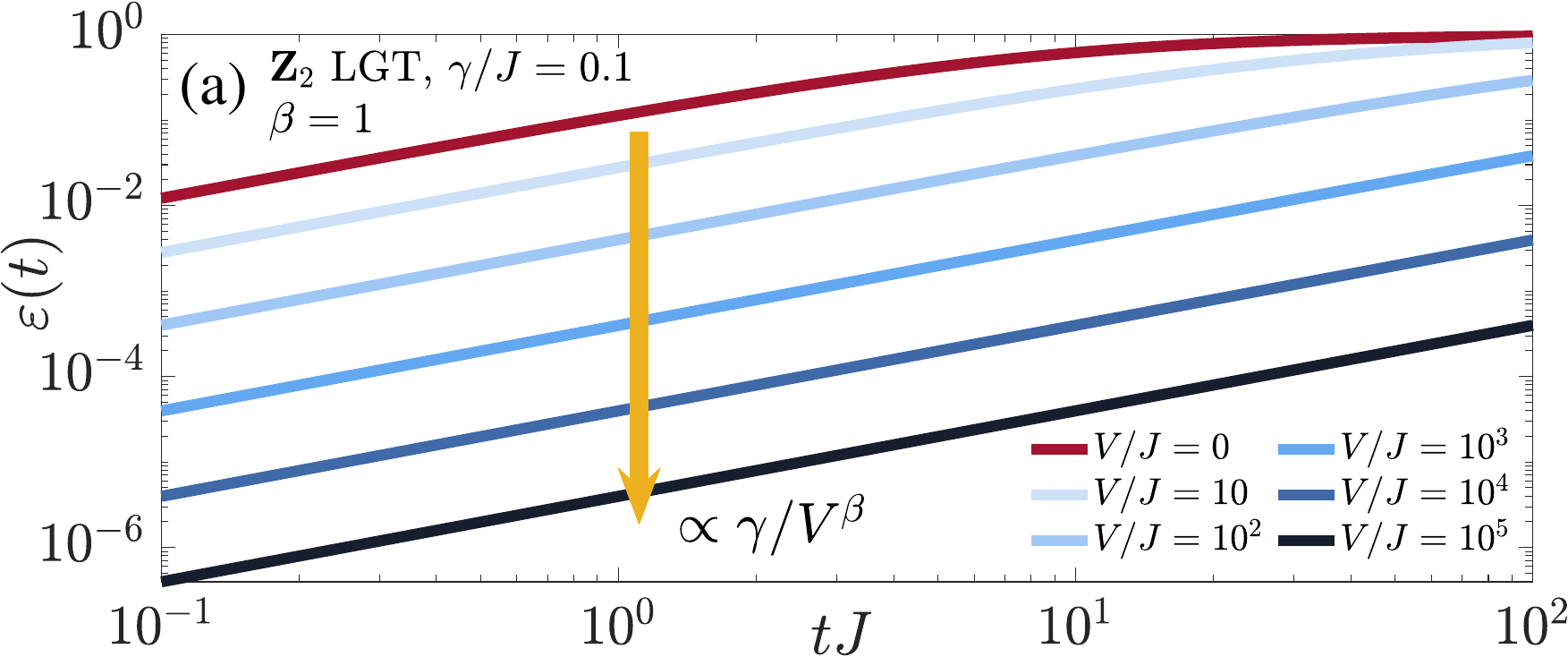}\\
	\vspace{1.1mm}
	\includegraphics[scale=0.45]{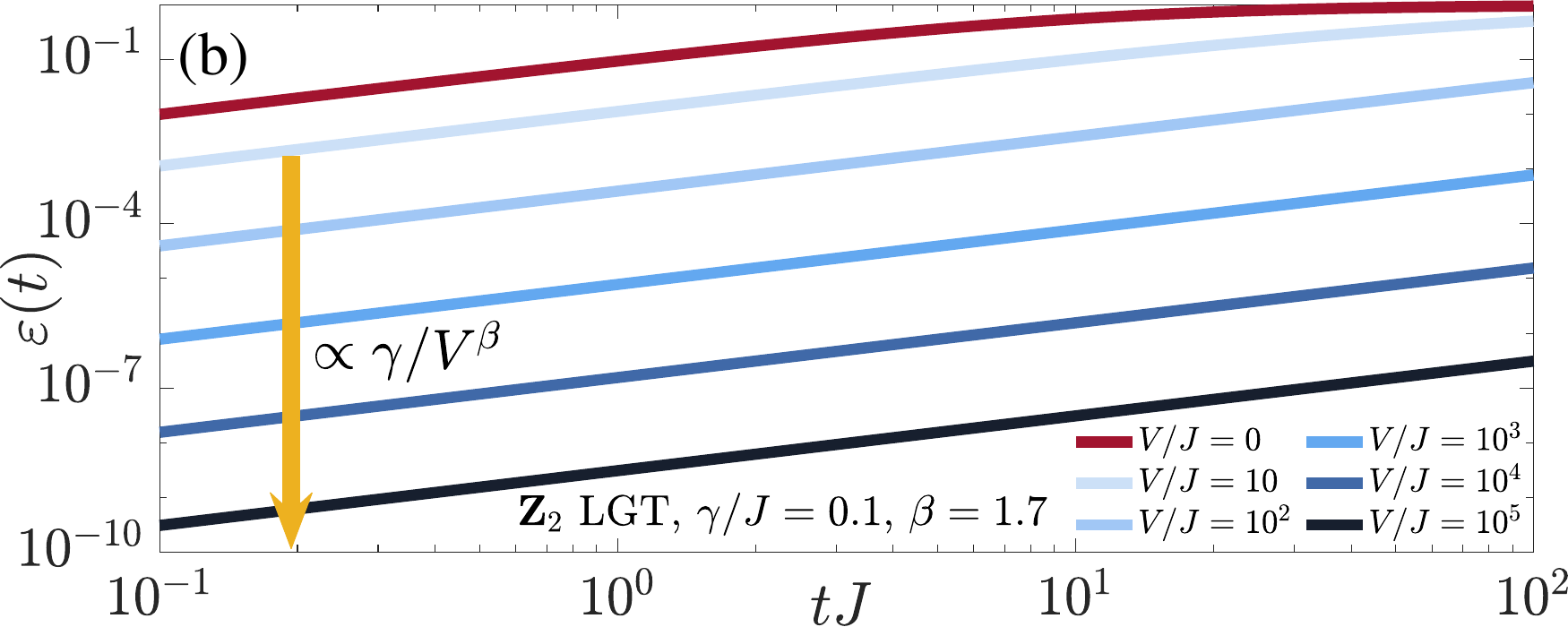}
	\caption{(Color online). Same as Figs.~\ref{fig:U1_beta1_gamma0.1}(a) and \ref{fig:U1_beta1.7_gamma0.1}(a), but for the $\mathbb{Z_2}$ lattice gauge theory~\eqref{eq:Z2LGT} and with the linear gauge protection term~\eqref{eq:HW} in the local pseudo generator, Eq.~\eqref{eq:Lpg}. The results are shown for the noncompliant sequence $[(-6)^j+5]/11$ with quantum jump operators operators $\hat{A}_j^{m}=\hat{a}_{j}+\hat{a}_{j}^\dagger$ and $\hat{A}_{j,j+1}^{g}=\hat{\tau}^{z}_{j,j+1}$. The qualitative conclusions are identical to the corresponding cases of the $\mathrm{U}(1)$ quantum link model where errors evolve $\propto\gamma t/V^\beta$, showcasing the generality of our findings.}
	\label{fig:Z2_gamma0.1} 
\end{figure}
To check the generality of the above findings, we now turn our attention to a different model, namely a $\mathbb{Z}_2$ lattice gauge theory that has been of recent theoretical  and experimental relevance \cite{Schweizer2019}.

We can now employ the concept of linear protection in terms of the local pseudogenerator according to Eq.~\eqref{eq:HW} in order to protect against $1/f$ noise in the $\mathbb{Z}_2$ lattice gauge theory. We prepare our system in a charge-density wave state in terms of the matter fields, with the electric fields aligned such that the system resides in the target sector $g_j^\text{tar}=+1,\,\forall j$. We quench this state with Hamiltonian~\eqref{eq:Z2LGT} at $h=0.54J$ in the presence of $1/f$ noise with the spectral function~\eqref{eq:spectral} and jump operators $\hat{A}_j^{m}=\hat{a}_{j}+\hat{a}_{j}^\dagger$ and $\hat{A}_{j,j+1}^{g}=\hat{\tau}^{z}_{j,j+1}$ coupling the matter and gauge fields, respectively, to the environment at a fixed value of $\gamma=0.1J$ and for several values of the protection strength $V$. The corresponding dynamics of the gauge violation is shown in Fig.~\ref{fig:Z2_gamma0.1}(a,b) for $\beta=1$ and $1.7$, respectively. The qualitative picture is identical to that of the $\mathrm{U}(1)$ quantum link model, where we find that at sufficiently large $V$ the gauge violation evolves $\propto\gamma t/V^\beta$ at short to intermediate times, before eventually plateauing at a maximal value that is delayed from a timescale $\propto 1/\gamma$ in the unprotected case to a timescale $\propto V^\beta/\gamma$ under linear gauge protection. 

We have also checked that these findings are valid for different initial states, quench parameters, and properly chosen sequenches $c_j$. As such, our conclusions are not specific to a given model, and we expect our findings to be general and applicable to any Abelian gauge theory.

\subsection{Supplemental numerical results for different initial state and sequence}\label{app:supp}
\begin{figure}[t!]
	\centering
	\includegraphics[scale=0.45]{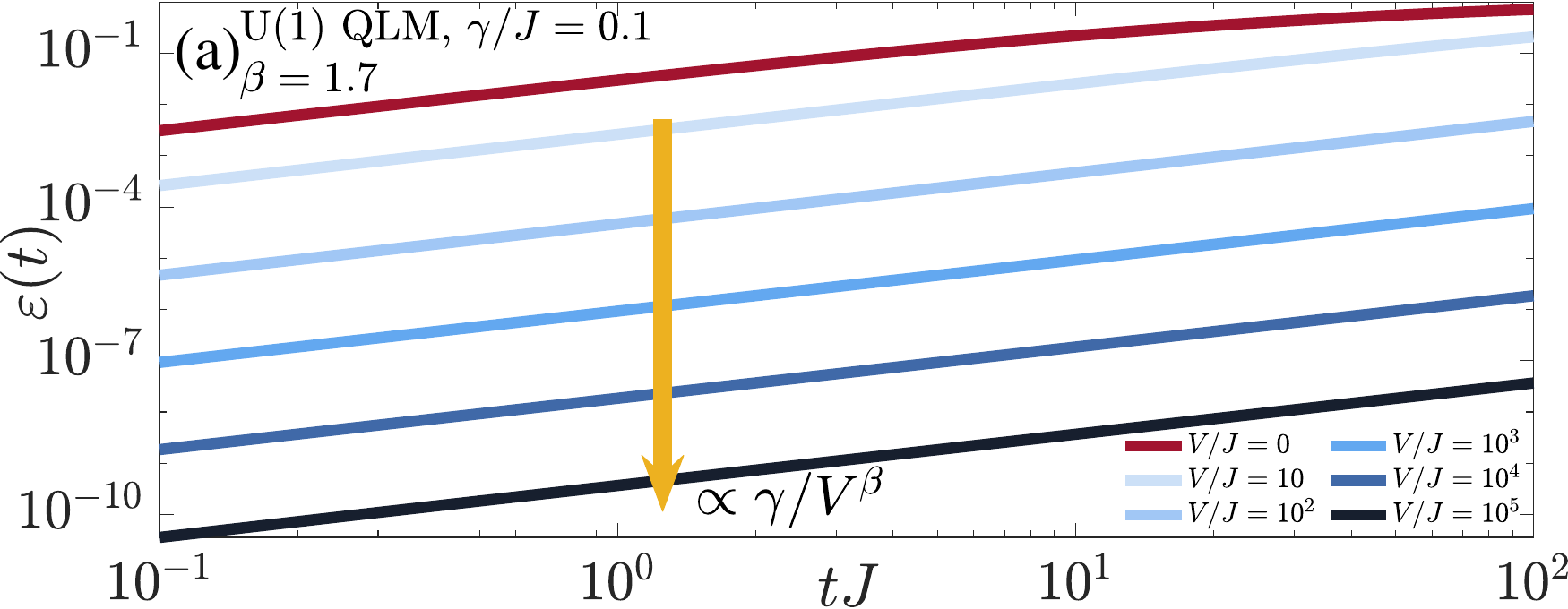}\\
	\vspace{1.1mm}
	\includegraphics[scale=0.45]{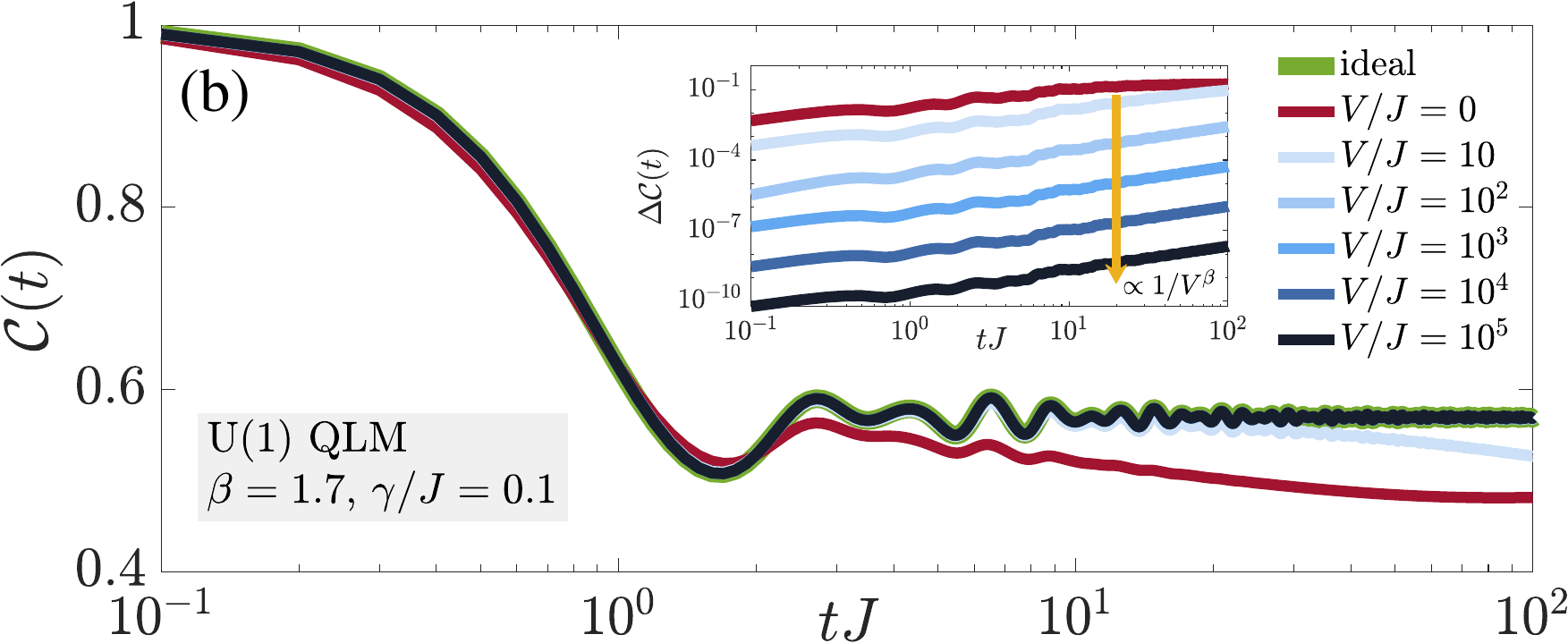}
 
	\caption{(Color online). Same as Fig.~\ref{fig:U1_beta1.7_gamma0.1}, but for a different gauge-invariant initial state, namely the charge-proliferated state where all matter sites are occupied and all local electric fields point down on their links. This state is also in the target sector $g_j^\text{tar}=0,\,\forall j$. The qualitative picture drawn in the main text is also valid here, where we see that the growth of gauge violation and errors in the local observables are both suppressed as $\propto{\gamma /V^{\beta}}$, indicating the independence of our conclusions from the choice of initial state.}
	\label{fig:U1_beta1.7_gamma0.1_CP} 
\end{figure}

The linear gauge protection scheme does not depend on the initial state, and will work effectively so long as the initial state is in the correct gauge sector(s) to be protected. We demonstrate this by repeating the results of Fig.~\ref{fig:U1_beta1.7_gamma0.1} but for a charge-proliferated state, which has every site occupied with matter, and all its local electric fields pointing down. The corresponding dynamics of the gauge violation and chiral condensate are shown in Fig.~\ref{fig:U1_beta1.7_gamma0.1_CP}(a,b), respectively, and the qualitative behavior is identical to that of the vacuum initial state in Fig.~\ref{fig:U1_beta1.7_gamma0.1}, with an error $\propto\gamma t/V^\beta$ in both cases.

\begin{figure}[t!]
	\centering
	\includegraphics[scale=0.45]{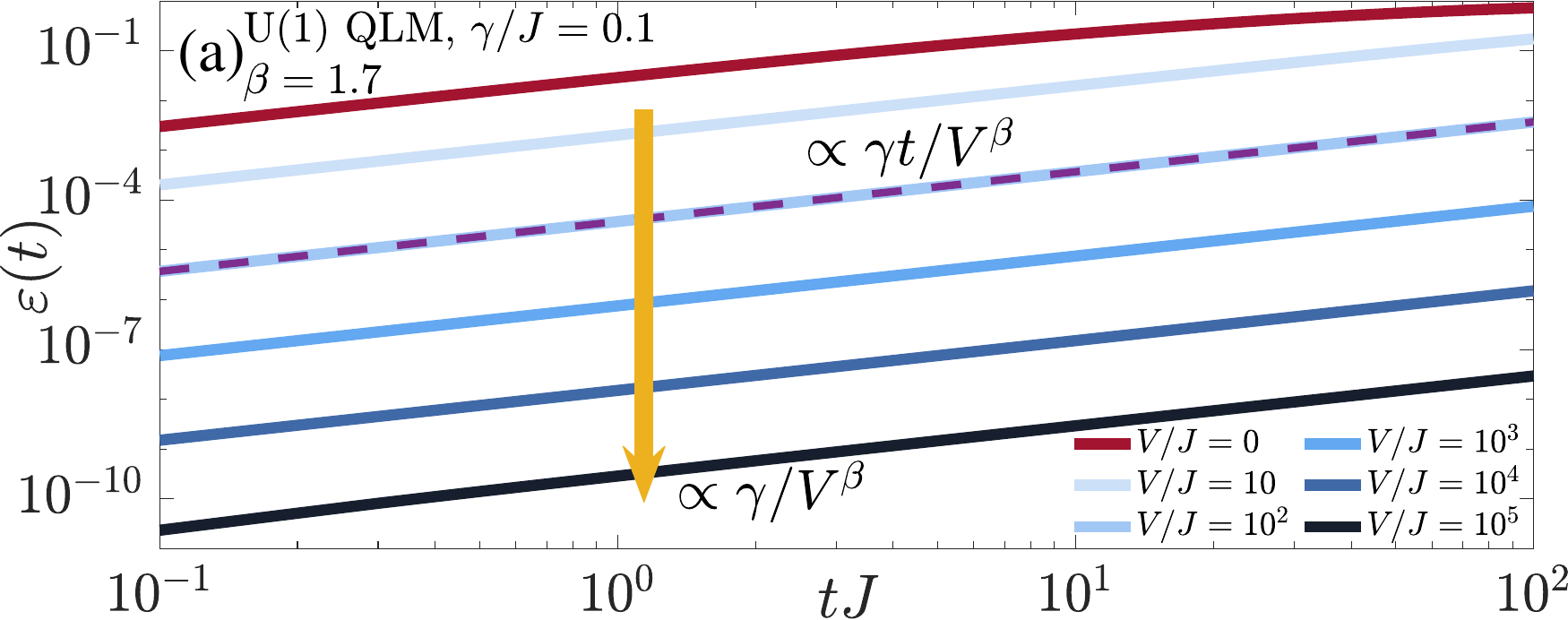}\\
	\vspace{1.1mm}
	\includegraphics[scale=0.45]{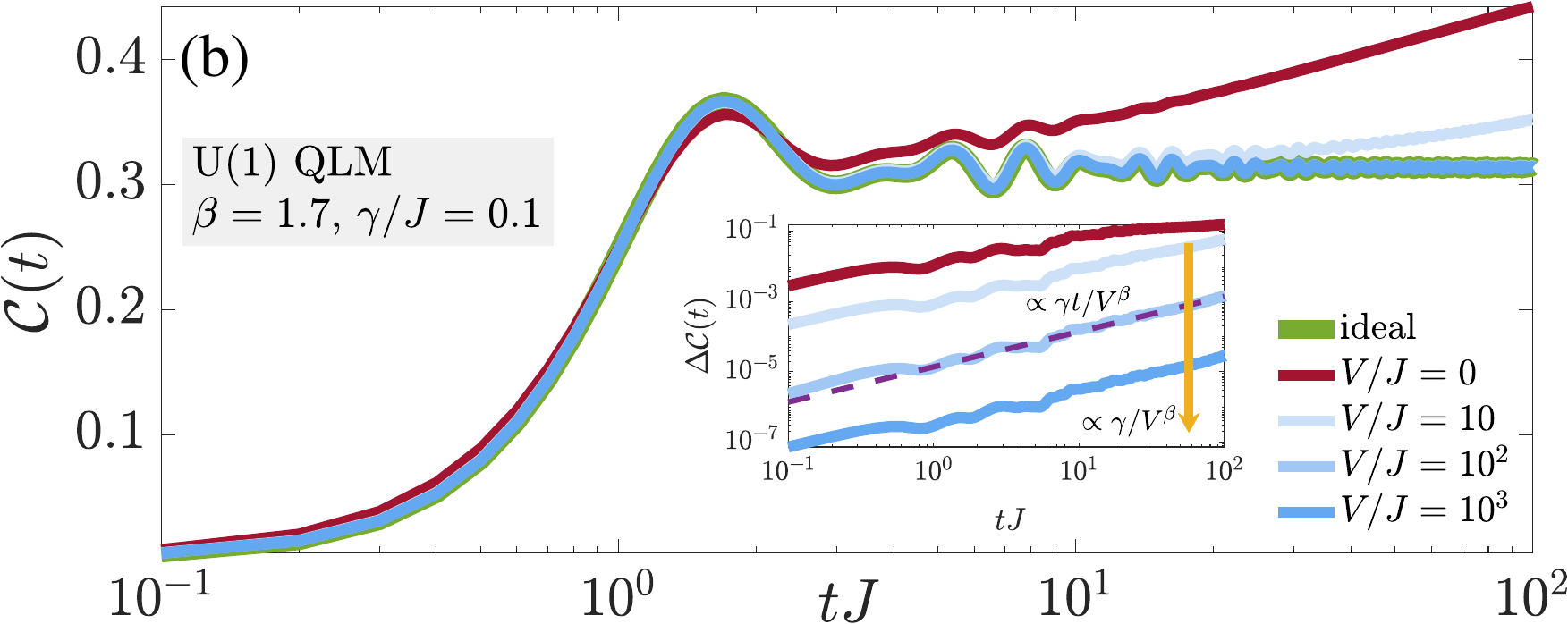}
	\caption{(Color online). Same as Fig.~\ref{fig:U1_beta1.7_gamma0.1} but for a noncompliant sequence $c_j=(-1)^j$, which is more experimentally feasible than its noncompliant counterpart. As seen in the quench dynamics of the (a) gauge violation and (b) the chiral condensate, the suppression of errors also evolves $\propto\gamma t/V^\beta$.}
	\label{fig:U1_beta1.7_gamma0.1_nc} 
\end{figure}
Due to numerical overhead, we are limited in our ED calculations to small system sizes. However, in modern cold-atom quantum simulators, much larger sizes can be attained . This makes it difficult to construct a compliant sequence for such state-of-the-art quantum simulators, as the coefficients of the latter grow roughly exponentially with system size. However, we can use a simpler noncompliant sequence such as $c_j=(-1)^j$. We repeat the results of Fig.~\ref{fig:U1_beta1.7_gamma0.1} using such a sequence, where the corresponding dynamics is shown in Fig.~\ref{fig:U1_beta1.7_gamma0.1_nc}. We see that both the gauge violation and the chiral condensate show qualitatively identical behavior to the case of the compliant sequence of Fig.~\ref{fig:U1_beta1.7_gamma0.1_nc}, with an error $\propto\gamma t/V^\beta$ in both cases.

\subsection{Perturbative explanation of numerical results}\label{app:TDPT}
We can explain the initial growth of gauge violation under $1/f$ noise in our numerical results by perturbatively expanding the Bloch--Redfield master equation. It can be shown that Eq.~\eqref{eq:vn} can be written in the familiar Lindblad form \cite{https://doi.org/10.48550/arxiv.1902.00967}, after employing the secular approximation and transforming back to the Schr\"odinger picture, as
\begin{align}\nonumber
d_t\hat{\rho}=&-i\left[\hat{H}_0+V\hat{H}_G, \hat{\rho}\right]+ \sum_{\omega}\sum_{j} S_{j}(\omega)\\
&\times\Big[\hat{A}_{j}(\omega)\hat{\rho} \hat{A}_{j}^{\dagger}(\omega)-\frac{1}{2}\big\{\hat{A}_{j}^{\dagger}(\omega)\hat{A}_{j}(\omega) ,\hat{\rho} \big\} \Big].
\end{align}
One can write the above in the concise form
\begin{align}\label{eq:a2}
d_t\hat{\rho}=(\mathcal{S}+\mathcal{D})\hat{\rho},
\end{align}
where 
\begin{subequations}
\begin{align}
\mathcal{S}[\hat{\rho}]&=-i\left[\hat{H}_0+V\hat{H}_G, \hat{\rho}\right],\\\nonumber
\mathcal{D}[\hat{\rho}]&=\sum_{\omega}\sum_{j} S_{j}(\omega)
\Big[\hat{A}_{j}(\omega)\hat{\rho} \hat{A}_{j}^{\dagger}(\omega)\\
&-\frac{1}{2}\big\{\hat{A}_{j}^{\dagger}(\omega)\hat{A}_{j}(\omega) ,\hat{\rho} \big\} \Big].
\end{align}
\end{subequations}
By Taylor expanding the solution to Eq.~\eqref{eq:a2}, we can find the leading order incoherent term to explain the growth of the gauge violation in the regimes $V=0$ and $V\gg J$. 
Choosing a target sector $g_j^\mathrm{tar}$, the gauge violation is  $\varepsilon(t)=\mathrm{Tr}\big\{\hat{\mathcal{G}}\hat{\rho}(t)\big\}$ where we have introduced the abbreviation  $\hat{\mathcal{G}}=\sum_{j}(\hat{G}_{j}-g_j^\mathrm{tar})^{2}/L$. 
The contribution of the first-order term in the absence of gauge protection (i.e., $V=0$) is
\begin{align}\nonumber
    t \mathrm{Tr}\{\hat{\mathcal{G}}\mathcal{D}\hat{\rho}_{0}\}= &t\sum_{\omega}\sum_{j} S_{j }(\omega)
\mathrm{Tr}\Big[\hat{\mathcal{G}}\hat{A}_{j}(\omega)\hat{\rho}_{0} \hat{A}_{j}^{\dagger}(\omega)\\\label{eq:A3}
&-\frac{1}{2}\big\{\hat{\mathcal{G}}\hat{A}_{j}^{\dagger}(\omega)\hat{A}_{j}(\omega) ,\hat{\rho}_{0} \big\}\Big]\sim {\gamma } t\,,
\end{align}
where we have utilized the fact that $\mathrm{Tr}\{\hat{\mathcal{G}}\mathcal{S}\hat{\rho}_{0}\}=i\mathrm{Tr}\{\left[\hat{H}_0+V\hat{H}_G, \hat{\mathcal{G}}\right]\hat{\rho}_{0}\}=0$ and the initial value $\mathrm{Tr} \{\hat{\mathcal{G}}\hat{\rho}_{0}\}=0$.

Once the gauge protection is switched on, in the limit $V\gg J$ the dominating coherent term is $\hat{H}_G={\sum_{m}V\epsilon_{m}^{g}|\epsilon_{m}^{g}\rangle\langle\epsilon_{m}^{g}|}$, where $\epsilon_{m}^{g}=\mathbf{c}^\intercal\mathbf{g}$, where $\mathbf{g}$ is a gauge sector. 
The relevant transition frequencies thus scale as $\omega_{mn}\sim V$. 
Taking this into account, neglecting corrections proportional to the energy scales of $\hat{H}_0$, and using the definition of the spectral function in Eq.~\eqref{eq:A3}, we obtain $\epsilon(t)\sim\gamma t/V^{\beta}$, hence explaining the corresponding scaling in the numerical results  up to first order. 
These results can be further corroborated by drawing conclusions from other contexts. E.g., in applications of error correction in adiabatic quantum computing, it was shown that increasing the energy gap to the excited states can suppress the transition rate out of the code space if the noise power spectrum is decreasing with frequency \cite{PhysRevA.74.052322}.

\section{Experimental implementation of linear gauge protection in cold atom experiments}
\begin{figure}[!htb]
    \centering
    \includegraphics[scale=0.95]{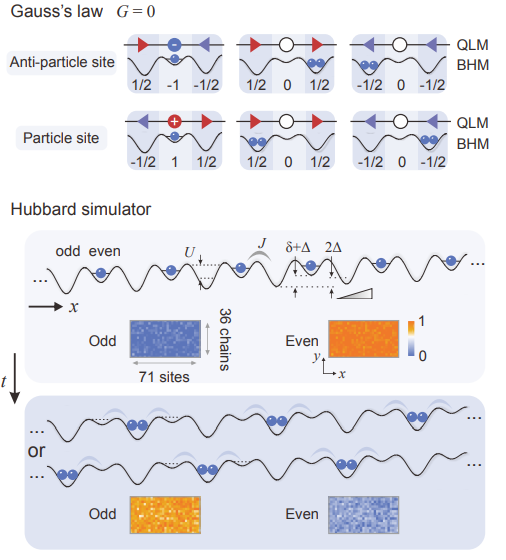}
    \caption{The optical-lattice quantum simulator of the spin-$1/2$ $\mathrm{U}(1)$ quantum link model is depicted in a diagram, which showcases the first experimental demonstration of gauge invariance quantification. The demonstration was made possible through linear gauge protection, and the schematic shows shallow and deep sites representing matter and gauge sites, respectively. By incorporating a small tilt $\Delta$, a strong on-site interaction strength $U$, and a staggered potential $\delta$, an effective linear gauge protection term was achieved in this implementation. Fig. from \cite{Zhou2021}}
    \label{fig:my_label}
\end{figure}
Recently, the U(1) quantum link model was successfully implemented in a quantum simulator that uses cold atoms \cite{Zhou2021,Yang:2020Science}. The experiments performed  using this setup and mapping showed  coherent times beyond expected timescales. One can speculate that it might be due to effective suppression of the growth of gauge violations due to $1 / f$ noise sources. This implementation was made possible by mapping the U(1) quantum link model onto the tilted staggered Bose-Hubbard model, which is represented by the Hamiltonian,

$$
\begin{aligned}
\hat{H}_{\mathrm{BHM}}= & -\kappa \sum_{\ell=1}^{N-1}\left(\hat{b}_{\ell}^{\dagger} \hat{b}_{\ell+1}+\text { H.c. }\right)+\frac{U}{2} \sum_{\ell=1}^{N} \hat{n}_{\ell}\left(\hat{n}_{\ell}-1\right) \\
& +\sum_{\ell=1}^{N}\left[(-1)^{\ell} \frac{\delta}{2}+\ell \Delta\right] \hat{n}_{\ell},
\end{aligned}
$$

The total number of sites in the system is denoted by $N$, and the even (odd) sites correspond to matter (gauge) sites. Referring to the index $j$ in Eq. \eqref{eq:U1QLM}, we can express it as $j=2\ell$, which implies that $N$ is approximately equal to $2L$. The bosonic ladder operators on site $\ell$ are represented by $\hat{b}_{\ell}$ and $\hat{b}_{\ell}^{\dagger}$.
The on-site boson occupation is given by $\hat{n}_{\ell}=\hat{b}_{\ell}^{\dagger} \hat{b}_{\ell}$. In the Hamiltonian, the tunneling constant is denoted by $\kappa$, the on-site interaction strength by $U$, the chemical potential by $\delta$ (which renders even/odd sites shallow/deep), and the tilt due to gravity is represented by $\Delta$.

The key aspect of the mapping involves enforcing a restricted local Hilbert space, $\operatorname{span}\left\{|0\rangle_{\ell},|1\rangle_{\ell}\right\}$, on even sites, which corresponds to the two eigenstates of the matter-field operator.

On odd sites, the restricted local Hilbert space is $\operatorname{span}\left\{|0\rangle_{\ell},|2\rangle_{\ell}\right\}$, which represents the two eigenstates of the local flux operator $\hat{s}_{j,j+1}^{z}$. 

In the limit of $U, \delta \gg \kappa, \mu$, the U(1) quantum link model can be derived up to second order in perturbation theory. By setting $\mu=\delta-U / 2$, a "Gauss's law constraint" is imposed in the form of a linear gauge protection term in perturbation theory.

$$
\begin{aligned}
\hat{H}_{G} & =\sum_{j} c_{j} \hat{\mathcal{G}}_{j}, \\
c_{j} & =2(-1)^{j} j \Delta, \\
\hat{\mathcal{G}}_{j} & =(-1)^{\ell}\left[\frac{1}{2}\left(\hat{n}_{j-1, j}+\hat{n}_{j, j+1}\right)+\hat{n}_{j}-1\right] .
\end{aligned}
$$

Using the same indexing as the QLM, designating sites on the bosonic superlattice as either odd (gauge) or even (matter). Specifically, sites labeled as $j, j+1 \equiv 2 \ell+1$ are considered odd, while those labeled as $j \equiv 2 \ell$ are considered even.


\chapter{Stabilizing Disorder-free localization and Quantum many-body scars against $1/f$ noise}\label{chap4}

\section{Disorder-free localization}
In the previous chapter, we showed numerical results demonstrating that linear gauge protection can suppress the growth of gauge violations in the physical sector or the sector obeying gauss law, $\textbf{g}_{\text{tar}}=0$. This occurs if the initial state is homogeneously prepared in the target gauge invariant sector or any other target superselection sector for both $U(1)$ and $\mathbb{Z}_{2}$ LGTs. As a result, the dynamics of local observables remain close to the ideal gauge theory, thanks to the enhanced coherent timescales as a result of suppression of $1/f$ noise due to linear gauge protection.

However, it is also possible to propagate the dynamics in multiple superselection sectors simultaneously and suppress any intersector dynamics. This leads to a recently discovered phenomenon known as disorder-free localization \cite{Smith2017}, a new paradigm in strong ergodicity breaking in the non-equilibrium dynamics gauge theories, which arises from quenches starting in initial states that form a superposition of a vast number of gauge superselection sectors. This distinctive feature of DFL(Disorder-free localization) can occur in nonintegrable translation-invariant models, such as the ones we investigate throughout this thesis, without any presence of quenched disorder.

In contrast, quench dynamics propagated by integrable models will not thermalize but instead relax to a generalized Gibbs ensemble arising from the plethora of local integrals of motion. On the other hand, generic nonintegrable systems are expected to thermalize according to the eigenstate thermalization hypothesis (ETH), a conjecture which explains the process of thermalization at the level of the system’s energy eigenstates\cite{Srednicki1994,Deutsch1991}. In the presence of quenched disorder in interacting models, many-body localization (disorder-MBL) arises, which violates ETH and leads to localized dynamics in local observables\cite{Nandkishore_review}. Disorder-MBL has been the subject of intense theoretical investigation recently and has been experimentally probed in various quantum synthetic matter setups.

DFL is interpreted as localization arising as a result of the dynamical emergence of an effective disorder over the background charges associated with the superselection sectors $\textbf{g}$ involved in the superposition \cite{PhysRevLett.120.030601}. In other words, the local constraints of the gauge symmetry take on the role of local conserved quantities in the form of fixed background charges. These charges appear as a discrete disorder potential in a typical sector, leading to an indefinite delay of thermalization and a strong violation of ETH, since it states that individual eigenstates of quantum-ergodic
systems act as thermal ensembles, thus the system’s relaxation should not depend strongly on the initial conditions.

Interestingly, even though the initial state is a superposition of an extensive number of superselection sectors, it can be prepared in a product state, making it easy to realize in experiments involving quantum simulations of lattice gauge theories in synthetic quantum matter platforms. 

However, as we have stressed before, due to the presence of unavoidable gauge breaking errors in these setups due to experimental imperfections, it turns out that there is a great deal of fine-tuning that is required to witness DFL since these gauge violations cause transitions between different gauge superselection sectors in which we would like to restrict our dynamics, quickly destroying features of DFL, immediately leading to thermalization to a canonical gibbs ensemble\cite{Halimeh2021stabilizing}.

\subsection{Stabilizing Disorder-free localization}
It has been recently shown that linear gauge protection can restore DFL in $U(1)$ QLM \cite{Halimeh2021stabilizingDFL} and even enhance it in the case of $\mathbb{Z}_{2}$ LGT\cite{Halimeh2021enhancing} This is due to the fact that the addition of translation-invariant alternating sum of generators of gauge symmetry suppresses intersector dynamics for at least times polynomial in the protection strength based on the quantum-Zeno effect while leaving intrasector dynamics untouched. As a result, one obtains a dynamical emergence of a renormalized gauge theory which is perturbatively close to our ideal gauge theory.

To probe localization in both models, one prepares the initial state, which is a domain wall in the matter fields, i.e, the left is occupied while the right half is empty. The electric fields are prepared in such a manner that they are the superposition of both local eigenstates. This results in the formation of a superposition over an extensive number of gauge superselection sectors. Ideal gauge theory dynamics dictate that the dynamics of imbalance does not decay to zero(as one would expect from thermalization), which is nothing but the difference between the matter occupation of the left and right halves of the chain. This can also be written as follows:
\begin{equation}\label{Ibl}
\mathcal{I}(t)=\frac{1}{L t} \int_0^t d s \sum_{j=1}^L p_j\left\langle\psi(s)\left|\hat{n}_j\right| \psi(s)\right\rangle
\end{equation}
where $p_j=\left\langle\psi_0\left|\hat{\sigma^{z}_{j}}\right| \psi_0\right\rangle$ and $n_j=\frac{\hat{\sigma}^{z}_{j}+\mathbb{I}}{2}$(for the case of U(1) QLM), with $|\psi(t)\rangle=e^{-i\hat{H}t}|\psi_{0}\rangle$ and $\hat{H}=\hat{H}_{0}+\lambda \hat{H}_{1}+V \hat{H}_{G}$ in the regime of unitary gauge breaking dynamics.

\begin{figure}[t!]
    \centering
    \begin{subfigure}[b]{0.49\textwidth}
        \centering
        \includegraphics[width=\textwidth]{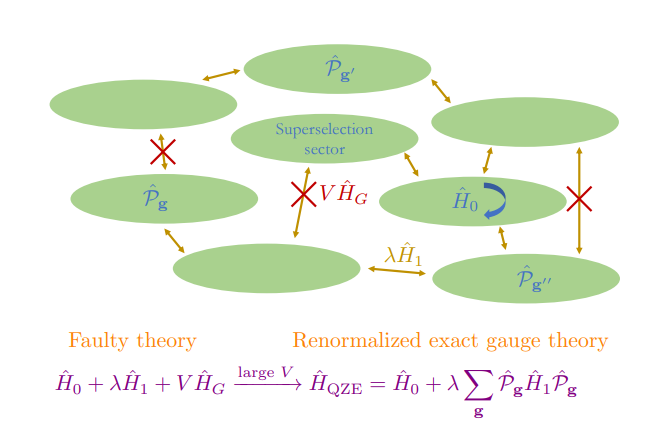}
         \caption{ Stabilization of disorder-free localization. $\hat{H}_0$ does not couple different superselection sectors, leading to DFL when the initial state is a superposition over many sectors. Coherent gauge-breaking errors $\lambda\hat{H}_1$ lead to transitions between the different superselection sectors, destroying DFL. Upon switching on gauge protection $V\hat{H}_G$ induces quantum Zeno dynamicsthat suppresses inter-sector processes and stabilizes DFL up to timescales at least polynomial in $V$ Fig adapted from \cite{Halimeh2021stabilizingDFL}.}
         \label{DFL1}
    \end{subfigure}
    \begin{subfigure}[b]{0.49\textwidth}
        \centering
        \includegraphics[width=\textwidth]  {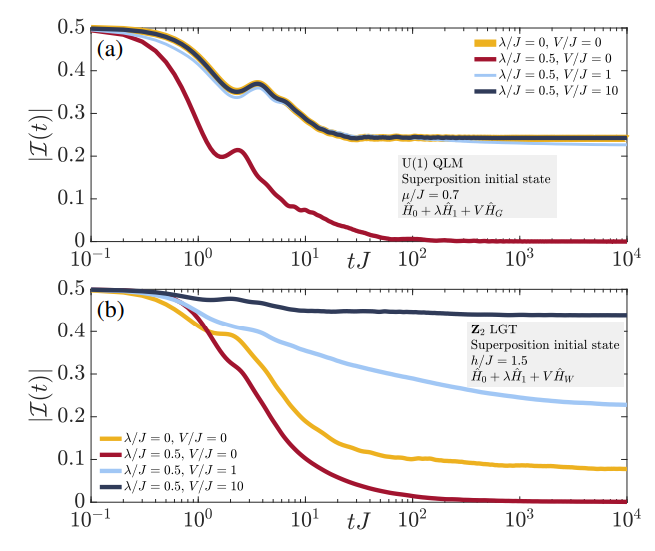}
   
        \caption{Restoration of disorder-free localization. Top; In the case of the $\mathrm{U}(1)$ QLM, we protect against gauge-breaking errors are protected $\lambda\hat{H}_1$ using SGP(stark gauge protection), involving $\hat{G}_j$, with $c_j=j(-1)^j$.  Even moderate $V$ stabilizes DFL to the same plateau as in the case of the ideal theory. Bottom; In the case of the $\mathbb{Z}_2$ LGT, linear gauge protection based on the local pseudo-generator is employed $\hat{W}_j$, with $c_j=j$. This leads to an emergent gauge theory with an \textit{enriched} local symmetry containing the $\mathbb{Z}_2$, inducing a greater effective disorder and hence enhanced DFL. Figure is adapted from Ref.~\cite{Lang2022stark}}
        \label{DFL2}
    \end{subfigure}

\end{figure}
\FloatBarrier


The recent efforts to stabilize disorder-free localization (DFL) in this regime can be succinctly summarized \cite{Halimeh_BriefReview}, some of which are illustrated in Figures \ref{DFL1} and \ref{DFL2}.
In the case of the U(1) quantum link model (QLM), one can also employ a linear gauge protection mechanism, wherein $c_j = j(-1)^j$, to counter coherent gauge breaking errors under periodic boundary conditions. This Stark gauge protection approach \cite{Lang2022stark}, named due to the linear staggered potential incorporated in $c_j$, goes beyond the concept of quantum Zeno dynamics and ensures DFL stability for all accessible evolution times. In particular, it was demonstrated, using a Magnus expansion, that when the protection strength $V$ is large, an effective Hamiltonian emerges. This Hamiltonian suppresses gauge-breaking terms not only by $V$ but also by the matter site index present in the Stark potential. It was also found that unitary gauge errors were suppressed $\propto{\frac{1}{(2j+1)V}}$, indicating that larger system sizes lead to improved performance. The emergent gauge theory retains the same U(1) gauge symmetry as the U(1) QLM, and DFL is restored to the same plateau as that of the ideal case. This can also be illustrated in the imbalance dynamics illustrated in Figure \ref{DFL1}, demonstrating a near-perfect quantitative agreement even for moderate values of $V = 10J$.
If one chooses translation-invariant sequences $c_j=(-1)^j$  that alternate between odd and even sites instead of an SGP(stark gauge protection) sequence, stabilization of DFL is achieved  up to times at least linear
in the protection strength in agreement with the worst-case prediction of the QZE, and at best, up to timescales
quadratic in the protection strength. Particularly, as one increases $V$ , the timescale of the DFL plateau was numerically found to be $\propto V^2/(\lambda^2 J)$, the quench dynamics under $\hat{H}$ is faithfully reproduced in the large-$V$ limit by the effective Hamiltonian $\hat{H}_\mathrm{QZE}=\hat{H}_0+\lambda\sum_\mathbf{g}\hat{\mathcal{P}}_\mathbf{g}\hat{H}_1\hat{\mathcal{P}}_\mathbf{g}$ \cite{Halimeh2021stabilizing}

For the $\mathbb{Z}_{2}$ LGT, the linear gauge protection scheme based on local pseudo generators (LPG) was employed, with SGP sequence $c_j = j$, leading to not only the stabilization of DFL but also its enhancement\cite{Lang2022stark}. As shown in Fig. \ref{DFL2}, the imbalance settling into a plateau at $V = 10J$ has a significantly higher value than that of the ideal case. This enhancement arises due to the LPG terms, which give rise to an enlarged local symmetry containing the original $\mathbb{Z}_{2}$ gauge symmetry. As a result, the initial state becomes a superposition over a greater number of local-symmetry sectors, leading to an effective disorder over a larger number of background charges associated with all these sectors(also illustrated in Fig \ref{fig gZ2} of the previous chapter). It was numerically demonstrated that an emergent gauge theory, represented by the equation $\hat{H}_\mathrm{QZE}=\sum_\mathbf{w}\hat{\mathcal{P}}_\mathbf{w}\big(\hat{H}_0+\lambda\hat{H}_1\big)\hat{\mathcal{P}}_\mathbf{w}$, can effectively describe the dynamics up to a timescale proportional to $V/J^2$, another effective $\mathbb{Z}_2$ gauge theory emerges, which lasts up to a timescale proportional to $V^2/(\lambda^2 J)$ in the presence of gauge-breaking errors. In the absence of errors, the DFL prethermal plateau due to this effective gauge theory persists indefinitely \cite{Halimeh2021enhancing}. It is also worth mentioning that this enhancement is independent of the protection sequence $c_j$, and one can also choose the sequence $[(-6)^j+5]/11$.

Also, if one begins with a gauge invariant state and undergoes a quench using the hamiltonian $\hat{H}$ for both models, it will not result in localized dynamics. This means that the imbalance will always decay to zero, and the system will ultimately thermalize and reach the canonical Gibbs ensemble, regardless of the values of $\lambda$ and $V$. Hence, it was shown to be evident that single-body gauge protection alone cannot produce disorder-free localization. Instead, achieving disorder-free localization necessitates an initial state of superposition as in the ideal case. In other words, linear gauge protection merely restores disorder-free localization for U(1) QLM and $\mathbb{Z}_2$ LGT (In the case when full local generator is used) .

\renewcommand\thefigure{\thechapter.\arabic{figure}}    
\renewcommand{\theequation}{\thechapter.\arabic{equation}}
\setcounter{equation}{0}
\setcounter{figure}{0}

\subsection{Stabilization of DFL against $1/f$ noise: Numerical results}

We now present our numerical results on the quench dynamics to probe DFL in gauge theories subjected to $1/f$ noise in a manner similar to what we described in the previous chapter. In all cases, we take $\beta=1$ (the qualitative conclusions remain the same irrespective of $\beta$), we  monitor the quench dynamics in the presence of $1/f$ noise with and without linear gauge protection. In particular, we will focus on the dynamics of the imbalance according to Eq.\eqref{Ibl}-

\begin{align}\label{eq:Ibl1}
\mathcal{I}(t)&=\frac{1}{L t}\int_0^t d s\sum_{j=1}^{L}p_j\Tr\Big\{\hat{\rho}(s)\hat{n}_j\Big\},
\end{align}
where $\hat{\rho}(s)$ is the time-evolved density operator of the system at time $s$, and we take a temporal average. Due to the large evolution times we investigate, we restrict our system size to $L=4$ sites due to computational overhead, and we employ periodic boundary conditions.
    \subsubsection{U(1) Quantum link model}

First, we look at the spin-$1/2$ U(1) QLM. Considering initial states that are matter domain walls, with the electric fields aligned either along the positive $x$-direction or along the $z$-direction, which we call $\hat{\rho}^{x,z}_{0}=\ket{\psi^{x,z}_0}\bra{\psi^{x,z}_0}$. Where $\hat{\rho}^{z}_{0} $ satisfies  $\Tr\{\hat{\rho}^{z}_{0}\hat{G}_j\}=0,\,\forall j$ . Numerically, this is represented as a product state with the fields taken as eigenstates of the $\hat{\sigma}^{x}$ for $\hat{\rho}^{x}_{0}$ and in the case of $\hat{\rho}^{z}_{0}$ the fields are represented by the eigenstates of $\hat{\sigma}^{z}$. Also, $\hat{\rho}^{x}_{0} $ is not gauge-invariant but rather is in a superposition of superselection sectors.

\begin{figure}[t!]
    \centering
    \begin{subfigure}[b]{0.49\textwidth}
        \centering
        \includegraphics[width=\textwidth]    {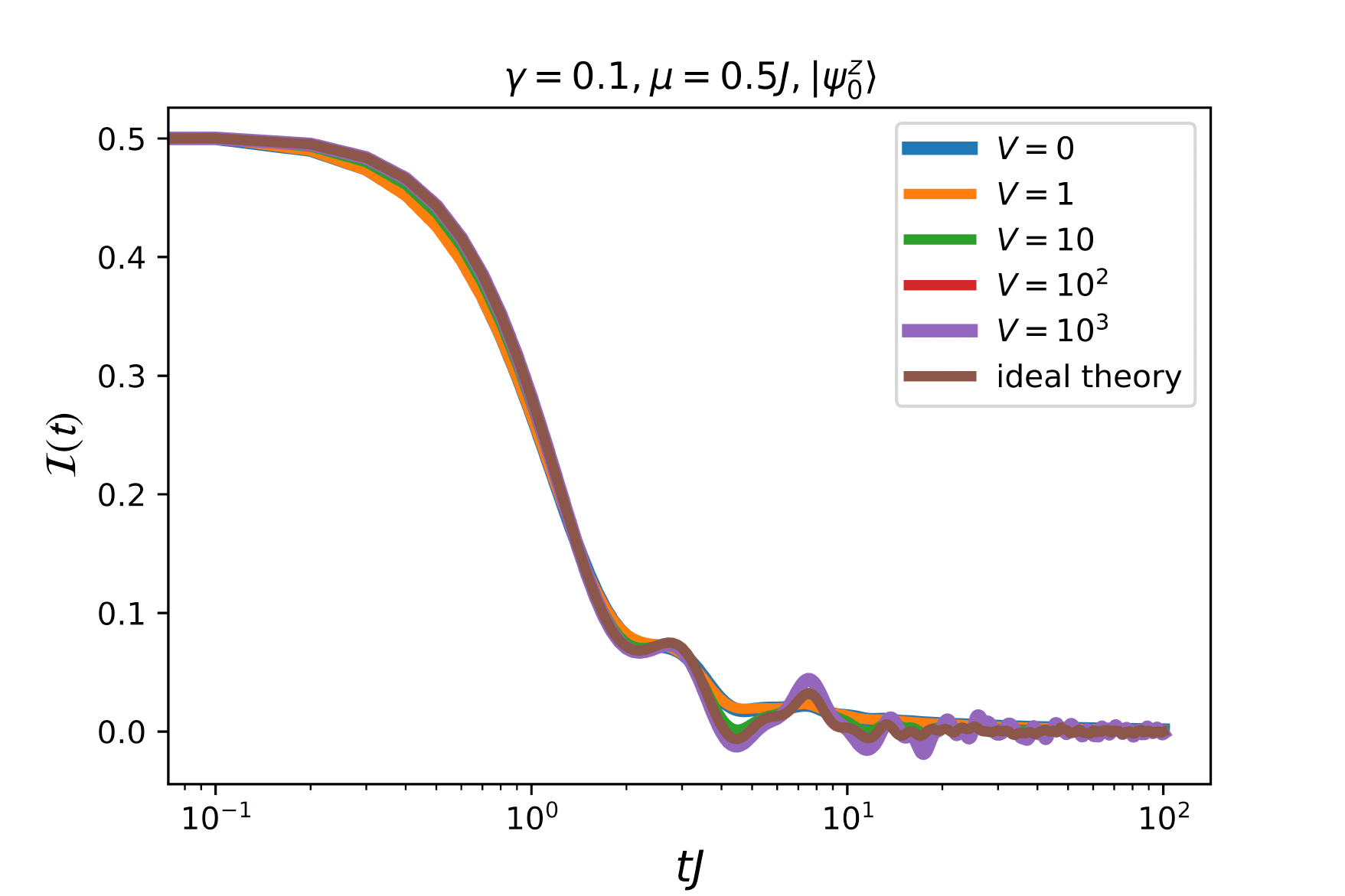}
         \caption{Dynamics of imbalance according to Eqn \ref{eq:Ibl1}, when starting in a homogenously prepared gauge invariant state $\hat{\rho}^{z}_{0}$.  The system is expected to thermalize in the superselection sector $\mathbf{g}=(0,0,0,0)$ irrespective of $\gamma$ and $V$}
         \label{DFL3}
    \end{subfigure}
    \begin{subfigure}[b]{0.49\textwidth}
        \centering
        \includegraphics[width=\textwidth]  {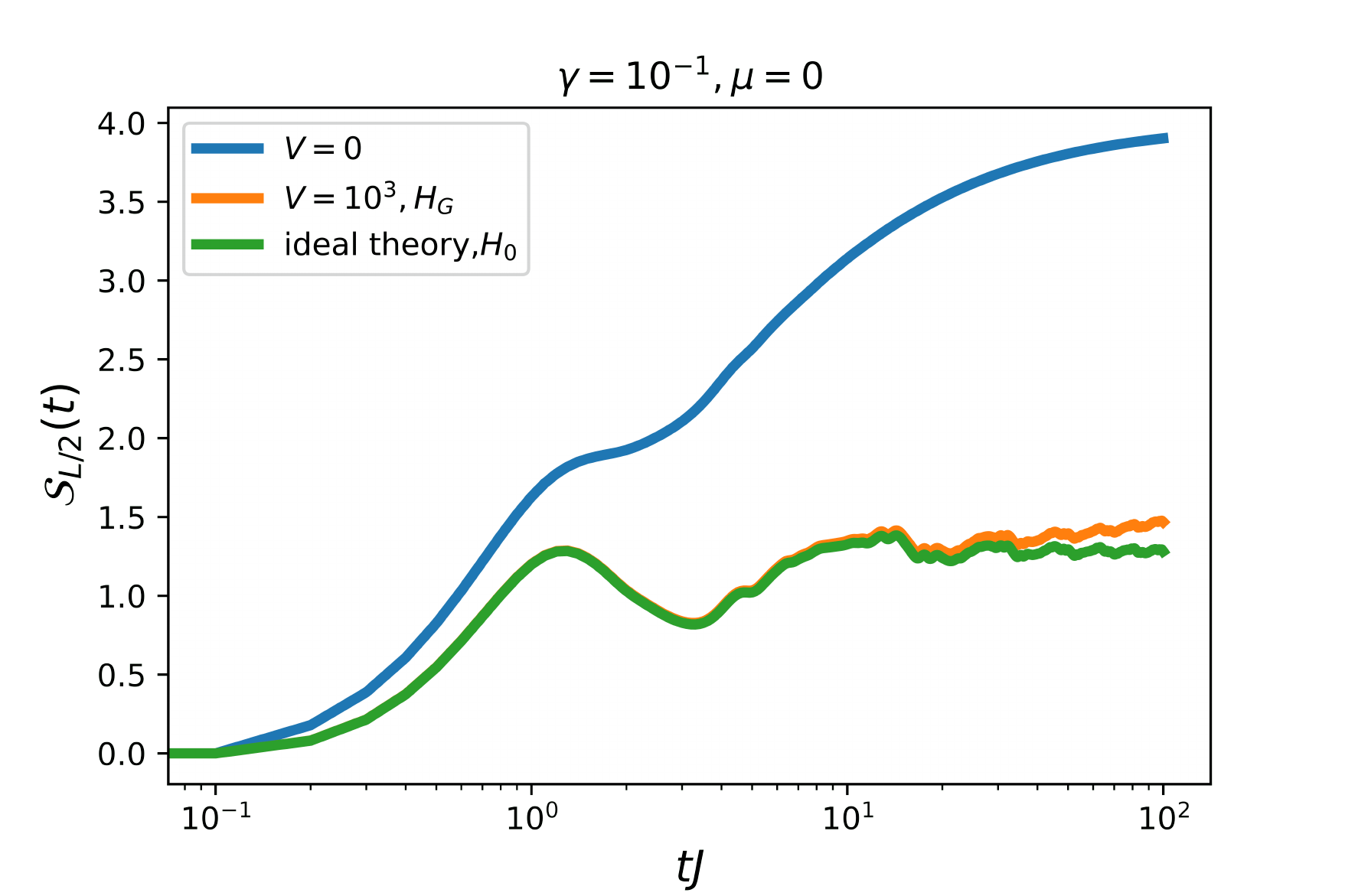}
   
        \caption{Controlled restoration of DFL in presence of $1/f$ noise as gauge protection is switched on upto timescales $\propto{V/\gamma}$}.
        \label{DFL4}
    \end{subfigure}    
\end{figure}

 We subject to a quench with $\hat{H}_0+V\hat{H}_G$ at $\mu/J=0.5$ in the presence of $1/f$ noise with power spectrum~\eqref{eq:spectral} and jump operators $\hat{A}_j^{m}=\hat{\sigma}^{x}_{j}$ and $\hat{A}_{j,j+1}^{g}=\hat{s}^{x}_{j,j+1}$ corresponding to an initial state $\hat{\rho}^{z}_{0}$. Since this state was prepared in a homogenous gauge sector, the system is expected to thermalize in the long-time limit in the superselection sector $\mathbf{g}=(0,0,0,0)$. This is indeed the case, as can be seen in Fig.~\ref{DFL3} for $\gamma=V=0$ (brown curve) and the timescale for thermalization is independent of $\gamma$ and $V$ similar to its coherent error counterpart described in the previous section. In contrast, the quench dynamics of $\hat{\rho}^{z}_{0}$ under $\gamma=V=0$ displays localized behavior in the imbalance (brown) Fig.~\ref{DFL4} , with the latter settling into a nonzero plateau for all numerically accessible times. Upon turning on $\gamma>0$, but still, with $V=0$, this localized phase starts to decay. In particular, the error with respect to the ideal case grows $\propto{\gamma t}$, after which the imbalance goes to zero, indicating thermalization and the absence of localized behavior, as shown by the blue curve. 

\begin{table}[t!]
    \centering
    \begin{tabular}{||c||c|c||}
    \hline $\mathbf{g}=\left(g_1, g_2, g_3, g_4\right)$ & $\Tr\{\hat{\rho}^{z}_{0}\hat{\mathcal{P}}_{\mathbf{g}}\}$ & $\Tr\{\hat{\rho}^{x}_{0}\hat{\mathcal{P}}_{\mathbf{g}}\}$ \\
    \hline \hline$(-2,1,0,0)$ & 0 & 0.0625 \\
    \hline$(-2,1,1,0)$ & 0 & 0.0625 \\
    \hline$(-2,2,-1,1)$ & 0 & 0.0625 \\
    \hline$(-2,2,0,0)$ & 0 & 0.0625 \\
    \hline$(-1,0,0,1)$ & 0 & 0.0625 \\
    \hline$(-1,0,1,0)$ & 0 & 0.0625 \\
    \hline$(-1,1,-1,1)$ & 0 & 0.0625 \\
    \hline$(-1,1,0,0)$ & 0 & 0.125 \\
    \hline$(-1,1,1,-1)$ & 0 & 0.0625 \\
    \hline$(-1,2,-1,0)$ & 0 & 0.0625 \\
    \hline$(-1,2,0,-1)$ & 0 & 0.0625 \\
    \hline$(0,0,0,0)$ & 1 & 0.0625 \\
    \hline$(0,0,1,-1)$ & 0 & 0.0625 \\
    \hline$(0,1,-1,0)$ & 0 & 0.0625 \\
    \hline$(0,1,0,-1)$ & 0 & 0.0625 \\
    \hline
\end{tabular}
    \caption{Possible Superselection sectors $\mathbf{g}$ and the expectation values of their projectors $\hat{\mathcal{P}}_\mathbf{g}$ relative to the initial states  for quenches corresponding to U(1) QLM.}
    \label{tab:my_label}
\end{table}

We now employ the single-body gauge protection~\eqref{eq:HG} to stabilize DFL with fixed $\gamma{=}0.1J$ and compliant sequence chosen to be $c_j=\{-115,116,-118,122\}/122$ . Starting in the superposition initial state $\hat{\rho}^{x}_{0}$, we show in Fig.~\ref{DFL4}(b) the controlled restoration of the DFL phase with increasing $V$. As $V$ is increased, the DFL plateau is numerically found to be restored upto a timescale $\propto V/J\gamma$, which also corroborates the results pertaining to the growth of gauge violation in the presence of $1/f$ noise obtained in the previous chapter, henceforth establishing the connection between gauge invariance and disorder-free localization.

\FloatBarrier

The protection term can also be translation-invariant.Since the results remain qualitatively the same for other protection sequences such as $c_j=(-1)^{j}$. Hence, the restoration of the DFL plateau through this expression cannot be solely credited to either disorder-MBL or Stark-MBL.


\subsubsection{$\mathbb{Z}_{2}$ LGT}

\begin{figure}[t!]
    \centering
    \begin{subfigure}[b]{0.49\textwidth}
        \centering
        \includegraphics[width=\textwidth]    {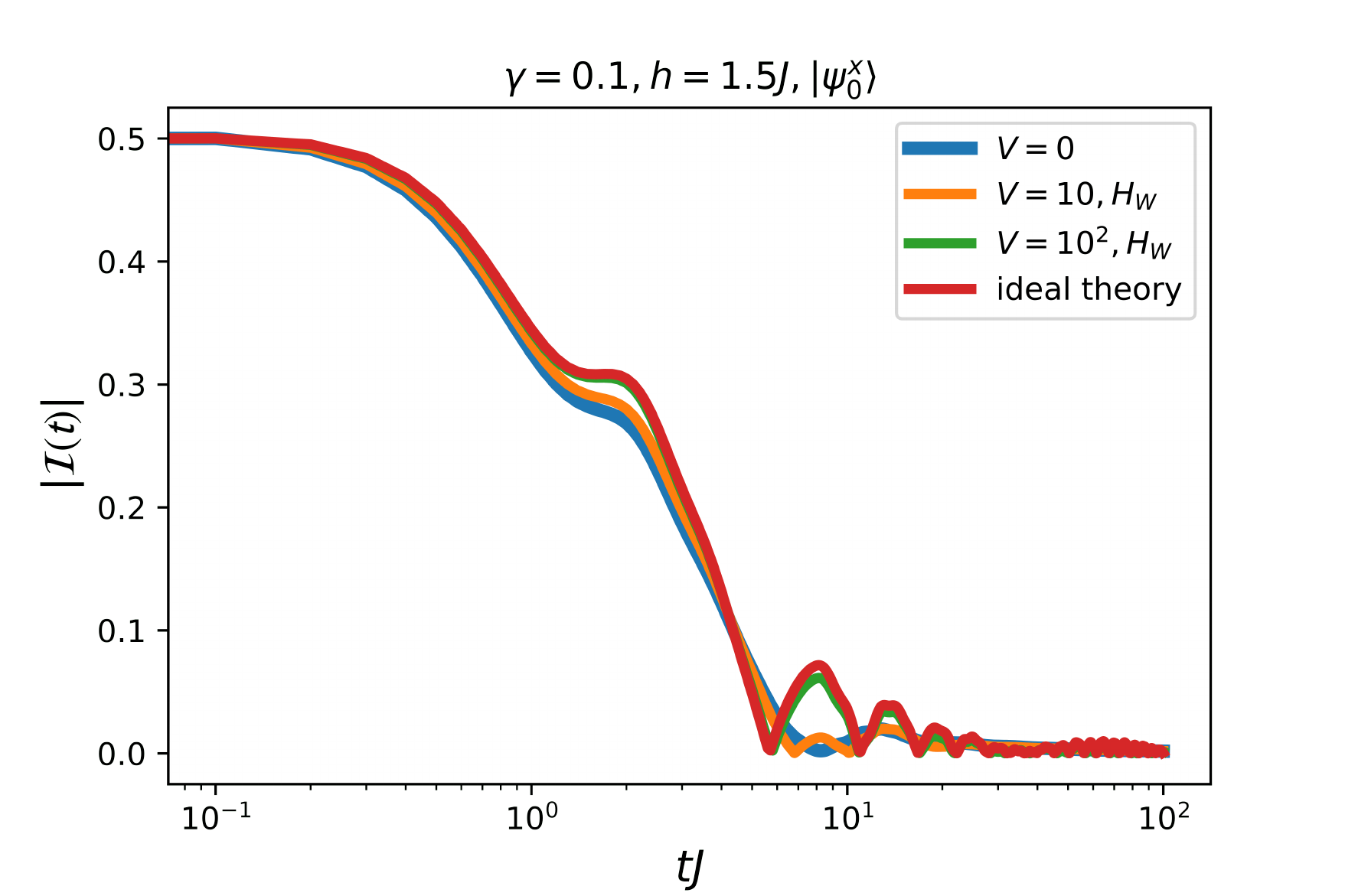}
         \caption{Quench dynamics of the $\mathbb{Z}_{2}$ LGT as described by \eqref{eq:Z2LGT} with $\hat{H}_0+V\hat{H}_{W}$ at $h=1.5J$ in the initial state  $\hat{\rho}^{x}_{0}$ in the homogenous superselection sector. The dynamics is identical to that of the case of U(1) QLM quenched in the gauge invariant sector which showed thermalization. }
         \label{DFL5}
    \end{subfigure}
    \begin{subfigure}[b]{0.49\textwidth}
        \centering
        \includegraphics[width=\textwidth] {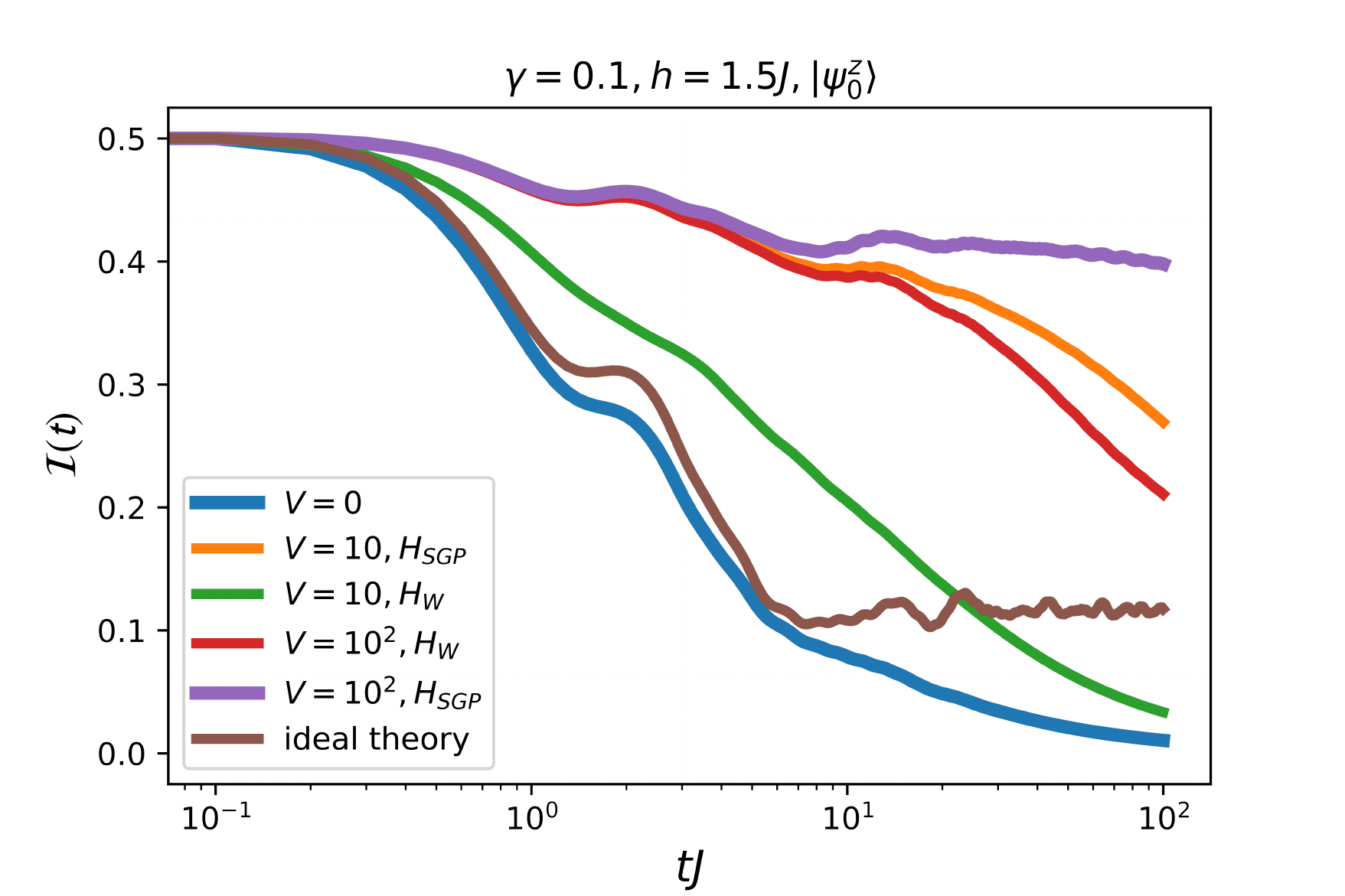}
        
        \caption{Imbalance for the case, when the initial state $\hat{\rho}^{z}_{0}$ is in a superposition of various gauge sectors . One can clearly see that DFL is not only restored but also enhanced upon employing LPG protection against noise. Greater performance is also achieved when using the SGP sequence, which behaves like a linear potential in the respective gauge sectors.}
        \label{DFL6}
    \end{subfigure}    
\end{figure}

Again, to establish the generality of our findings, we initiate quench dynamics of the $\mathbb{Z}_{2}$ LGT as described by \eqref{eq:Z2LGT} with $\hat{H}_0+V\hat{H}_W$ at $h=1.5J$.We prepare our system in the initial state  $\hat{\rho}^{x}_{0}$ in the homogenous superselection sector, which for the $\mathbb{Z}_{2}$ satisfies the following condition  $\Tr\{\hat{\rho}^{x}_{0}\hat{G}_j\}=1,\,\forall j$ .This is a domain wall state at half-filling from the perspective of hard-core bosons and is also gauge invariant; see Table~\ref{DFLTa}. In the wake of a quench in the presence of $1/f$ noise  with the jump operators $\hat{A}_j^{m}=\hat{a}_{j}+\hat{a}_{j}^\dagger$ and $\hat{A}_{j,j+1}^{g}=\hat{\tau}^{z}_{j,j+1}$ coupling the matter and gauge fields, respectively, to the environment at a fixed value of $\gamma=0.1J$ and for several values of the protection strength $V$, the system is expected to thermalize in the superselection sector $\mathbf{g}=(1,1,1,1)$, and as shown in Fig-\ref{DFL6} we find that the imbalance does relax to zero to the canonical Gibbs ensemble with the timescales being independent of $\gamma$ and $V$. When the system is initialized in $\hat{\rho}^{z}_{0}$, which is also a domain-wall state in the hard-core bosons, but since its electric fields all point in the positive $z$-direction, for the case of $\mathbb{Z}_{2}$ LGT an equal-weight superposition over all physical superselection sectors corresponding to $\hat{G}_j$ is prepared, shown in the Table~\ref{DFLTa}.

\begin{table}[t!]
    \centering
   \begin{tabular}{||c||c|c||}
    \hline $\mathbf{g}=\left(g_1, g_2, g_3, g_4\right)$ & $\Tr\{\hat{\rho}^{x}_{0}\hat{\mathcal{P}}_{\mathbf{g}}\}$ & $\Tr\{\hat{\rho}^{z}_{0}\hat{\mathcal{P}}_{\mathbf{g}}\}$ \\
    \hline \hline$(-1,-1,-1,-1)$ & 0 & 0.125 \\
    \hline$(-1,-1,+1,+1)$ & 0 & 0.125 \\
    \hline$(-1,+1,-1,+1)$ & 0 & 0.125 \\
    \hline$(-1,+1,+1,-1)$ & 0 & 0.125 \\
    \hline$(+1,-1,-1,+1)$ & 0 & 0.125 \\
    \hline$(+1,-1,+1,-1)$ & 0 & 0.125 \\
    \hline$(+1,+1,-1,-1)$ & 0 & 0.125 \\
    \hline$(+1,+1,+1,+1)$ & 1 & 0.125 \\
    \hline
\end{tabular}
    \caption{Superselection sectors $\mathbf{g}$ and the expectation values of their projectors $\hat{\mathcal{P}}_\mathbf{g}$ relative to the initial states  for quenches corresponding to $\mathbb{Z}_{2}$ LGT.}
    \label{DFLTa}
\end{table}

\FloatBarrier
 
As previously stated, starting with this particular initial state results in DFL. Fig.~\ref{DFL6} demonstrates that if there is no noise and $V=0$, the system will preserve the initial state memory for all accessible evolution times. The imbalance will eventually relax to a value close to $0.1$. However, the presence of $1/f$ noise results in gauge-breaking errors, which destroys disorder-free localization. Specifically, for a given value of $\gamma$, the imbalance will decay to zero, indicating thermalization. This behavior is qualitatively identical to that of the U(1) QLM.

\begin{figure}[t!]
    \centering
    \includegraphics[scale=0.4]{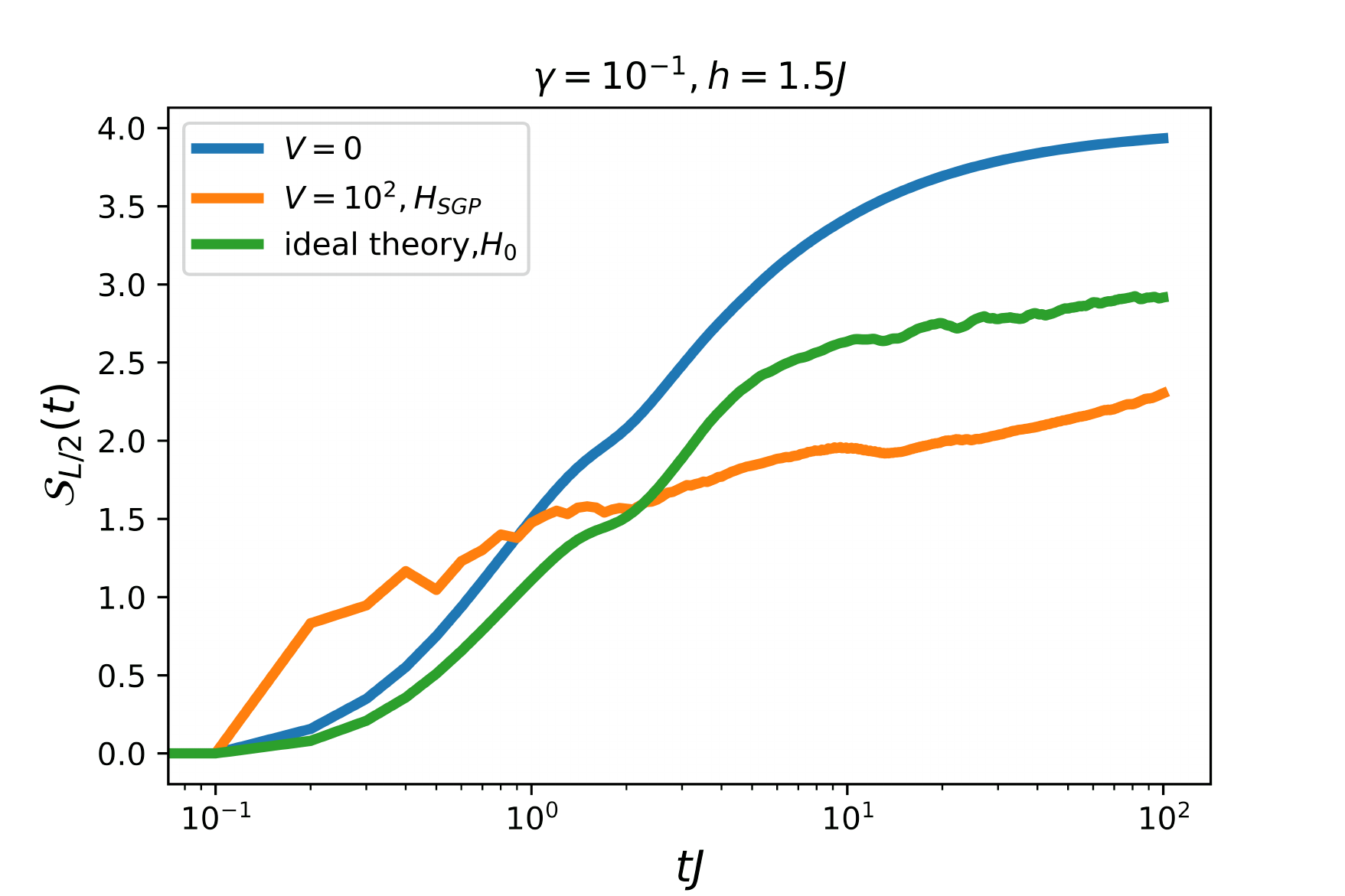}
    \caption{Localization behavior probed from the lens of the mid-chain entanglement entropy according to eq.\eqref{vne1}. This quantity shows a similar qualitative behavior as in the case of imbalance dynamics, which shows suppression at later times, indicating restoration and enhancement of localized behavior for large enough values of $V$}. 
    \label{DFL7}
\end{figure}
Taking $c_j=[(-6)^j+5]/11$ and employing LPG gauge protection to stabilize DFL, as illustrated in Fig.~\ref{DFL6}(b), the localized phase can last longer with increased $V$ and is strengthened up to a timescale proportional to $V/J\gamma$. The prethermal disorder-free localization plateaus also retain more memory of the initial state by exhibiting larger values than the ideal case before undergoing thermalization.

This is possible because the initial state can be viewed as a superposition over the superselection sectors of $\hat{W}_j$ as well as those of $\hat{G}_j$ leading to a greater effective disorder over the background charges.

To further enhance this performance, we have also used the stark gauge protection term $H_{SGP}$, which uses the site-dependent sequence $c_j$, hence leading to even greater stabilization of DFL for even moderate protection strengths.

To gain another perspective of the localized dynamics and further affirm this picture, we can calculate the mid-entanglement entropy(temporally averaged), which can be calculated by the Von-Neumann entropy of one of the subsystems by partitioning the chain into two equal halves,
$\mathcal{S}_{L/2}(t)$, shown in Fig.~\ref{DFL7}.
\begin{equation}\label{vne1}
    \mathcal{S}_{L/2}(t)=-\Tr_{L/2}[\hat{\rho}_{L/2}(t)\ln{\hat{\rho}_{L/2}(t)}]
\end{equation}

The dynamics of this quantity at $V=0$ show volume law growth quickly saturating to its maximally mixed gibbs state, indicative of thermalization. 

When LPG gauge protection is switched on, the entanglement entropy grows much slower. It hence is also suppressed even more so than that of the ideal case at intermediate timescales as one takes a sufficiently large value of $V$, indicating both enhancement and restoration of localized dynamics, showing qualitative behavior which is identical to imbalance dynamics. 

We can also observe that the entropy grows much faster at very early timescales, even more so than in the case of $\hat{H}_{0}$. This is due to the fact that in the ideal case, $\hat{H}_0$ can only drive intra-sector dynamics within each sector $\mathbf{g}$; in the enhanced model, it additionally drives intra- and inter-sector dynamics in the emergent sectors due to $\hat{W}_{j}$, which leads to faster growth of $\mathcal{S}_{L/2}(t)$ at very early times. However, this is only a small price to pay since the dynamics clearly exhibit more localized  behavior than under the ideal theory at later times. 

\section{Weak ergodicity breaking: Quantum many-body scars in the U(1) quantum link model}

As mentioned in the previous section, DFL is a paradigm that involves strong ergodicity breaking. However, a new phenomenon called \textit{quantum many-body scars} \cite{Turner2018} has been observed recently in a Rydberg atom experiment. This phenomenon complements the DFL paradigm, showing a weak form of ergodicity breaking. Quantum many-body scars have been identified in many spin chains, particularly in non-integrable models. Essentially, these scars are unique states within a system that have low entanglement entropy despite being located far from the ground state and embedded in an otherwise thermalizing spectrum. They are evenly spaced in energy and exist within a small subset of the Hilbert space that is weakly connected to the rest of the system. These properties suggest that the system may only weakly break ergodicity, as only a few states (on the order of $O(L)$) out of many (on the order of $O(e^{cL})$) do not follow the Eigenstate Thermalization Hypothesis (ETH). 

Therefore, if one starts with an initial state in the small subset of the Hilbert space mentioned earlier and then quenches the system, the dynamics of the system will show persistent oscillations that last longer than expected, effectively slowing down the thermalization process. Notably, these "quantum many-body scars" do not have any association with the model's specific symmetry, and they do not necessitate a superposition across symmetry sectors, as in the case of DFL.

The first observation of quantum many-body scars happened in a Rydberg-atom setup that used an Ising-type spin model \cite{Bernien2017}, which was later shown to be equivalent to the spin-$1/2$ $\mathrm{U}(1)$ quantum link model (QLM) \cite{Surace2020}.
\begin{figure}[!htb]
    \centering
    \begin{subfigure}[b]{0.49\textwidth}
        \centering
        \includegraphics[width=\textwidth]    {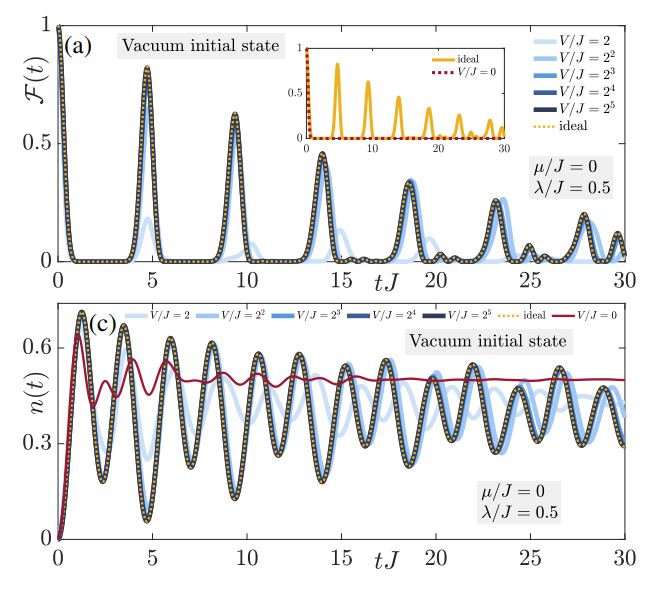}
        
         \caption{Top; Periodic revivals of fidelity in the ideal case and it's quick decay in the presence of coherent errors ($\lambda=0.5J$ here). Upon adding the linear gauge protection term, the fidelity is restored to that of the ideal case at sufficiently large protection strength $V$, these simulations were obtained in Ref.\cite{halimeh2022robust} using Krylov-based methods for 12 sites. Bottom; Characteristic of many-body scarring is persistent oscillations in local observables such as the  chiral condensate. Such oscillations are quickly damped in the presence of unprotected errors (red solid curve) but are reliably restored upon employing linear gauge protection (different shades of blue).}
    \end{subfigure}
    \begin{subfigure}[b]{0.49\textwidth}
        \centering
        \includegraphics[width=\textwidth]
        {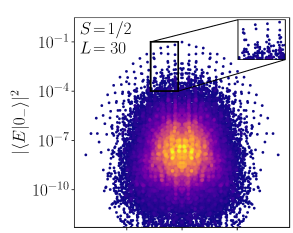}
   
        \caption{Atypical nonthermal eigenstates of quantum many-body scars – which are distinguished by their
anomalously enhanced overlaps with the initial vacuum state of QED. Each dot represents an eigenstate of energy $E$ in the U(1) QLM with $\mu=0$ known to exhibit resonant scarring. Fig. adapted from \cite{Desaules2022weak}.}
    \label{fig:my_label}
    \end{subfigure}    
\end{figure}

Given that quantum many-body scarring is also known not to be stable against perturbations in case of no stabilization scheme\cite{Surace2020} and the recent success of linear gauge protection in stabilizing gauge symmetry against gauge breaking errors, it was a natural question whether this scheme can protect scarred dynamics in quantum simulations of gauge theories. Therefore, recently it was established that such a  scheme also protects scarring dynamics against coherent gauge breaking errors\cite{halimeh2022robust} Specifically, it was numerically shown, using Krylov subspace-based methods for $L=12$ sites,  that the revivals in fidelity, which is a characteristic of scarring, were perfectly restored over all investigated timescales at experimentally accessible protection strengths $V\gtrsim8J$. Moreover, this was also reflected in the dynamics of persistent oscillations well beyond the relevant timescales in local observables, such as chiral condensate/electric flux, which are another characteristic of scarring. Upon employing gauge protection, these oscillations were restored at moderate values of $V$ when subjected to gauge breaking errors. Hence, these results also showed an intimate connection between QMBS and gauge invariance.

\subsection{Scarred dynamics in the presence of incoherent errors due to 1/f noise}

Proceeding along the same lines, we investigate the quench dynamics of the vacuum state of QED (denoted by $\ket{\psi_{0}^{\text{vac}}}$) under the influence of $1/f$ noise. We then perform a quench of this state using $\hat{H}_0+V\hat{H}_G$ at $\mu=0$ in the presence of $1/f$ noise, which is known to cause scarring behavior, also called resonant scarring, first observed as long-lived oscillations in the Rydberg-atom setup. Our objective is to explore how decohering errors due to noise affect scarring dynamics, and whether the single-body protection scheme can restore its characteristic features.

Specifically, we first compute the dynamics of a global quantity which is the fidelity, particularly sensitive to errors. This can be written as 
\begin{align}\label{eq:fidelity}
    \mathcal{F}(t)=\left\langle\psi_{0}^{\text{vac}}\left|\hat{\rho}(t)\right| \psi_{0}^{\text{vac}}\right\rangle,
\end{align}
where $\hat{\rho}(t)$ is the time evolved density matrix . 

In the absence of errors, we can see a consistent revival in fidelity that lasts for all numerically accessible timescales. This indicates nonthermal behavior in an ideal situation, as shown by the purple curve. The revival period depends on the energy spacing between the scar eigenstates of the quench Hamiltonian, which are almost equally spaced.~\cite{Turner2018}.

Upon introducing noise (results shown for $\gamma=0.1J$) without protection ($V=0$), the fidelity exhibits ergodic behavior, quickly decaying to zero; see thick black curve in Fig~\ref{scar2}. Upon employing the gauge protection scheme~\eqref{eq:HG}, however, we find that nonthermal behavior is restored in the fidelity $\propto{V}$.

One can also compute another prominent feature of scarred dynamics, which are local observables. We calculate the familiar chiral condensate $n(t)=\frac{1}{2}+\frac{1}{2L} \sum_{j=1}^{L}\Tr\big\{\hat{\rho}(t)\hat{\sigma}^z_j\big\}$, whose dynamics were also computed in the previous chapter when quantifying gauge violation, measuring how strongly the chiral symmetry related to fermions in the model is spontaneously broken-

   As shown in Figure~\ref{scar3}, for unprotected errors ($V=0$), the system quickly relaxes to its thermal value. However, under linear gauge protection, oscillations are restored, and at higher values of $V$, the system shows excellent quantitative agreement with the ideal case (yellow dotted curves). This behavior is observed for all investigated evolution times, indicating the restoration of scarring dynamics.
   
\begin{figure}[t!]
    \centering
    \includegraphics[scale=0.4]{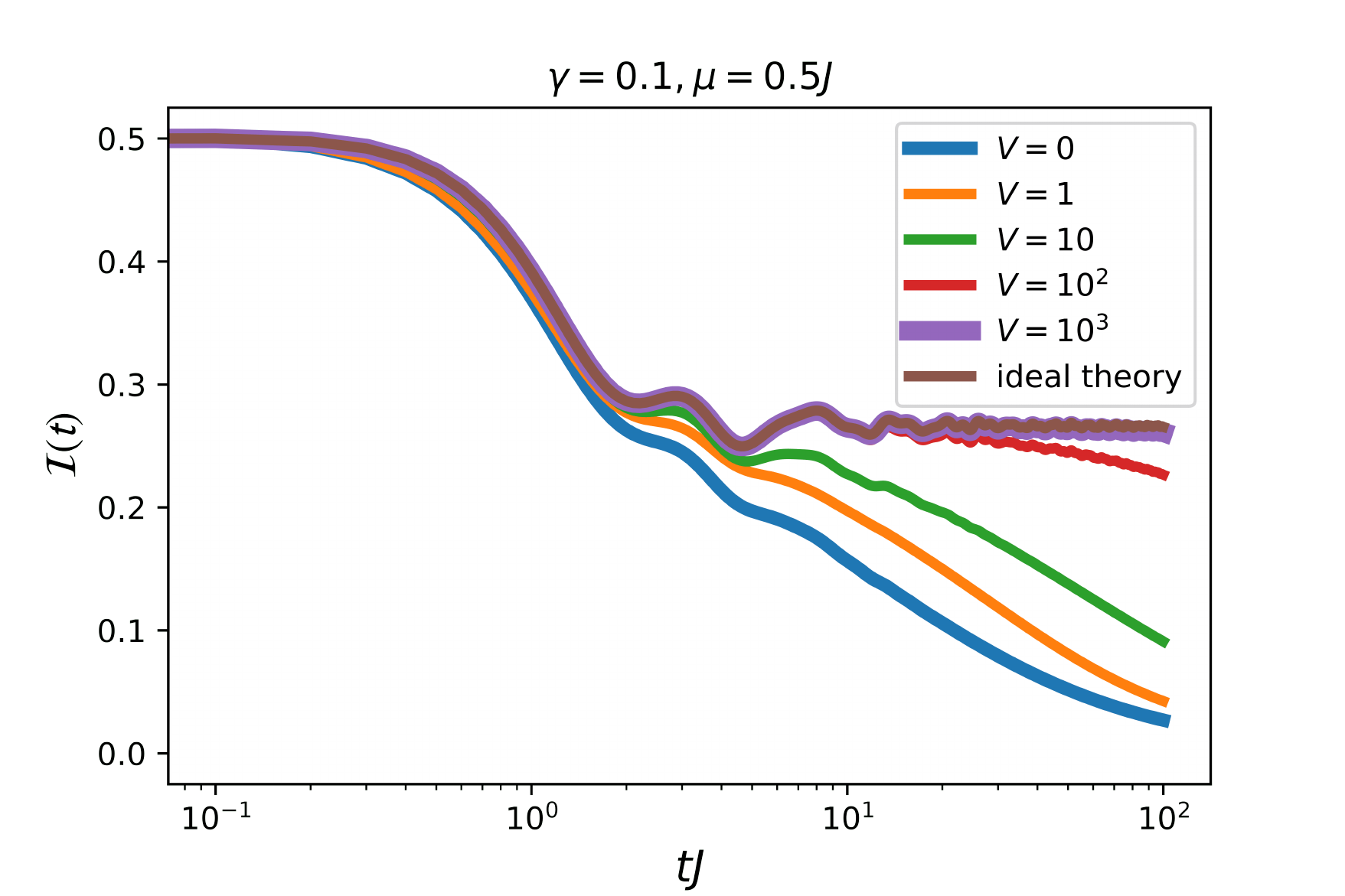}
   
    \caption{Dynamics of the mid-chain entanglement in presence of $1/f$ noise upon employing linear gauge protection. For sufficiently large $V$ the entanglement entropy is suppressed, signifying that the dynamics stay longer in the low-entropy subspace, which is a characteristic of scarred dynamics.}
    \label{Scar1}
\end{figure}

\begin{figure}[!htb]
    \centering
    \begin{subfigure}[b]{0.49\textwidth}
        \centering
        \includegraphics[width=\textwidth]    {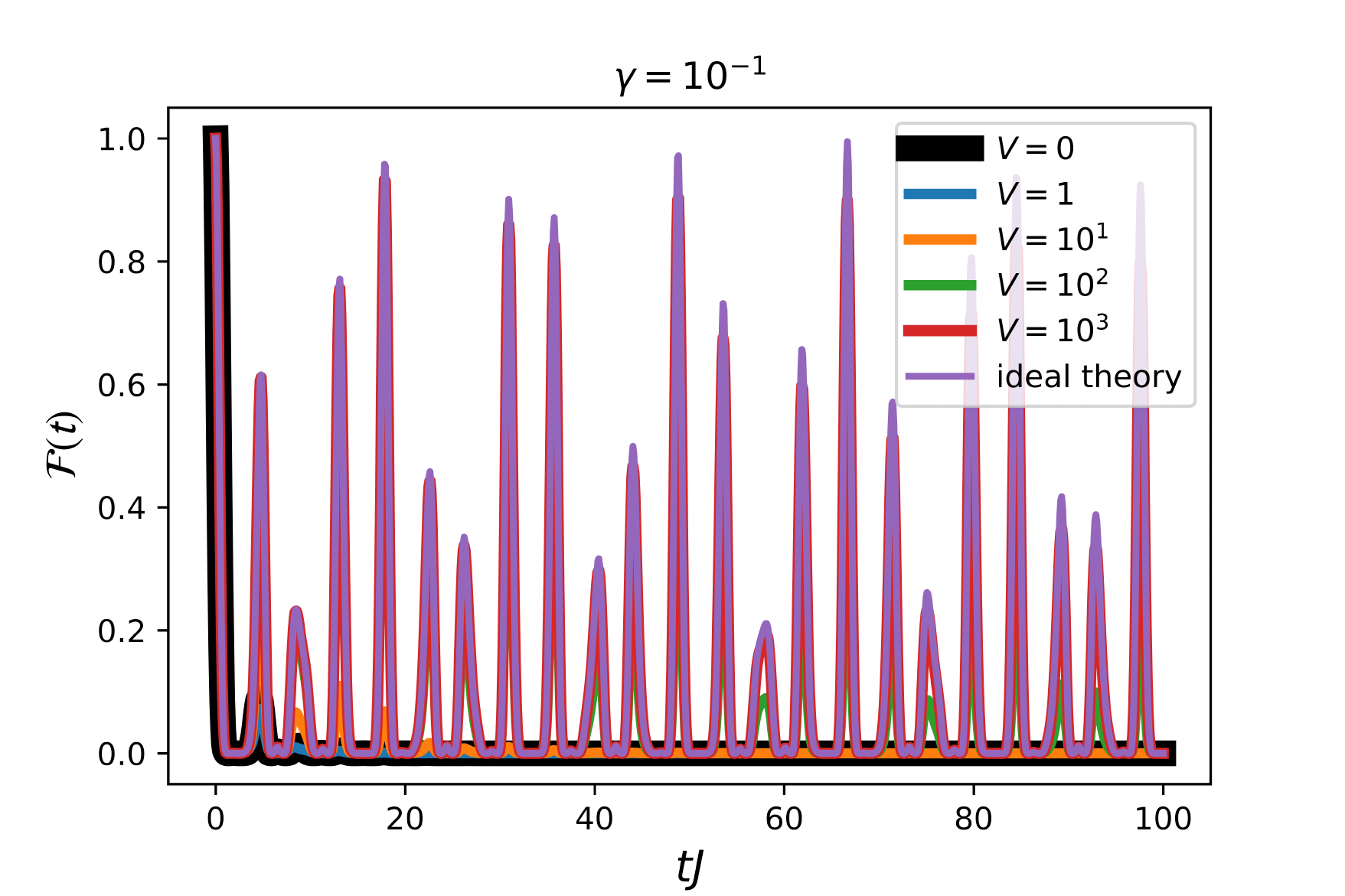}
         \caption{Periodic revivals in fidelity in the presence of noise and linear gauge protection for several values of $V$ when quenched with the vacuum state $ \ket{\psi_{0}^{\text{vac}}}$. The fidelity is restored to its ideal value as we progressively increase the protection strength. }
        \label{scar2}
    \end{subfigure}
    \begin{subfigure}[b]{0.49\textwidth}
        \centering
        \includegraphics[width=\textwidth]  {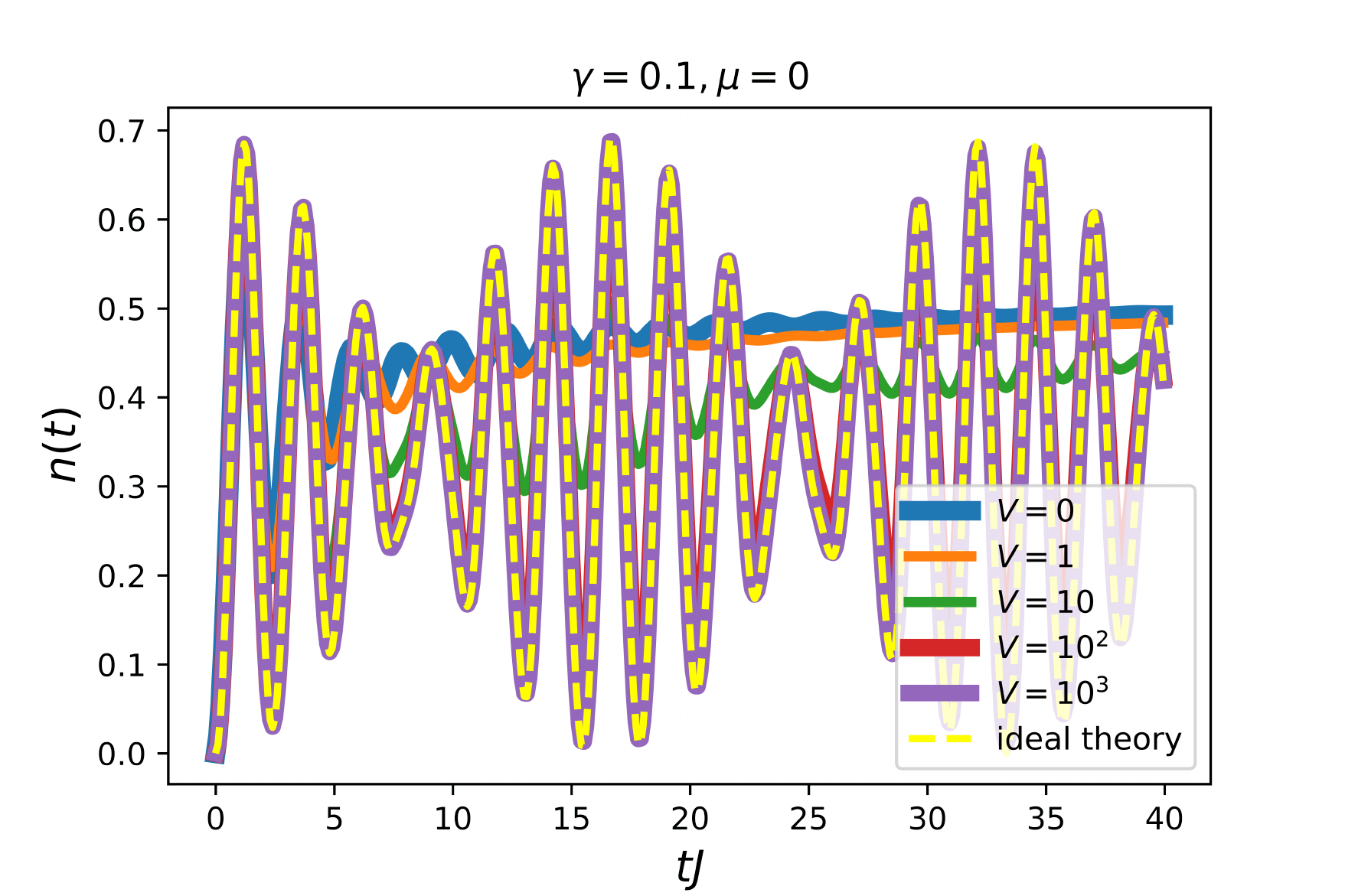}
   
        \caption{Scarring captured through a local observable, the chiral condensate. The persistent oscillations are revived upon employing gauge protection, which otherwise thermalizes in presence of noise and $V=0$}
        \label{scar3}
    \end{subfigure}    
\end{figure}

At this point, one can clearly foresee that there is a direct connection between the restoration of scarred dynamics and gauge invariance, due to the suppression of growth of gauge violations which goes as $\gamma/V$. This implies that leakage out of the target gauge sector exposes its quantum many-body scars to other subspaces in the total Hilbert space that couple with the scars. This exposure leads to thermal behavior when no protection is employed. This behavior is identical to that of its coherent counterpart \cite{halimeh2022robust}.

Additionally taking into account the mid-chain entanglement entropy $\mathcal{S}_{L/2}(t)$ behavior, as depicted in Fig.~\ref{Scar1}, we observe that in the ideal case  $\gamma=V=0$, the scarred dynamics display a remarkably low mid-chain entanglement entropy with a slower growth rate when compared to the same quench in the presence of $1/f$ noise. This difference in growth rate indicates that the dynamics are not thermalizing, as the latter case shows volume law growth, a characteristic of thermalization.

However, when linear gauge protection is employed, the entanglement entropy is suppressed and its growth rate is slowed down. As a result, the dynamics remain in the low-entropy subspace for an extended period. This observation indicates that the linear gauge protection mechanism plays a crucial role in maintaining the scarred dynamics by inhibiting the system from thermalizing for the case of $1/f$ noise too.

\chapter{Conclusions and outlook}\label{chap5}

\renewcommand\thefigure{\thechapter.\arabic{figure}}    
\renewcommand{\theequation}{\thechapter.\arabic{equation}}
\setcounter{equation}{0}
\setcounter{figure}{0}
In this thesis, we have studied an experimentally feasible scheme to effectively engineer gauge symmetries in the quantum simulation of lattice gauge theories in current and near-term NISQ devices.

In particular, we have demonstrated numerically how linear gauge protection schemes based on the local full generator or on the local pseudogenerator can suppress the growth of gauge violations due to $1/f$ noise with power spectrum $S(\omega)=\gamma/\lvert\omega\rvert^\beta$ as     $\varepsilon(t)\propto\gamma t/V^\beta$ in gauge-theory quantum simulations, where $V$ is the protection strength. This extends coherent lifetimes by $V^\beta$ in experiments where $1/f$ noise is the dominant source of decoherence. As examples, we have used two paradigmatic Abelian systems: the $\mathrm{U}(1)$ quantum link model and the $\mathbb{Z}_2$ lattice gauge theory. We have shown numerically, and argued analytically through time-dependent perturbation theory, that whereas without protection the gauge violation and errors in local observables evolve $\propto \gamma t$ in the presence of $1/f$ noise, under linear gauge protection this dynamics changes to $\propto\gamma t/V^\beta$.

Linear gauge protection may also help in suppressing $1/f$ noise sources in recent cold-atom experiments, where long coherent evolution times have been demonstrated \cite{Zhou2021}. This is due to the fact that the perturbative mapping of the $\mathrm{U}(1)$ quantum link model onto the Bose--Hubbard quantum simulator of Refs.~\cite{Zhou2021} gives rise to a leading order term that can be rearranged into a term equivalent to Eq.~\eqref{eq:HG}, with a site-dependent sequence $c_j$ \cite{Lang2022stark}.
\\
\newline
Exploring the potential of further suppressing the gauge violations can be an intriguing direction of research by introducing time-dependent variations to the time-independent sequence $c_j$. Notably, previous studies have demonstrated that, in the case of uncorrelated white noise sources, the leakage out of the target subspace can be delayed \cite{Stannigel2014,Hauke2013}. This opens the possibility of formulating the linear gauge protection scheme in the context of dynamical decoupling, which has been known to induce the Quantum-Zeno effect in open-quantum systems. By applying appropriate control pulses, this approach can help mitigate the undesirable effects of environmental interactions and average them out, as an alternative to the conventional formulation based on energy gap protection (EGP).
\\
\newline
Another intriguing extension of our work could involve modeling a more realistic system-bath coupling, taking into account non-Markovian effects and/or different bath spectral functions. It has been noted in the literature that different microscopic configurations of the environment, even with the same spectra, may correspond to distinct physical phenomena. Hence, relying solely on the knowledge of the noise spectrum is insufficient to fully describe the impact of the environment on the quantum dynamics of the system. To address this, more detailed models for the noise source are needed, which may require advanced and computationally intensive numerical tools  beyond the standard Bloch-Redfield approach outlined in Chapter 3. One potential approach could involve solving the exact master equation and associated non-equilibrium Green's functions for the open quantum many-body system. This method has been demonstrated to accurately capture the decoherence dynamics of a superconducting resonator coupled to an electromagnetic reservoir with $1/f$ noise at finite temperature, where a comprehensive quantum description of the environment was presented.
\\
\newline
Throughout this thesis, we have employed a description of a bath that results in a noise power spectrum of the environment, allowing us to account for gauge-breaking errors arising from dissipative dynamics in real systems. However, an intriguing approach would be to reverse this perspective and manipulate the dissipative dynamics to prepare gauge-symmetric quantum states as steady states of open systems. This could involve designing the environment intentionally, with dissipators that are fully controllable and can even be driven in time. This research direction is commonly referred to as "reservoir engineering," where the goal is to engineer the properties of the environment to achieve desired quantum states and dynamics in open systems \cite{PhysRevLett.77.4728,MULLER20121}.
\\
\newline
Furthermore, in this work, we also demonstrated the power of linear gauge protection  in stabilizing exotic far-from-equilibrium phenomena occurring in gauge theories, namely quantum-many body scars and disorder-free localization against incoherent errors arising from $1/f^\beta$ noise. Taking quantum simulations of gauge theories from the realm of high-energy physics to condensed matter. Especially, for the case of Quantum Many-Body scar states which are highly non-thermal, it is noteworthy that the system will retain its information without losing coherence if it is initially prepared in such states. This particular aspect is significant to quantum memory and information processing applications.
\\
\newline
Our findings offer the promising prospect of engineering gauge protection terms that can be feasibly implemented with fewer requirements than the ideal gauge theory itself and can suppress the growth of gauge violations due to $1/f$-like noise sources. We expect our conclusions to hold in higher spatial dimensions, as well as for other generic Abelian gauge theories. Another possible direction is the extension of these schemes for the case of non-abelian gauge theories\cite{Halimeh2020d,Halimeh_2022} to facilitate the quantum simulation of the holy grail of particle physics which is (3+1) QCD. Investigating dissipation in these settings presents significant challenges due to the exponentially-large Hilbert space and the requirement to compute the full-density matrix. However, there are more sophisticated numerical techniques available, such as quantum trajectories \cite{PhysRevLett.68.580,RevModPhys.70.101} and tensor networks \cite{Jaschke_2019}, which can partially address these challenges.

\end{document}